\documentclass[11pt,a4paper]{amsart}
\usepackage{amssymb, amsmath}
\usepackage{graphicx,color}
\usepackage{url}

\usepackage{mathrsfs}

\newtheorem{theorem}{Theorem}[section]
\newtheorem{lemma}[theorem]{Lemma}           
\newtheorem{cor}[theorem]{Corollary}
\newtheorem{prop}[theorem]{Proposition}

\theoremstyle{definition}
\newtheorem{definition}[theorem]{Definition}

\theoremstyle{remark}

\numberwithin{equation}{section}


\title[Resonant scattering for tunable QWs on graphs with tails]
{Resonant scattering for tunable quantum walks on graphs with tails}

\author[K. Higuchi]{Kenta Higuchi}
\address[K. Higuchi]{Faculty of Education,
Gifu University, Yanagido 1-1, Gifu, 501-1193, Japan}
\email{higuchi.kenta.b7@f.gifu-u.ac.jp}

\author[R. Ishikawa]{Ryuta Ishikawa}
\address[R. Ishikawa]{ABeam Systems Ltd., Ushijima-cho 6-1, Nishi-ku, Nagoya, Aichi, 451-6019, Japan}
\email{ryutaishikawary@gmail.com}

\author[H. Morioka]{Hisashi Morioka}
\address[H. Morioka]{Graduate School of Science and Engineering,
Ehime University, Bunkyo-cho 3, Matsuyama, Ehime, 790-8577, Japan}
\email{morioka.hisashi.ya@ehime-u.ac.jp}

\author[E. Segawa]{Etsuo Segawa}
\address[E. Segawa]{Graduate School of Environment and Information Sciences,
Yokohama National University, Tokiwadai 79-7, Hodogaya-ku, Yokohama, Kanagawa, 240-8501, Japan}
\email{segawa-etsuo-tb@ynu.ac.jp}

\author[E. Yoshimura]{Eijirou Yoshimura}
\address[E. Yoshimura]{SCSK Minori Solutions Ltd., Toyosu 3-2-20, Koto-ku, Tokyo 135-0061, Japan}
\email{eichanyoshimu@gmail.com}

\date{\today}

\begin{document}
\maketitle

\begin{abstract} 
We study the resonant scattering for discrete time quantum walks on graphs with some tails.
In our arguments, we reduce the study of resonances to the perturbation of eigenvalues of a finite rank matrix associated with the internal graph.
Then we can apply Kato's perturbation theory of matrices, and the reduction process of generalized eigenspaces allows us to derive an explicit asymptotic expansion of the scattering matrix.
As a consequence, we obtain the resonant scattering at resonant energies.
\end{abstract}

\section{Introduction}
\begin{figure}[t]
\centering
\includegraphics[width=9cm, bb=0 0 521 338]{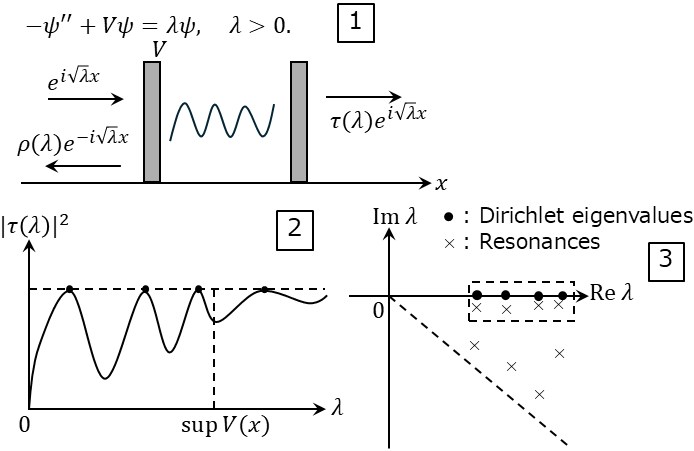}\\
\includegraphics[width=10cm, bb=0 0 813 382]{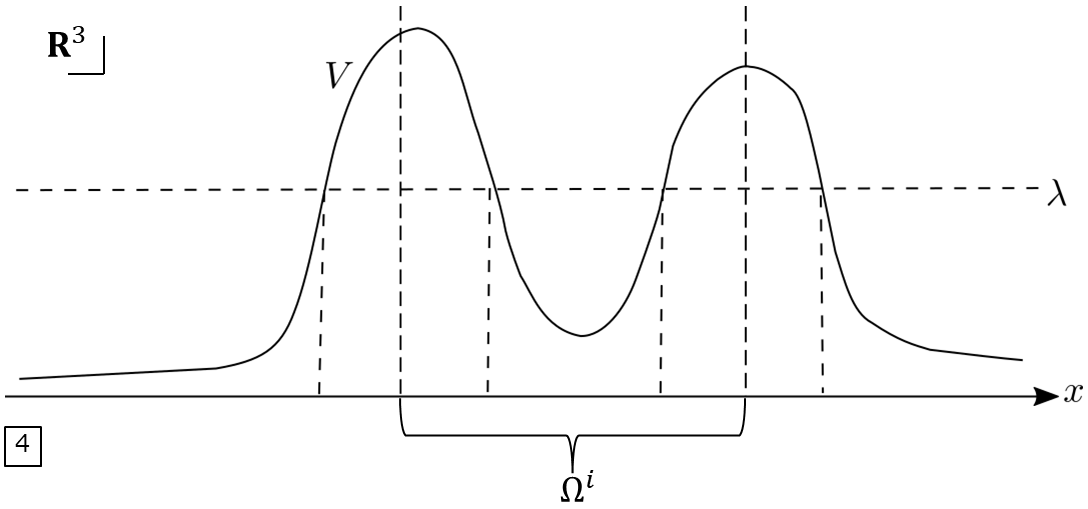}
\caption{(Figure \ref{fig_resonantscattering}-1) The generalized eigenfunction to $-\psi'' +V\psi =\lambda \psi$, $\lambda >0$, with the double barrier potential satisfies the asymptotic behavior $\psi (x)\sim e^{i\sqrt{\lambda}x}+\rho (\lambda) e^{-i\sqrt{\lambda} x}$ as $x\to -\infty$ and $\psi (x)\sim \tau (\lambda) e^{i\sqrt{\lambda}x}$ as $x\to \infty$.
(Figures \ref{fig_resonantscattering}-2, \ref{fig_resonantscattering}-3) It is well-known that $|\rho (\lambda)|^2 + |\tau (\lambda)|^2 =1$. Even though $\lambda < \sup_x V(x)$, the reflected wave vanishes for some $\lambda$.
(Figures \ref{fig_resonantscattering}-3, \ref{fig_resonantscattering}-4) A similar phenomenon is also known for the potential $V \in C^{\infty} ({\bf R}^3 )$ of the shape resonance model $H(h)= -(1/2)h^2 \Delta +V $ on ${\bf R}^3$ with small parameter $ h > 0$. If $ V (x) \to 0 $ rapidly as $|x| \to \infty$, $H(h)$ has no positive eigenvalue. However, there exist some resonances of $H(h)$ near the Dirichlet eigenvalues of $(-(1/2)h^2 \Delta +V)|_D$ in a bounded domain $\Omega^i$.
}
\label{fig_resonantscattering}
\end{figure}

In studies of scattering theory for Schr\"{o}dinger equations, the resonant scattering is one of well-known phenomena.
The most basic case appears in the double barrier model of the one-dimensional Schr\"{o}dinger equation $-\psi'' +V\psi =\lambda \psi $ with $\lambda >0$ as the resonant-tunneling effect. 
See Figure \ref{fig_resonantscattering}.
Namely, even if $0< \lambda < \sup_{x\in {\bf R}} V(x)$, there are some energies $\lambda$ such that the reflection coefficient $\rho (\lambda)$ vanishes as well as the transmission coefficient $\tau (\lambda)$ satisfies $|\tau (\lambda)|=1$ (\cite{Bo}, \cite{ChEsTsu}).
For multi-dimensional Schr\"{o}dinger operators $H(h)=-(1/2)h^2 \Delta +V$ with a small parameter $h>0$, similar phenomena are also known in the shape resonance models, and there are resonances in a small neighborhood of the corresponding Dirichlet eigenvalues associated with a suitable bounded domain (see \cite{CDKS}, \cite{Kl}, \cite{Na1,Na2}, \cite{GeMa}, \cite{Be}, \cite{DyZw}).
Resonances can be defined as a generalization of eigenvalues of the Hamiltonian $H(h)$.
Actually, the resonances of $H(h)$ are often defined by the eigenvalues of the corresponding complex distorted Hamiltonian (see \cite{AgCo}, \cite{Si}, \cite{Hu}).
These previous works imply that the resonant scattering is one of characteristics of quantum mechanics and the resonances of Hamiltonians act crucial roles.
These results suggest that the existence of resonant scattering reflects the existence of bounded classical trajectories as well as scattering classical trajectories.

In this paper, we study the resonant scattering for discrete time quantum walks (DTQWs or QWs for short) on infinite graphs which consist of a finite graph and some tails.
For DTQWs, one usually defines the time evolution of states by a unitary operator on a Hilbert space without corresponding self-adjoint Hamiltonian.
Then we consider the spectral theory for unitary operators.
A primitive form of the study of DTQWs in view of scattering theory can be seen in \cite{FeHi}.
The scattering theory for DTQWs has been developed in recent works \cite{Su}, \cite{RST,RST2}, \cite{Mo}, \cite{HiSe,HiSe2}, \cite{Ti}, \cite{KKMS}, and \cite{KKMS2}. 
As an earlier work \cite{KaKu} than these papers, we can see an abstract theory of wave operators in view of perturbations of unitary operators rather than that of self-adjoint operators. 
This is a theory of the discrete unitary group associated with the unitary operators rather than the continuous one-parameter unitary group associated with self-adjoint operators.
Resonances for DTQWs were also studied in \cite{HiMo1}, \cite{HMS} and \cite{kHig}.
An analogue of the resonant-tunneling effect was introduced by \cite{MMOS}.

As a main subject of this paper, we introduce tunable QWs which are perturbed Grover walks.
Throughout this paper, we assume that Assumptions 1 and 2 (introduced in \S \ref{section_QWwithtail}) hold true.
We add Assumption 3 in \S \ref{section_maintheorem}.
For tunable QWs, one can control the strength of the interaction between tails and the internal finite graph by a parameter.
Roughly speaking, there are controllable gates on the boundary between tails and the internal graph, and we are interested in the situation that the gates are slightly open.
As has been seen in the previous works in the resonant scattering for Schr\"{o}dinger operators, we show that the resonances of QWs have strong contributions for the behavior of scattered waves.

In our arguments, we reduce the study of resonances for QWs to that of eigenvalues for a finite rank matrices associated with the internal finite graph.
After that, we apply the perturbation theory of matrices (\cite{Ka2,Ka}) in order to derive the asymptotic expansion of resonances and the corresponding eigenprojections.
As a consequence, we can show an explicit asymptotic expansion of the scattering matrix, and this implies the resonant scattering at the resonant energies.

This paper is organized as follows.
In \S \ref{section_QWwithtail}, we introduce the details of graphs and the definition of tunable QWs on graphs.
After that, we recall the basic properties of spectra of the time evolution operator.
Since the time evolution operator of QWs is unitary, we see that its spectrum is included in $S^1$.
Here we show that the essential spectrum of the time evolution operator coincides with $S^1$ and the eigenfunctions of eigenvalues have finite supports.
In \S \ref{section_resQW}, we introduce the resonances rigorously, following the arguments of the complex translation method (\cite{HiMo1}).
The equivalence of various definintions of resonances (stated as Propositions \ref{S2_prop_exponentiallyincreasingsolution} and \ref{S3_prop_Eepev_resonance}) is proved here.
The scattering matrix is introduced in \S \ref{section_scatteringmatrix}.
In order to define the scattering matrix, we construct generalized eigenfunctions of the time evolution operator in $\ell^{\infty}$ sense.
Here we adopt the construction of the stationary state given by \cite{HiSe} instead of the method of limiting absorption principle for the resolvent operator.
This construction allows us to consider the scattering theory even if the spectral parameter coincides with an eigenvalue embedded in the essential spectrum.
Then the resonance expansion formula of the scattering matrix is given here.
In \S \ref{SS_tunableqw0000}, we reconsider the tunable QW in an auxiliary parameter and we review some fundamental results of the spectral mapping theorem for QWs (\cite{SeSu}, \cite{MOS}).
A precise exposition of our main subject is given at the end of \S \ref{SS_tunableqw0000}.
\S \ref{section_tkato} is a summary of perturbation theory and the reduction process for generalized eigenspace of matrices (\cite{Ka2,Ka}).
The asymptotic expansion of the eigenprojections are derived in an application of the perturbation theory.
In particular, the reduction process acts a crucial role.
\S \ref{section_maintheorem} is the proofs of our main theorems.
We study the tunable quantum walk in view of the resonances and the resonant scattering.
In order to show the asymptotic behavior of resonances.
The main results are Theorems \ref{S5_thm_nonresonantscattering} and \ref{S5_thm_resonantscattering}.
Some concrete examples and visualizations are given in \S \ref{section_numerical}.

The notations which are used throughout this paper are as follows.
The Kronecker delta is denoted by $\delta _{p,q} $ for some indices $p,q$.
For a unitary operator $U$, $\sigma (U)$, $\sigma_{ess} (U)$, $\sigma_p (U)$ and $\sigma_{ac} (U)$ denote the spectrum, the essential spectrum, the set of eigenvalues, and the absolutely continuous spectrum, respectively.
We use the same notations for each part of the spectrum of a self-adjoint operator.
For a linear operator $A$, $\sigma_p (A)$ is also used to represent the set of eigenvalues of $A$.
For a Hilbert space $\mathcal{H}$, ${\bf B} (\mathcal{H})$ denotes the space of bounded linear operators on $\mathcal{H}$.
The flat torus and the complex flat torus are defined by ${\bf T} : = {\bf R} /2\pi {\bf Z}$ and ${\bf T} _{{\bf C}} := {\bf C} /2\pi {\bf Z}$, respectively.
For $a>0$, we put $\mathcal{O}_a^{\pm} = \{ \kappa \in {\bf T}_{{\bf C}} \ ; \ \pm ( \mathrm{Im} \, \kappa -a )>0 \} $.


\section{Quantum walks on graphs with tails} \label{section_QWwithtail}
\subsection{Infinite graphs with tails}
Let $\Gamma = ( V,A ) $ be a connected and infinite graph with the set $V$ of vertices and the set $A$ of oriented edges.
For $a\in A$, the origin and the terminus are denoted by $o(a)$ and $t(a)$, respectively, i.e., $a=(o(a),t(a)) $.
Throughout this paper, we assume that $a\in A$ if and only if $\overline{a} \in A$ where $\overline{a} = (t(a),o(a))$ for $a=(o(a),t(a))$.

\begin{figure}[t]
\centering
\includegraphics[width=11cm, bb= 0 0 671 364]{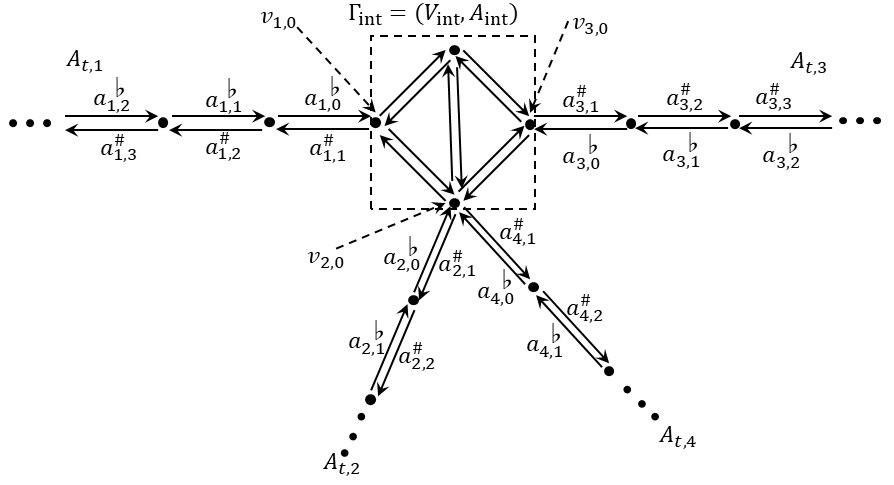}
\caption{An example of $\Gamma = (V,A)$ with four tails. The internal graph acts as a perturbation. The closed paths of the internal graph are analogues of the bounded classical trajectories for Schr\"{o}dinger operators.
Note that some tails can have a common boundary vertex.
In this case, we have $v_{2,0} = v_{4,0} $.}
\label{fig_graphwithtails}
\end{figure}

Let $X \sqcup Y$ denote the disjoint union of sets $X$ and $Y$ such that $X\cap Y=\emptyset $.
Suppose that 
$$
V=V_{\mathsf{int}} \cup V_{\mathsf{t},1} \cup \cdots \cup V_{\mathsf{t},N} , \quad A=A_{\mathsf{int}} \sqcup A_{\mathsf{t},1} \sqcup \cdots \sqcup A_{\mathsf{t},N} ,
$$ 
for a positive integer $N$ where the following properties hold.
\begin{enumerate}
\item The subgraph $\Gamma_{\mathsf{int}} =( V_{\mathsf{int} },A_{\mathsf{int}} )$ is connected and finite.

\item
$V_{\mathsf{t},j} = \{ v_{j,k} \ ; \ k=0,1,2,\ldots \} $ and $v_{j,0} \in V_{\mathsf{t},j}\cap V_{\mathsf{int}} $ for $j=1,\ldots,N$. 

\item
$A_{\mathsf{t},j} = \{ a_{j,k}^{\flat} , a_{j,l}^{\sharp} \ ; \ k=0,1,2,\ldots , \ l=1,2,\ldots \} $ with $a_{j,k}^{\flat} = (v_{j,k+1},v_{j,k} ) $ and $a_{j,l}^{\sharp} = (v_{j,l-1},v_{j,l} )$.
By this definition, we have $\overline{a^{\flat}_{j,l-1} }=a_{j,l}^{\sharp} $.

\item
For some $j,k\in \{ 1,\ldots,N\}$, $v_{j,0}=v_{k,0}$ is allowed. 
Namely, some tails can connect to a common boundary vertex.
\end{enumerate}
In view of the above properties, we call $a_{j,k}^{\flat}$ and $a_{j,l}^{\sharp}$ an \textit{incoming edge} and an \textit{outgoing edge}, respectively.
The infinite subgraph $\Gamma_{\mathsf{t},j} = ( V_{\mathsf{t},j} , A_{\mathsf{t},j} ) $ for every $j=1,\ldots,N$ is a \textit{tail}.
See Figure \ref{fig_graphwithtails}.

For every vertex $v\in V$, we define the subsets $A _v^{\flat} $ and $ A_v^{\sharp} $ by 
$$
A_v^{\flat} = \{ a\in A \ ;  \ t(a)=v\} , \quad A_v^{\sharp} = \{ a\in A \ ; \ o(a)=v \} .
$$
The assumption for $\Gamma$ guarantees $\# A_v^{\flat} $ and $ \# A_v^{\sharp} $ to coincide with each other.
Thus we put 
$$
\mathrm{deg} (v)= \# A_v^{\flat} = \# A_v^{\sharp} , \quad v\in V .
$$

\subsection{Quantum walks on graphs}
The totality of ${\bf C}$-valued functions on $A$ is denoted by ${\bf C}^A$.
We define the Hilbert space 
$$
\mathcal{H}= \ell^2 (A) = \{ \psi \in {\bf C}^A \ ; \ \| \psi \| _{\mathcal{H}} < \infty \} ,
$$ 
by the norm $\| \psi \| _{\mathcal{H}}^2 = \langle \psi , \psi \rangle_{\mathcal{H}} $ associated with the inner product $\langle \psi , \phi \rangle _{\mathcal{H}} $ for $\psi = \{ \psi (a)\}_{a\in A} , \phi = \{ \phi (a) \} _{a\in A} \in {\bf C}^A$ given by 
$$
\langle \psi , \phi \rangle _{\mathcal{H}} = \sum_{a\in A} \psi (a) \overline{\phi (a)} .
$$
The sets ${\bf C}^{A_{\mathsf{int}}}$, ${\bf C}^{A_{\mathsf{t},j}}$ and the Hilbert spaces $\ell^2 (A_{\mathsf{int}} )$, $\ell^2 (A_{\mathsf{t},j})$ are defined in the similar way.
In the following arguments, we also often use the notation 
$$
\mathcal{H}_{\mathsf{tail}} = \oplus_{j=1}^N \ell^2 (A_{\mathsf{t},j} ),
$$
and we naturally identify ${\bf C}^{A_{\mathsf{int}}}$ with $\ell^2 (A_{\mathsf{int}})$.

For $\psi \in {\bf C}^A$, we define the shift operator $S$ by 
\begin{equation}
S \psi (a)= \psi ( \overline{a} ), \quad a\in A.
\label{S2_eq_shiftoperator}
\end{equation}
Obviously, $S$ is self-adjoint and unitary on $\ell^2 (A)$.
The coin operator $C$ is defined by 
\begin{equation}
C \psi (a)= \sum _{b\in A_{t(a)}^{\flat}} c_{a,b}^{t(a)} \psi (b) , \quad a\in A, \quad \psi \in {\bf C}^A ,
\label{S2_eq_coinoperator}
\end{equation}
where the $\mathrm{deg} (v) \times \mathrm{deg} (v)$ matrix $c^v = [c_{a,b}^v ] _{a,b\in A_v^{\flat} }$ is unitary for every vertex $v\in V$.
Now we define the time evolution operator of the QW by 
\begin{equation}
U:= SC
\label{S2_eq_qwoperator}
\end{equation} 
for the QW on $\Gamma$.
By definition, we have 
$$
U\psi (a)= \sum_{b\in A_{t(\overline{a})}^{\flat}} c_{\overline{a},b}^{t(\overline{a})} \psi (b) = \sum_{b\in A_{o(a)}^{\flat} } c_{\overline{a},b}^{o(a)} \psi (b),  \quad a\in A, \quad \psi \in {\bf C}^A .
$$

Now we impose an assumption for $U$ on the tails $A_{\mathsf{t},j}$. 

\medskip

{\bf Assumption 1.} The operator $U$ satisfies 
\begin{equation}
U\psi (a_{j,k}^{\flat} )= \psi (a_{j,k+1}^{\flat} ), \quad U\psi (a_{j,l+1}^{\sharp} )=\psi (a_{j,l}^{\sharp} ), \quad k\geq 0 , \quad l\geq 1 , \quad j=1,\ldots,N,
\label{S2_eq_freewalk}
\end{equation}
which means 
$$ 
c^{v_{j,k}} = \begin{bmatrix} 0 & 1 \\ 1 & 0 \end{bmatrix} ,\quad k\geq 1 .
$$

\medskip

In order to study the scattering theory for QWs, we consider $U$ as a perturbation of the free QW $U_0$.
Namely, we introduce the following assumption.

\medskip

%
%
%
%
{\bf Assumption 2.}
The operators $U$ and $U_0$ satisfy the following properties.
\begin{enumerate}
\item
$U $ and $U_0 $ are unitary operators on $\mathcal{H}$.

\item
$U$ and $U_0$ satisfy Assumption 1.

\item 
$U_0 = \widetilde{U} _0 \oplus U _{\mathsf{int}} $ on $\mathcal{H}_{\mathsf{tail}} \oplus {\bf C}^{A_{\mathsf{int}} }$ where $\widetilde{U}_0$ satisfies 
$$
\widetilde{U}_0 \psi (a_{j,1}^{\sharp} )=\psi (a_{j,0}^{\flat} ) , \quad j=1,\ldots,N,
$$ 
and $U_{\mathsf{int}}$ is a unitary operator on ${\bf C}^{A_{\mathsf{int}}}$.

\end{enumerate}

\medskip

\textit{Remark.}
Assumption 2-(3) implies that there is no interaction between $ \psi |_{A_{t,j}}$ and $\psi|_{A_0}$ by $U_0$.
On the other hand, for $U $, we consider a case where there is an interaction between $ \psi |_{A_{\mathsf{t},j}}$ and $\psi|_{A_{\mathsf{int}}}$ at $v_{j,0} \in V_{\mathsf{t},j} \cap V_{\mathsf{int}}$.
The QW given by $\widetilde{U}_{0}$ is purely free QW on every tail $A_{\mathsf{t},j}$.
The operator $U_{\mathsf{int}}$ defines a unitary time evolution on the $  \# A_{\mathsf{int}} $-dimensional vector space.
Then we can identify $U_{\mathsf{int}}$ with a $\# A_{\mathsf{int}} \times \# A_{\mathsf{int}}$ unitary matrix.
Obviously, $U_0$ has some eigenvalues embedded in $S^1$ which are also eigenvalues of $U_{\mathsf{int}} $.
The assertion (2) in Assumption 2 implies that $U -U_0$ is of finite rank.

\medskip

Let us introduce a tunable QW as an example of QWs under Assumptions 1-2. 
Here tunable QWs are given as $U=U_{\epsilon}$ for the parameter $\epsilon \in [0,1]$ based on the Grover walk.
This is the main subject of consideration in this paper. 
The scattering theory of tunable quantum walks are going to be discussed in \S \ref{SS_tunableqw0000}-\S \ref{section_numerical}.
The Grover walk is defined by the coin operator $C$ with the Grover matrix 
\begin{equation}
g^v = \frac{2}{\mathrm{deg} (v)} J _{\mathrm{deg}(v)} -I_{\mathrm{deg }(v)}, \quad v\in V,
\label{S2_eq_grovermatrix000}
\end{equation}
where $J_n$ is the $n\times n$ all-ones matrix and $I_n$ is the $n\times n$ identity matrix.
We construct $ U_{\epsilon} $ as a perturbation of the Grover walk at vertices $v_{j,0} $, $j=1,\ldots,N$.
For $\epsilon \in [0,1] $ and a positive integer $n$, we put
$$
G_{n,\epsilon} = \frac{1}{n} J_n - e^{-i\pi \epsilon} \left( I_n -\frac{1}{n} J_n \right) .
$$
Note that $ G_{n,0} = 2n^{-1} J_n - I_n$ and $G_{n,1} = I_n $.
For any $\epsilon \in [0,1]$, $G_{n,\epsilon} $ is an $n\times n$ unitary matrix.
For every $ v_{1,0} , \ldots , v_{N,0} $, we replace the matrix of the Grover walk by  
\begin{equation}
g_{\epsilon}^{v_{j,0}} = \begin{bmatrix} G_{n(v_{j,0})-N_j ,\epsilon} & 0 \\  0 & I_{N_j} \end{bmatrix} G_{n(v_{j,0}),1-\epsilon} , 
\label{S2_eq_tunablegrovermatrix00}
\end{equation}
where $n(v)=\mathrm{deg} (v)$ for $v\in V$ and $N_j $ is the number of tails connected to $v_{j,0}$.
For the case $v_{j,0} = v_{k,0}$, we have $g_{\epsilon}^{v_{j,0}} = g_{\epsilon}^{v_{k,0}}$.
Here we choose the basis of ${\bf C}^{ A_{v_{j,0}}^{\flat}} $ so that 
$$
[\psi (a_1) , \ldots , \psi (a_{n(v)-N_j}) , \psi (a_{n(v)-N_j +1}) , \ldots , \psi (a_{n(v)} )]^{\mathsf{T}} 
$$ 
with $a_1 ,\ldots , a_{n(v)-N_j} \in A_{\mathsf{int}} \cap A_{v_{j,0}}^{\flat}$ and $a_{n(v)-N_j +1},\ldots,a_{n(v)} \in  A_{v_{j,0}}^{\flat} \setminus A_{\mathsf{int}}$.
Then we have 
\begin{gather}
\lim_{\epsilon \downarrow 0} g_{\epsilon}^{v_{j,0}}  = \begin{bmatrix} 2(n(v)-N_j)^{-1} J_{n(v)-N_j} -I_{n(v)-N_j} & 0 \\ 0 & I_{N_j}  \end{bmatrix} =: g_0^{v_{j,0}} ,  \label{S2_eq_tunablegrovermatrixo01} \\ 
\lim_{\epsilon \uparrow 1} g_{\epsilon}^{v_{j,0}} = \begin{bmatrix} I_{n(v)-N_j} & 0 \\ 0 &  I_{N_j} \end{bmatrix} (2n(v)^{-1} J_{n(v)} -I_{n(v)}) =g^{v_{j,0}} . \label{S2_eq_tunablegrovermatrixo02}
\end{gather}
We see that the QW $U_0$ defined by the coin operator $C_0$ associated with the matrices
\begin{gather*}
c^v = \left\{
\begin{split}
g^v &, \quad v\in V \setminus \{ v_{1,0} , \ldots , v_{N,0} \} , \\ 
g_0^v &, \quad v\in \{ v_{1,0} , \ldots , v_{N,0} \} ,
\end{split}
\right.
\end{gather*}
satisfies (3) in Assumption 2.
Then $U_0 = SC_0$ is the free QW based on the Grover walk.
For $\epsilon >0$, $U_{\epsilon} = SC_{\epsilon}$ with the matrices 
\begin{gather}
c_{\epsilon}^v = \left\{
\begin{split}
g^v &, \quad v\in V \setminus \{ v_{1,0} , \ldots , v_{N,0} \} , \\ 
g_{\epsilon}^v &, \quad v\in \{ v_{1,0} , \ldots , v_{N,0} \} ,
\end{split}
\right.
\label{S2_eq_tunablegrovermatrix01}
\end{gather}
is a perturbation of $U_0$ which has a small interaction at every $v_{j,0}$.
The strength of the interaction at $v_{j,0}$ is tuned by the parameter $\epsilon\in (0,1]$.

\subsection{Spectra} \label{subsection_spectrum}
As has been seen in the general theory of unitary operators, we have $\sigma (U)\subset S^1$.
The spectrum of $\widetilde{U}_0 $ on $\mathcal{H}_{\mathsf{tail}} $ can be studied in a method of Fourier analysis as follows.
Let $\mathcal{J}_j :\ell^2 (A_{\mathsf{t},j})\to \ell^2 ({\bf Z})$ be the unitary operator defined by 
\begin{gather*}
\mathcal{J}_j \psi (x)= \left\{
\begin{split}
\psi (a_{j,-x}^{\sharp} ) &, \quad x\leq -1 , \\
\psi (a_{j,x}^{\flat} ) &, \quad x\geq 0 ,
\end{split}
\right. \quad \psi \in \ell^2 (A_{\mathsf{t},j}).
\end{gather*}
Then $\mathcal{H}_{\mathsf{tail}} $ can be identified with $ \oplus_{j=1}^N \ell^2 ({\bf Z} )$. 
The operator $\widetilde{U}_0 |_{A_{\mathsf{t},j}} $ is unitary equivalent to the shift operator $L$ on ${\bf Z}$  defined by $L u (x)=u(x+1)$ for $u\in \ell^2 ({\bf Z} )$.

\begin{lemma}
We have $\sigma (\widetilde{U}_0 )=\sigma_{ac} (\widetilde{U}_0)=S^1 $.
\label{S2_lem_spectrumfreeQW0}
\end{lemma}

Proof.
By using the Fourier series on ${\bf Z}$
$$
\mathcal{F}u (\xi )= (2\pi )^{-1/2} \sum_{x\in {\bf Z}} e^{-ix\xi} u(x), \quad \xi \in {\bf T},
$$ 
we see that $\widehat{L}=\mathcal{F}L\mathcal{F}^{-1}$ is the operator of multiplication by the function $e^{i\xi}$. 
It follows that $\sigma (L)=\sigma_{ac} (L)=S^1$.
Due to $\widetilde{U}_0 |_{A_{\mathsf{t},j}} = \mathcal{J}_j ^{-1} L  \mathcal{J}_j $, we have $\sigma (U_0 |_{A_{\mathsf{t},j}})=\sigma_{ac} (U_0 |_{A_{\mathsf{t},j}})=S^1$.
In view of $\widetilde{U}_0 = \oplus_{j=1}^N \widetilde{U}_0 |_{A_{\mathsf{t},j}} $, we obtain the lemma.
\qed

\medskip

The absolute continuity of spectra is unstable under compact perturbations.
However, we can see the stability of the essential spectrum of $U$.
\begin{lemma}
We have $\sigma_{ess} (U)=\sigma_{ess} (U_0)=\sigma_{ess} (\widetilde{U}_0 )=S^1$ and $\sigma_p (U_0 ) = \sigma_p (U_{\mathsf{int}} )$.

\label{S2_lem_spectrumperturbed}
\end{lemma}

Proof.
We take $\lambda \in {\bf T}$ arbitrarily.
In view of Lemma \ref{S2_lem_spectrumfreeQW0}, there exists a sequence $\psi _n$ in $\mathcal{H}_{\mathsf{tail}}$ such that $\| \psi_n \| _{\mathcal{H}_{\mathsf{tail}}} =1$, $\| (\widetilde{U}_0 -e^{-i\lambda} )\psi_n \| _{\mathcal{H}_{\mathsf{tail}}} \to 0$, $\psi_n \to 0$ weakly in $\mathcal{H}_{\mathsf{tail}}$ as $n\to \infty $.
Let $\widetilde{\chi}^* : \mathcal{H}_{\mathsf{tail}} \to \mathcal{H}$ be defined by 
\begin{gather*}
\widetilde{\chi}^* \psi (a)= \left\{
\begin{split}
\psi (a)&, \quad a\in A_{\mathsf{t},1} \sqcup \cdots \sqcup A_{\mathsf{t},N} , \\ 
0 &, \quad a\in A_{\mathsf{int}} ,
\end{split}
\right. \quad \psi \in \mathcal{H}_{\mathsf{tail}} .
\end{gather*}
Then we have 
$$
( U_0 -e^{-i\lambda} ) \widetilde{\chi}^* \psi _n = \widetilde{\chi}^* (\widetilde{U}_0 -e^{-i\lambda} ) \psi_n + ( U_0 \widetilde{\chi}^* -  \widetilde{\chi}^*  \widetilde{U}_0  )\psi_n \to 0
$$
in $\mathcal{H}$ as $n\to \infty$, since the operator $ U_0 \widetilde{\chi}^* -  \widetilde{\chi}^*  \widetilde{U}_0 $ is of finite rank on $\mathcal{H}$.
By taking a suitable normalization, we can assume $\| \widetilde{\chi}^* \psi_n \|_{\mathcal{H}} =1 $.
It follows $e^{-i\lambda} \in \sigma_{ess} (U_0)$ as well as $\sigma_{ess} (U_0)=S^1 $ from the above argument.
We can show $\sigma_{ess} (U)=S^1$ in the same way.
$\sigma_p (U_0 ) = \sigma_p (U_{\mathsf{int}} )$ is trivial.
\qed

\medskip

The operator $U$ possibly has some eigenvalues embedded in $S^1$.
The number of eigenvalues and its multiplicities are as follows.

\begin{lemma}
The number of eigenvalues of $U$ is at most $\# A_{\mathsf{int}}$ taking into account the multiplicities.
The associated eigenfunctions vanish in $A\setminus A_{\mathsf{int}}$.
\label{S2_lem_eigenvaluesU}
\end{lemma}

Proof.
Suppose that $e^{-i\lambda} \in \sigma_p (U)$ for $\lambda \in {\bf T}$.
Let $\psi \in \mathcal{H}$ be an eigenfunction of $U$ associated with $e^{-i\lambda}$.
In view of Assumption 1 and the equation $U\psi = e^{-i\lambda} \psi$, we have 
\begin{gather}
|\psi (a_{j,k}^{\flat} )| = |U \psi (a_{j,k}^{\flat})|=| \psi (a_{j,k+1}^{\flat} )|,\quad k\geq 0 , \label{S2_eq_eigentail01} \\
|\psi (a_{j,l}^{\sharp})|=|U \psi (a_{j,l}^{\sharp})|=| \psi (a_{j,l-1}^{\sharp} )|,\quad l\geq 2, \label{S2_eq_eigentail02}
\end{gather} 
for any $j=1,\ldots ,N$.
Since $\psi \in \mathcal{H}$, we have $\psi (a_{j,k}^{\flat})\to 0$ as $k\to \infty$ and $\psi (a_{j,l}^{\sharp} )\to 0$ as $l\to \infty $.
The equalities (\ref{S2_eq_eigentail01}) and (\ref{S2_eq_eigentail02}) imply $\psi (a_{j,k}^{\flat} )=0$ for $k\geq 0$ and $\psi (a_{j,l}^{\sharp} )$ for $l\geq 1$.
It follows that $\mathrm{supp} \, \psi \subset A_{\mathsf{int}} $.
We obtain the lemma.
\qed

%

\section{Resonances for quantum walks}
\label{section_resQW}
\subsection{Complex distortion}
In order to define resonances for $U$, there are some equivalent definitions.
One is to introduce a complex distortion of $U$.
A complex distortion of QWs on the square lattice has been introduced in \cite{HiMo1}.
In this paper, we generalize it for QWs on graphs with tails as follows.
On ${\bf Z}$, we define
$$
\widetilde{Q} (\theta)u(x)= e^{i\theta x} u(x), \quad x\in {\bf Z} , \quad \theta \in {\bf T} .
$$
By the definition, $\widetilde{Q} (\theta)$ is a unitary operator on $\ell^2 ({\bf Z} )$.
The shift operator $L$ on ${\bf Z}$ introduced in \S\ref{subsection_spectrum} satisfies 
$$
L(\theta) = \widetilde{Q} (\theta ) L \widetilde{Q} (\theta )^{-1} = e^{-i\theta}  L \quad \text{on} \quad {\bf Z} .
$$
By using $\mathcal{J}_j$, $j=1,\ldots,N$, we define the unitary operator $Q(\theta)$ for $\theta \in {\bf T} $ by
\begin{gather}
Q(\theta)\psi (a)= \left\{
\begin{split}
\mathcal{J}_j^{-1} \widetilde{Q}(\theta) \mathcal{J}_j \psi (a) &, \quad a\in A_{\mathsf{t},j}, \quad j=1,\ldots,N, \\
\psi (a) &, \quad a\in A_{\mathsf{int}} ,
\end{split}
\right.  \quad   \psi \in \mathcal{H}.
\label{S2_eq_realdistortion}
\end{gather}
Now we introduce the real distortion $U (\theta)$ by 
\begin{equation}
U (\theta) = Q(\theta) U Q(\theta)^{-1} , \quad \theta \in {\bf T} .
\label{S2_eq_realdistortion}
\end{equation}
In particular, we have 
\begin{equation}
U_0 (\theta)= e^{-i\theta} \widetilde{U}_0 \oplus U_{\mathsf{int}} \quad \text{on} \quad \mathcal{H}_{\mathsf{tail}} \oplus {\bf C}^{A_{\mathsf{int}}} .
\label{S2_eq_realdistortionU0}
\end{equation}
In view of (\ref{S2_eq_realdistortion}) and (\ref{S2_eq_realdistortionU0}), the operator $U (\theta)$ can be extended to $\theta \in \mathcal{O}^-_0$ in ${\bf B} (\mathcal{H})$.
Since $U (\theta)$ is a normal operator for $\theta \in \mathcal{O}^-_0$ in ${\bf B} (\mathcal{H})$, we immediately see the following fact.
\begin{lemma}
For $\theta\in \mathcal{O}_0^-$, we have $\sigma_{ess} (U (\theta))= e^{\mathrm{Im} \, \theta} S^1 $.
\label{S2_lem_essentialspectrum_complexdistortion}
\end{lemma}

Let
$$
R_0 (z)= (U_0 -e^{-iz } )^{-1} , \quad R (z)= (U -e^{-iz} )^{-1} ,
$$
for $z\in \mathcal{O}^+_0 \cup \mathcal{O}^-_0$.
The standard resolvent equations hold true as 
\begin{equation}
R (z)=R_0 (z)(1-W R (z))=(1-W R (z))R_0 (z), \quad W = U -U_0 .
\label{S2_eq_resolventeq}
\end{equation}
In the similar way, we put
$$
R(z,\theta) =(U (\theta) -e^{-iz} )^{-1} , \quad R_0 (z,\theta)= ( U_0 (\theta)-e^{-iz} )^{-1} , \quad e^{-iz} \not\in \sigma (U (\theta))\cup \sigma (U_0 (\theta)),
$$
for $\theta\in \mathcal{O}_0^-$ and we can see that the resolvent equations
\begin{gather}
\begin{split}
&R (z,\theta)=R_0 (z,\theta)(1-W (\theta) R (z,\theta))=(1-W (\theta) R (z,\theta))R_0 (z,\theta), \\
&W (\theta)= U (\theta) -U_0 (\theta),
\end{split}
\label{S2_eq_resolventeq2}
\end{gather}
hold true. 
Note that $W (\theta)$ is a finite rank operator for any $\theta \in {\bf T}\cup \mathcal{O}_0^-$.
In view of (\ref{S2_eq_realdistortionU0}), we also have 
$$
R_0 (z,\theta)= (e^{-i\theta} \widetilde{U}_0 \oplus U_{\mathsf{int}} -e^{-iz} )^{-1} =e^{i\theta} (\widetilde{U}_0 \oplus  e^{i\theta} U_{\mathsf{int}} -e^{-i(z-\theta)})^{-1} =: e^{i\theta} R_{0,1} ( z,\theta ) .
$$
By some direct calculations with (\ref{S2_eq_resolventeq}) and (\ref{S2_eq_resolventeq2}), we can see 
$$
(1-WR (z))(1+WR_0 (z)) = (1+W R_0 (z))(1-W R (z))=1,
$$
for $z\in \mathcal{O}_0^+ \cup \mathcal{O}_0^- $, and
$$
(1-W (\theta) R (z,\theta ))(1+W (\theta ) R_0 (z,\theta )) = (1+W (\theta) R_0 (z,\theta))(1-W (\theta) R(z,\theta))=1,
$$
for $z\in \mathcal{O}_0^+ \cup \mathcal{O}_0^- $ and $\theta \in {\bf T}$.
Then it follows that $1-WR(z)$ and $1-W (\theta)R (z,\theta) $ are invertible for $z\in \mathcal{O}_0^+ \cup \mathcal{O}_0^-$ and $\theta \in {\bf T}$.
The equations (\ref{S2_eq_resolventeq}) and (\ref{S2_eq_resolventeq2}) are replaced by 
\begin{gather}
R (z)=R_0 (z)(1+W R_0 (z))^{-1}  , \label{S2_eq_resolventeq3} \\ 
R (z,\theta )=R_0 (z,\theta )(1+W(\theta) R_0 (z,\theta ))^{-1}  , \label{S2_eq_resolventeq4}
\end{gather}
for $z\in \mathcal{O}_0^+ \cup \mathcal{O}_0^-$ and $\theta \in {\bf T}$.

Now we can see that the complex distortion $Q(\theta)$ define the meromorphic extention of the resolvent operator $R (z)$ in the following sense.
Let
\begin{gather*}
\mathcal{D} = \cap_{\gamma >0} \left\{ \psi \in {\bf C}^A \ ; \ \begin{split} &e^{\tau k } |\psi (a_{j,k}^{\flat}) | \to 0 , \ k\to \infty , \\ &e^{\tau l } |\psi (a_{j,l}^{\sharp}) | \to 0, \ l\to \infty, \end{split} \ \tau\in (0,\gamma), \ j=1,\ldots,N \right\} .
\end{gather*}
We define the function $F_{\phi} (z) $ for $z\in \mathcal{O}_0^+$ and $\phi \in \mathcal{D}$ by 
$$
F_{\phi} (z)= \langle R (z)\phi ,\phi \rangle_{\mathcal{H}} .
$$
We put $\Omega_{\theta}^+ = \mathrm{exp} (-i\mathcal{O}^+_{\mathrm{Im} \, \theta}) =\{ e^{-iz} \ ; \ z\in\mathcal{O}^+_{\mathrm{Im} \, \theta}\} $. 

\begin{lemma} 
The subset $\mathcal{D}$ is dense in $\mathcal{H}$.
For any $\phi\in \mathcal{D}$ and any $a<0$, the function $F_{\phi} (z)$ has a meromorphic extension to $\mathcal{O}_a^+$ with poles of finite rank.
Poles of $F_{\phi} (z)$ lie on ${\bf T}$ or in $\mathcal{O}_0^- \cap \mathcal{O}_a^+$ and are invariant with respect to $\theta \in \mathcal{O}_0^-$ as long as poles lie in the region $\Omega ^+_{\theta} $.

\label{S2_lem_meromorphicextension}
\end{lemma}

Proof.
The proof is parallel to \cite[Lemma 2.5]{HiMo1}.
For the sake of completeness of our argument, we reproduce the proof here.
Since $Q(\theta)$ defined by (\ref{S2_eq_realdistortion}) is unitary on $\mathcal{H}$ for $\theta \in {\bf T}$, we have 
\begin{gather}
\begin{split}
F_{\phi} (z)&= \langle R (z)\phi ,\phi \rangle_{\mathcal{H}} \\
&= \langle R (z,\theta )Q(\theta)\phi, Q(\theta)\phi \rangle_{\mathcal{H}} \\
&= \langle R_0 (z,\theta)(1+W (\theta)R_0 (z,\theta))^{-1} Q(\theta)\phi,Q(\theta)\phi \rangle_{\mathcal{H}} ,
\end{split}
\label{S2_eq_meromorphicproof00}
\end{gather}
for $z\in \mathcal{O}_0^+ $, due to the equation (\ref{S2_eq_resolventeq4}).

We fix $z_0 \in \mathcal{O}_0^+$ and $\phi \in \mathcal{D}$.
Note that $\mathrm{Im} (z_0 - \theta )>0$ for $\theta \in \mathcal{O}_{\mathrm{Im} \, z_0 }^-$.
Then $W (\theta) R_0 (z_0 , \theta )=e^{i\theta} W (\theta) R_{0,1} (z_0 , \theta )$ is of finite rank and analytic with respect to $\theta \in \mathcal{O}_{\mathrm{Im} \, z_0}^- $.
The existence of the inverse $(1+W (\theta)R_0 (z_0 ,\theta ))^{-1}$ for $\theta\in {\bf T} \subset \mathcal{O} _{\mathrm{Im} \, z_0 }^- $ (see (\ref{S2_eq_resolventeq4})) and the analytic Fredholm theory \cite[Theorem C.8]{DyZw} imply the existence of the meromorphic extension of $(1+W (\theta)R_0 (z_0 ,\theta ))^{-1}$ to $\theta \in \mathcal{O} _{\mathrm{Im} \, z_0}^- $ with poles of finite rank.
Thus the function
$$
G_{\phi} (z_0 ,\theta )= \langle R_0 (z_0 ,\theta)(1+W (\theta)R_0 (z_0 ,\theta))^{-1} Q(\theta)\phi,Q(\theta)\phi \rangle_{\mathcal{H}} 
$$
of $\theta$ is meromorphic in $\mathcal{O}_{\mathrm{Im} \, z_0}^- $.
Furthermore, the equality (\ref{S2_eq_meromorphicproof00}) for $\theta \in {\bf T} $ implies $G_{\phi} (z_0,\theta )=F_{\phi} (z_0 , \theta )$ for $\theta \in {\bf T} $.  
Then it follows that $ G_{\phi} (z_0 , \theta )$ is a constant for $\theta \in \mathcal{O}_{\mathrm{Im} \, z_0}^- $.
Now we obtain $ G_{\phi} (z_0 , \theta )=F_{\phi} (z_0)$ for any $\theta \in \mathcal{O}_{\mathrm{Im} \, z_0}^- $.
As a consequence, we see that $(1+W(\theta) R_0 (z,\theta))^{-1}$ is analytic with respect to $\theta \in \mathcal{O}^-_{\mathrm{Im} \, z_0}$.

Next we fix $\theta_0 \in \mathcal{O}_0^- $ and $\phi \in \mathcal{D}$.
As has been seen in the above argument, we have $G_{\phi} (z,\theta_0 )=F_{\phi } (z)$ for any $z\in \mathcal{O}_0^+ $.
Note that $\mathrm{Im} (z-\theta_0 ) >0$ for $z\in \mathcal{O}_{\mathrm{Im} \, \theta_0}^+  $.
We see that $R_0 (z,\theta_0 )=e^{i\theta_0} R_{0,1} ( z,\theta_0 ) $ and $W (\theta_0 ) R_{0} (z,\theta_0 )=e^{i\theta_0}  W (\theta_0 ) R_{0,1} ( z,\theta_0)$ are analytic with respect to $z\in \mathcal{O}_{\mathrm{Im} \, \theta_0}^+ $.  
The existence of the inverse $(1+W (\theta_0 )R_0 (z,\theta_0 ))^{-1} $ for some $z\in \mathcal{O}_0^+ $ has been shown in the above arguments, since $(1+W (\theta )R_0 (z,\theta ))^{-1} $ is analytic with respect to $\theta \in \mathcal{O}_{\mathrm{Im} \, z}^-$.
The analytic Fredholm theory (\cite[Theorem C.8]{DyZw}) implies that $(1+W (\theta_0 )R_0 (z,\theta_0 ))^{-1} $ has the meromorphic extension to $z\in \mathcal{O}_{\mathrm{Im} \, \theta_0}^+ $ with poles of finite rank.
Finally, we obtain the meromorphic extension of $F_{\phi} (z)$ to $z\in \mathcal{O}_{\mathrm{Im} \, \theta_0}^+$ with poles of finite rank.
Since $\theta_0 \in \mathcal{O}_0^-$ is arbitrary, we obtain the meromorphic extension of $F_{\phi} (z)$ to $z\in \mathcal{O}_a^+$ for any $a<0$.

Let us show the invariance of poles of $F_{\phi} (z)$ with respect to $\theta\in\mathcal{O}_0^- $.
Suppose that $\lambda (\theta)$ is a discrete eigenvalue of $U (\theta)$ in the region $\Omega ^+_{\theta} $.
The eigenvalue $\lambda (\theta)$ depends analytically on $\theta\in \mathcal{O}_a^+$ as long as $\lambda (\theta)$ belongs to $\Omega^+_{\theta}$.
For any $\alpha\in {\bf T}$, $U (\theta+\alpha )$ is unitary equivalent to $U (\theta)$.
Then we have $\lambda (\theta+\alpha )=\lambda (\theta)$ for any $\alpha \in {\bf T}$.
By the analyticity of $\lambda (\theta)$, we see that $\lambda (\theta)$ is constant as long as $\lambda (\theta )$ belongs to $\Omega_{\theta}^+$.
\qed

\medskip

In view of Lemma \ref{S2_lem_meromorphicextension}, now we can introduce a rigorous definition of \textit{resonances} of $U$.
Note that the resonances are independent of choice of $\theta\in \mathcal{O}^-_0$ as long as they are in the subset $\Omega_{\theta}^+$ even though they are defined as eigenvlues of $U (\theta)$.

\begin{definition}
Suppose that $\theta\in \mathcal{O}_0^-$ is fixed.
We call $e^{-i\lambda} $ for $\lambda \in \mathcal{O}_{\mathrm{Im} \, \theta}^+$ a \textit{resonance} of $U$ if $e^{-i\lambda} \in \sigma_p (U(\theta)) $.

\label{S2_def_resonance}
\end{definition}

\textit{Remark.}
In \cite{HMS}, the resonances are defined as poles of the cut-off resolvent operator $\chi  R(z) \chi $ where $\chi$ is the characteristic function of the subset $A_{\mathsf{int}}$.
We can see that this definition is equivalent to Definition \ref{S2_def_resonance}.

\subsection{Outgoing solutions}
The resonances are related to the existence of nontrivial outgoing solutions to the equation $U \psi = e^{-i\lambda} \psi $ as follows.
Here the solution $\psi\in {\bf C}^{A}$ is \textit{outgoing} if $\psi$ satisfies 
$$
\psi (a _{j,k}^{\flat} )=0, \quad k\geq 0, \quad j=1,\ldots,N.
$$

\begin{prop}
There exists a nontrivial outgoing solution $\psi \in {\bf C}^A$ to the equation $U \psi = e^{-i\lambda} \psi $ if and only if $e^{-i\lambda}  $ for $\lambda \in \mathcal{O}_0^-$ is a resonance of $U $.
%
%

\label{S2_prop_exponentiallyincreasingsolution}
\end{prop}

Proof.
If $e^{-i\lambda}$ is a resonance of $U$, there is an eigenfunction $\phi\in \mathcal{H}$ such that $U (\theta)\phi =e^{-i\lambda} \phi$ for $\theta \in  \mathcal{O}^-_{\mathrm{Im} \, \lambda}$.
In view of the definition of $U(\theta)$ by (\ref{S2_eq_realdistortion}) and Assumptions 1-2, we have 
$$
e^{-i\theta} \phi (a_{j,k+1}^{\flat} )= e^{-i\lambda} \phi (a_{j,k}^{\flat} ) , \quad e^{-i\theta} \phi (a_{j,l-1}^{\sharp} )= e^{-i\lambda} \phi (a_{j,l}^{\sharp} ),
$$
for $k\geq 0$, $l\geq 2$, and $j=1,\ldots,N$.
Precisely, we have $ \phi (a_{j,k}^{\flat} )= e^{-ik(\lambda -\theta)} \psi (a_{j,0}^{\flat} )$ and $\phi (a_{j,l}^{\sharp})=e^{i(l-1)(\lambda -\theta)}\phi (a_{j,1}^{\sharp} )$ for any $k\geq 0$ and $l\geq 1$, respectively. 
Note that $|e^{-ik(\lambda - \theta)}|$ is exponentially increasing as $k\to \infty$ and $|e^{i(l-1)(\lambda -\theta)}|$ is exponentially decreasing as $l\to \infty$, due to $\theta \in \mathcal{O}^-_{\mathrm{Im}\,\lambda}$.
Since $\phi\in\mathcal{H}$, we have $\lim_{k\to \infty} \phi (a_{j,k}^{\flat} )=0$. 
Then $\phi (a_{j,0}^{\flat} )$ has to vanish for all $j=1,\ldots,N$.
Letting $\psi = Q(\theta)^{-1} \phi$, we obtain $U\psi =e^{-i\lambda} \psi$, and $\psi$ is an outgoing solution.

Let us turn to the proof of the converse.
For an outgoing solution $\psi\in {\bf C}^A$ to the equation $U \psi = e^{-i\lambda} \psi$, we put $\phi =Q(\theta)\psi $.
Then $\phi$ satisfies $U (\theta)\phi = e^{-i\lambda} \phi$.
Let us show $\phi \in \mathcal{H}$.
Since $\psi $ is a solution to $U \psi = e^{-i\lambda} \psi$, we have 
$$
\psi (a_{j,l-1}^{\sharp} ) = e^{-i\lambda} \psi (a_{j,l}^{\sharp} ), \quad l\geq 2, \quad j=1,\ldots,N.
$$
Then we have $|\psi (a_{j,l}^{\sharp} )|= e^{- (l-1) \mathrm{Im} \, \lambda} |\psi (a_{j,1}^{\sharp} )| $ for any $l\geq 1$.
If we take $\theta \in \mathcal{O} _{\mathrm{Im} \, \lambda}^-$, we can see $\phi =Q(\theta)\psi \in \mathcal{H}$.
\qed

\medskip

Another characterization of resonances is given by the $\# A_{\mathsf{int}} \times \# A_{\mathsf{int}} $ matrices 
\begin{equation}
E = \chi_{\mathsf{int}} U \chi_{\mathsf{int}}^* , \quad E_0 = \chi_{\mathsf{int}} U_0 \chi_{\mathsf{int}}^* ,
\label{S3_eq_matrixE0}
\end{equation} 
where $\chi_{\mathsf{int}} : {\bf C}^A \to {\bf C}^{A_{\mathsf{int}}}$ is defined by $\chi_{\mathsf{int}} \psi (a)= \psi (a)$ for $a\in A_{\mathsf{int}}$ and $\chi_{\mathsf{int}}^* : {\bf C}^{A_{\mathsf{int}}} \to {\bf C}^A$ is defined by 
\begin{gather*}
\chi_{\mathsf{int}}^* \psi (a)= \left\{
\begin{split}
\psi (a) &, \quad a\in A_{\mathsf{int}} , \\
0 &, \quad a\in A\setminus A_{\mathsf{int}} .
\end{split}
\right.
\end{gather*}

\begin{lemma}
We have $|e^{-i\lambda}|\leq 1$ if $e^{-i\lambda} \in \sigma _p (E)  $.
Especially we have $\sigma_p (E_0 )\subset S^1 $.
\label{S3_lem_evinsideS1}
\end{lemma}

Proof.
Suppose that $e^{-i\lambda} \in \sigma_p (E )$.
Then there exists a normalized eigenvector $u\in {\bf C}^{A_{\mathsf{int}}}$ such that $E u=e^{-i\lambda} u$.
We have 
$$
|e^{-i\lambda}| ^2 =\| E u \|^2 _{\ell^2 (A_{\mathsf{int}} )} \leq \| U \chi_{\mathsf{int}}^* u \| _{\mathcal{H}}^2 = \| \chi_{\mathsf{int}}^* u \|^2 _{\ell^2 (A_{\mathsf{int}} )} =1 . 
$$
This inequality implies $|e^{-i\lambda}|\leq 1$.
Since $E_0$ is unitary on $\ell^2 (A_{\mathsf{int}})$, we obtain $\sigma_p (E_0 )\subset S^1 $.
\qed

\begin{prop}
For $\lambda\in \mathcal{O}_0^-$, $e^{-i\lambda}$ is a resonance of $U$ if and only if $e^{-i\lambda} \in \sigma_p (E)$.

\label{S3_prop_Eepev_resonance}
\end{prop}

Proof.
For the proof, we apply Proposition \ref{S2_prop_exponentiallyincreasingsolution} and Lemma \ref{S3_lem_evinsideS1}.
Suppose that $e^{-i\lambda}$ with $\lambda \in \mathcal{O}_0^- $ is a resonance of $U$.
Then there exists an outgoing solution $\psi \in {\bf C}^A$ to the equation $U \psi = e^{-i\lambda} \psi $.
Letting $u=\chi_{\mathsf{int}} \psi \in {\bf C}^{A_{\mathsf{int}}}$, we have 
$$
E u = \chi_{\mathsf{int}} U \psi + \chi_{\mathsf{int}} U (\chi_{\mathsf{int}}^* \chi_{\mathsf{int}} -1) \psi = e^{-i\lambda} u+ \chi_{\mathsf{int}} U (\chi_{\mathsf{int}}^* \chi_{\mathsf{int}} -1 )\psi .
$$
Since $\psi (a_{j,0}^{\flat} )=0$ for any $j=1,\ldots,N$, we obtain $\chi_{\mathsf{int}} U (\chi_{\mathsf{int}}^* \chi_{\mathsf{int}} -1 )\psi  =0$.
Thus $e^{-i\lambda} \in \sigma_p (E )$ and $u$ is an associated eigenvector of $E$.

Conversely, suppose that $e^{-i\lambda} \in \sigma_p (E )$ with $\lambda \in \mathcal{O}_0^-$.
Take an associated eigenvector $u\in {\bf C}^{A_{\mathsf{int}}} $.
We can construct $\psi \in {\bf C}^A$ such that $\psi$ is an outgoing solution to $U \psi = e^{-i\lambda} \psi $ and $\chi_{\mathsf{int}} \psi = u$.
Actually we put $\psi (a_{j,k}^{\flat})=0$ and $\psi (a_{j,l}^{\sharp} )=e^{i\lambda} U\psi (a_{j,l}^{\sharp} ) $ for all $k\geq 0$, $l\geq 1$ and $j=1,\ldots,N$.
Then we obtain the desired result.
\qed

\section{Scattering matrix} \label{section_scatteringmatrix}
\subsection{Generalized eigenfunctions associated with continuous spectra}
 There are generalized eigenfunctions $\psi \in \ell^{\infty} (A)$ satisfying $U \psi = e^{-i\lambda} \psi $ for $e^{-i\lambda} \in \sigma_{ess} (U ) $.
Usually, the generalized eigenfunction are constructed in terms of the limiting absorption method for the resolvent operator, i.e., $\lim_{\rho \downarrow 0}R (\lambda \pm i\rho )$ and the distorted Fourier transform associated with the spectrum of $U $.
However, in this paper, we adopt the construction of the stationary state introduced by \cite{HiSe}.
Their method allows us to construct generalized eigenfunctions for $ U $ even if $e^{-i\lambda} \in S^1$ is an eigenvalue of $U$.
Here we derive a sketch of the proof, refering to the results in \cite{HiSe}, for the sake of completeness of this paper, and in order to verify the consistency of related notations.

\begin{figure}[b]
\centering
\includegraphics[width=8cm, bb=0 0 561 276]{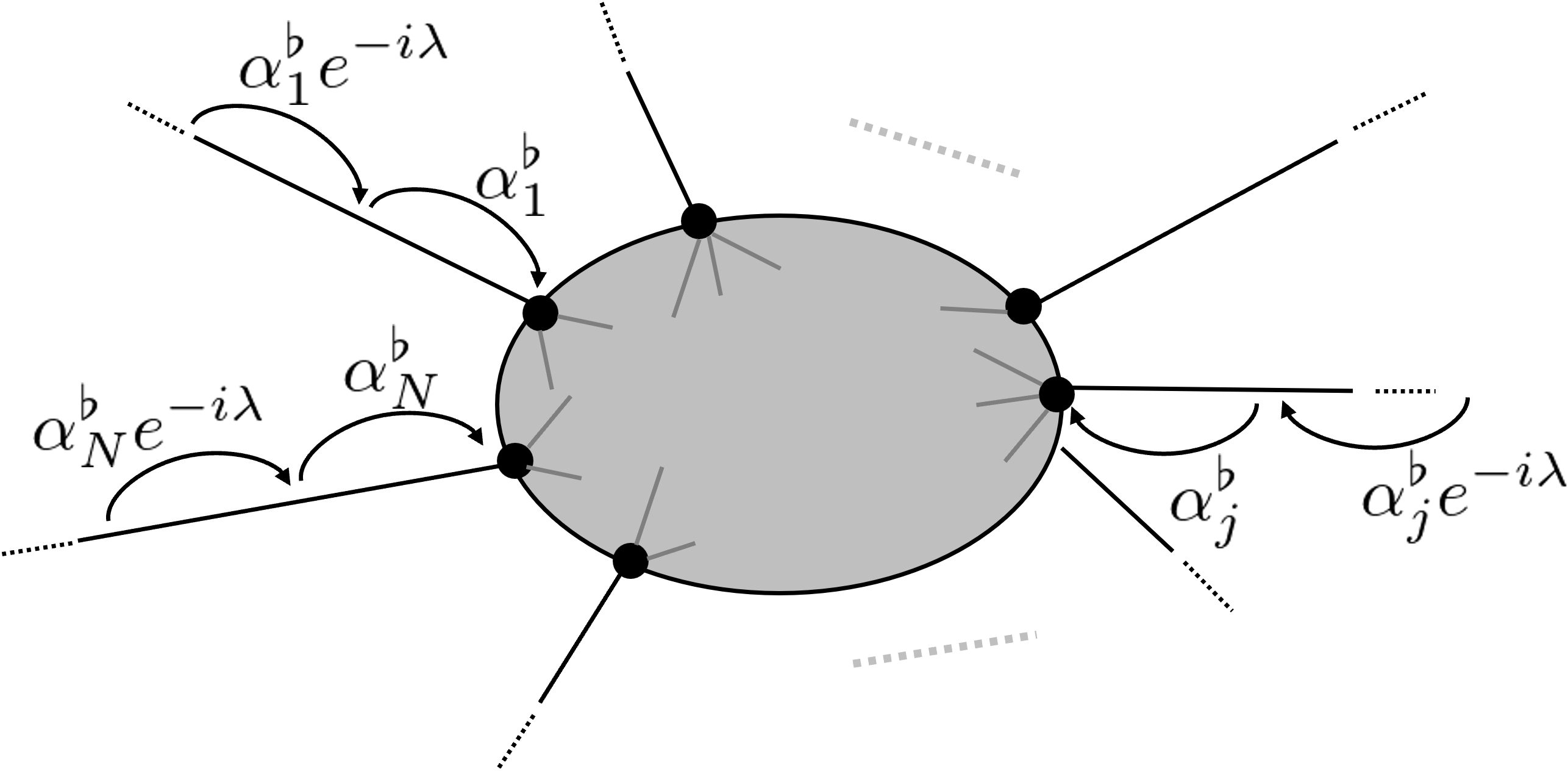}
\caption{The initial state $\varphi_0$ has its support only on incoming edges of tails.}
\label{fig_initialstate}
\end{figure}

Take a vector $[\alpha_1^{\flat} , \ldots, \alpha_N^{\flat} ]^{\mathsf{T}} \in {\bf C}^N$.
Let
\begin{gather}
\varphi_0 (a)= \left\{
\begin{split}
\alpha_j ^{\flat} e^{-i\lambda k} &, \quad a=a_{j,k}^{\flat} , \quad k\geq 0, \quad j=1,\ldots,N, \\ 
0 &, \quad \text{otherwise} ,
\end{split}
\right.
\label{S4_eq_initialincoming}
\end{gather}
for $\lambda \in {\bf T}$ and we take $\varphi_0$ as the initial state of the time evolution 
\begin{equation} 
\varphi_t = U \varphi _{t-1} , \quad t\geq 1. 
\label{S4_eq_timeevolution}
\end{equation} 
See Figure \ref{fig_initialstate}.
The following result is \cite[Theorem 3.1]{HiSe}.

\begin{prop} 
For the $t$-th iteration $\varphi_t$ defined by (\ref{S4_eq_initialincoming}) and (\ref{S4_eq_timeevolution}), we put $\psi_t = e^{i\lambda t} \varphi_t $.
There exists the limit $\psi _{\infty}(a):= \lim_{t\to \infty} \psi_t (a)$ for every $a\in A$ and $\psi_{\infty} \in \ell^{\infty} (A)$ is a solution to the equation $U\psi_{\infty} =e^{-i\lambda} \psi_{\infty} $.
\label{S4_prop_stationarystate}
\end{prop}

For the proof of Proposition \ref{S4_prop_stationarystate}, we need a lemma as follows.
This lemma has a crucial role also in Proposition \ref{S4_prop_uniqueness_data}.

\medskip

Sketch of the proof for Proposition \ref{S4_prop_stationarystate}.
Let $E$ be as has been introduced in (\ref{S3_eq_matrixE0}).
Letting $u_t = \chi_{\mathsf{int}} \varphi_t $, we have 
$$
u_t = \chi_{\mathsf{int}} U \varphi_{t-1} = E u_{t-1} + e^{-i\lambda (t-1)} \chi_{\mathsf{int}} U \varphi_0 .
$$
In the above equality, we have used the fact that $\chi_{\mathsf{int}} U (1-\chi_{\mathsf{int}}^* \chi_{\mathsf{int}} ) \varphi_{t-1} = e^{-i\lambda (t-1)} \chi_{\mathsf{int}} U \varphi_0 $ which is the inflow from every tail to the internal graph $\Gamma_{\mathsf{int}}$.
Since we have $u_0 =0$, $u_t$ satisfies 
\begin{equation}
u_{t+1} = E u_t +e^{-i\lambda t} f_0 , \quad f_0 =\chi _{\mathsf{int}} U \varphi_0 , \quad t\geq  0  , 
\label{S4_eq_iteration01}
\end{equation}
so that
\begin{equation}
u_t = e^{-i\lambda (t-1)} (1+e^{i\lambda} E + \cdots +e^{i\lambda (t-1)} E^{t-1} ) f_0 , \quad t \geq 1 .
\label{S4_eq_iteration02}
\end{equation}

For $\mu\in \sigma_p (E)$, the eigenprojection $P_{E,\mu}$ is defined by 
\begin{equation}
P_{E,\mu} = -\frac{1}{2\pi i}  \oint_{\mathcal{C} (\mu)} (E-z)^{-1} dz 
\label{S4_eq_defeigenprojection}
\end{equation}
where $\mathcal{C} (\mu)$ is the counterclockwise loop without self-intersection, and there is no other eigenvalue of $E$ inside $\mathcal{C} (\mu)$.
It is well-known that the equality $P_{E,\mu} E=EP_{E,\mu} $ holds true.
The eigennilpotent matrix $D_{E,\mu}$ is given by $D_{E,\mu}=(E-\mu)P_{E,\mu} $.
Then we can see $P_{E,\mu} D_{E,\mu} = D_{E,\mu} P_{E,\mu} = D_{E,\mu} $ and $D_{E,\mu}^{m(\mu)} =0$ for the algebraic multiplicity $m(\mu)=\mathrm{rank} (P_{E,\mu})$ of $\mu$.
In view of the spectral representation (\cite{Ka}), we have 
\begin{equation}
E = \sum _{\mu \in \sigma_p (E)} (\mu P_{E,\mu} + D_{E,\mu} ).
\label{S4_eq_spectralrepresentation00} 
\end{equation}
Precisely, we have for any $t\geq 1$
\begin{equation}
E^t = \sum_{\mu \in \sigma_p (E)} \sum_{s=0}^{\min \{ t,m(\mu)-1 \}} \begin{pmatrix} t \\ s \end{pmatrix} \mu^{t-s} P_{E,\mu}^{t-s} D_{E,\mu}^{s} , \quad \begin{pmatrix} t \\ s \end{pmatrix} = \frac{t!}{s! (t-s)!}.
\label{S4_eq_spectralrepresentation}
\end{equation}
Now we introduce the generalized eigenspaces of $E$ as 
\begin{gather*}
\mathcal{H}_c = \mathsf{Span} \{ u\in {\bf C}^{A_{\mathsf{int}}} \ ; \ u \text{ is a generalized eigenvector for an eigenvalue } \mu \text{ with } |\mu| =1 \} , \\ 
\mathcal{H}_s = \mathsf{Span} \{ u\in {\bf C}^{A_{\mathsf{int}}} \ ; \ u \text{ is a generalized eigenvector for an eigenvalue } \mu \text{ with } |\mu| <1 \} . 
\end{gather*}
Thus $\mathcal{H}_c$ and $\mathcal{H}_s$ are invariant subspaces of $E$.
Due to Lemma \ref{S3_lem_evinsideS1}, we have ${\bf C}^{A_{\mathsf{int}}} = \mathcal{H}_c \oplus \mathcal{H}_s$.
The following lemma is given as \cite[Lemmas 3.2, 3.3, 3.4, and 3.5]{HiSe}.
\begin{lemma}
The following assertions hold true.
\begin{enumerate}
\item
Let $|\mu |=1$.
A vector $u\in {\bf C}^{A_{\mathsf{int}}}$ satisfies $Eu=\mu u$ if and only if $E^* u= \mu^{-1} u $.

\item
$\mathcal{H}_c = \oplus _{|\mu|=1} \mathsf{Ker} (E -\mu )$.

\item
$\mathcal{H}_c \perp \mathcal{H}_s $.

\item
$f_0 = \chi_{\mathsf{int}} U \varphi_0  $ and $Ef_0 $ belong to $ \mathcal{H}_s$.

\end{enumerate}
\label{S4_lem_lemmaseigenfunction}
\end{lemma}

The equality (\ref{S4_eq_spectralrepresentation}), Lemmas \ref{S3_lem_evinsideS1} and \ref{S4_lem_lemmaseigenfunction} imply that the iteration (\ref{S4_eq_iteration01}) can be solved by (\ref{S4_eq_iteration02}) as 
$$
v_{\infty} = \lim_{t\to \infty} e^{i\lambda t} u_t = e^{i\lambda } \sum_{t=0}^{\infty} e^{i\lambda t} E^{t} f_0 .
$$
Then $v_{\infty}$ is the solution to the equation 
$$
(E - e^{-i\lambda} ) v_{\infty} = -f_0 .
$$

We have 
$$
\varphi_t (a_{j,1}^{\sharp} )= U \chi_{\mathsf{int}}^* u_{t-1} (a_{j,1}^{\sharp} )+e^{-i\lambda (t-1)} U \varphi_0 (a_{j,1}^{\sharp} ), \quad  j=1,\ldots,N.
$$
Then we obtain
\begin{gather}
\begin{split}
\lim_{t\to \infty} e^{i\lambda t} \varphi_t (a_{j,1}^{\sharp} ) &=\lim_{t\to \infty} \left( e^{i\lambda} U \chi_{\mathsf{int}}^* (e^{i\lambda (t-1)} u_{t-1} ) (a_{j,1}^{\sharp}  ) +e^{i\lambda} U \varphi_0 (a_{j,1}^{\sharp} ) \right) \\
&= e^{i\lambda} U \chi_{\mathsf{int}}^* v_{\infty} (a_{j,1}^{\sharp} )+ e^{i\lambda} U \varphi_0 (a_{j,1}^{\sharp} ).
\end{split}
\label{S4_eq_outgoingstateat1}
\end{gather}
Letting $\alpha_{j,t} ^{\sharp} = \varphi_t (a_{j,1}^{\sharp} )= U\chi_{\mathsf{int}}^* v_{t-1} (a_{j,1}^{\sharp} )+ e^{-i\lambda (t-1)} U\varphi_0 (a_{j,1}^{\sharp})$ for $t\geq 1$, we also have
$$ 
\varphi_t (a_{j,l}^{\sharp} )= \delta_{l,1} \alpha^{\sharp}_{j,t} + \delta_{l,2} \alpha^{\sharp}_{j,t-1} +\cdots +\delta_{l,t-1} \alpha_{j,2}^{\sharp} +\delta_{l,t} \alpha^{\sharp}_{j,1}= \sum_{k=1}^t \delta_{l,t-k+1} \alpha_{j,k}^{\sharp}=\alpha _{j,t-l+1}^{\sharp} . 
$$
Plugging (\ref{S4_eq_outgoingstateat1}) into the above equality, we have 
\begin{gather*} 
\begin{split}
\lim_{t\to \infty} e^{i\lambda t} \varphi_t (a_{j,l}^{\sharp} )&= \lim_{t\to \infty} e^{i\lambda t} \alpha _{j,t-l+1}^{\sharp}\\
&= \lim_{t\to \infty}  \left( e^{i\lambda l} U\varphi_0 (a_{j,1}^{\sharp})+ e^{i\lambda l}  U\chi_{\mathsf{int}}^* (e^{i\lambda (t-l)} v_{t-l} ) (a_{j,1}^{\sharp}) \right) \\
& = e^{i\lambda  l} \alpha_j^{\sharp} (\lambda),
\end{split} 
\end{gather*}
where 
$$ 
\alpha^{\sharp}_{j} (\lambda ) = e^{-i\lambda} \lim _{t\to \infty} e^{i\lambda t} \varphi _t (a_{j,1}^{\sharp} ) =   U \chi_{\mathsf{int}}^* v_{\infty} (a_{j,1}^{\sharp} )+   U \varphi_0 (a_{j,1}^{\sharp} ).
 $$

Now we put $\psi_{\infty} (a) = \lim _{t\to \infty} e^{i\lambda t} \varphi_{t} (a)$ for $a\in A$.
Precisely, we have 
\begin{gather}
\psi _{\infty} (a)= \left\{
\begin{split}
v_{\infty} (a) &, \quad a\in A_0 , \\ 
\alpha^{\flat}_j e^{-i\lambda k} &, \quad a=a_{j,k}^{\flat} , \quad k\geq 0 , \quad j=1,\ldots,N , \\ 
\alpha^{\sharp}_{j}(\lambda ) e^{i\lambda l} &, \quad a=a_{j,l}^{\sharp} , \quad l\geq 1, \quad j=1,\ldots,N.
\end{split}
\right.
\label{S4_eq_generalizedeigenfunction_formula}
\end{gather}
In order to verify the equation $ U \psi_{\infty } =e^{-i\lambda} \psi_{\infty} $, we compute 
\begin{gather*}
\begin{split}
U \psi_{\infty} 
= & \, \chi_{\mathsf{int}}^* E  v_{\infty} + (1-\chi_{\mathsf{int}} ^* \chi_{\mathsf{int}} )U \chi_{\mathsf{int}}^* \chi_{\mathsf{int}} \psi_{\infty} \\
&\, + \chi_{\mathsf{int}}^* \chi_{\mathsf{int}} U (1-\chi_{\mathsf{int}}^* \chi_{\mathsf{int}} )\psi_{\infty} +(1-\chi_{\mathsf{int}} ^* \chi_{\mathsf{int}} ) U (1-\chi_{\mathsf{int}}^* \chi_{\mathsf{int}} )\psi_{\infty} \\
=& \, e^{-i\lambda} \chi_{\mathsf{int}}^* v_{\infty} + (1-\chi_{\mathsf{int}}^* \chi_{\mathsf{int}} )U \psi_{\infty} ,
\end{split}
\end{gather*}
noting another equation with the outgoing part $\psi_{\infty}^{\sharp} = \psi_{\infty} - \varphi_0 -\chi_{\mathsf{int}}^* v_{\infty} $,
\begin{gather*}
\begin{split}
(1-\chi_{\mathsf{int}}^* \chi_{\mathsf{int}} )U\psi_{\infty} &= (1-\chi_{\mathsf{int}}^* \chi_{\mathsf{int}} )U (\chi_{\mathsf{int}}^* v_{\infty} + \varphi_0 + \psi_{\infty}^{\sharp} ) \\
&=(1-\chi_{\mathsf{int}}^* \chi_{\mathsf{int}} ) \left( \sum_{j=1}^N  \alpha^{\sharp}_j (\lambda )   \delta_{1}^{(\sharp,j)} + e^{-i\lambda} \varphi_0 + e^{-i\lambda} \psi_{\infty}^{\sharp} - \sum_{j=1}^N  \alpha^{\sharp}_j  (\lambda )  \delta_{1}^{(\sharp,j)} \right) \\
&= (1-\chi_{\mathsf{int}}^* \chi_{\mathsf{int}} )  e^{-i\lambda} (\psi_{\infty} - \chi_{\mathsf{int}}^* v_{\infty} ),
\end{split}
\end{gather*}
where $\delta _{1}^{(\sharp,j)} (a_{j,1}^{\sharp})=1 $ and $\delta _{1}^{(\sharp,j)} (a)=0$ for any $a\not= a_{j,1}^{\sharp} $.
Then we have 
$$
U \psi_{\infty} = e^{-i\lambda} \chi_{\mathsf{int}}^* v_{\infty} +  (1-\chi_{\mathsf{int}}^* \chi_{\mathsf{int}} )  e^{-i\lambda} (\psi_{\infty} - \chi_{\mathsf{int}}^* v_{\infty} ) = e^{-i\lambda} \psi_{\infty} .
$$
Now we obtain Proposition \ref{S4_prop_stationarystate}.
\qed

\subsection{Scattering matrix}
\begin{figure}[b]
\centering
\includegraphics[width=12cm, bb=0 0 672 353]{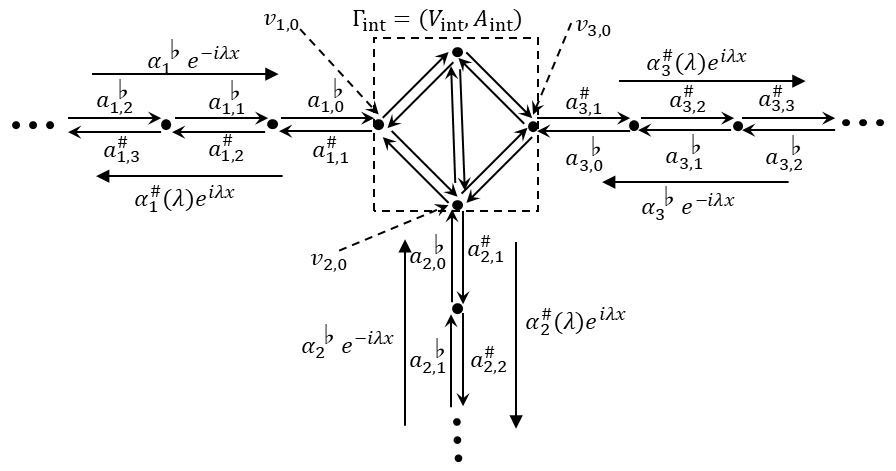}
\caption{In view of the scattering theory, we consider a solution $\psi \in \ell^{\infty} (A)$ to $U\psi = e^{-i\lambda} \psi$ such that $\psi$ satisfies the situation of this figure.
$\alpha^{\flat}$ is the vector of complex intensities of incoming waves, and $\alpha^{\sharp} (\lambda)$ is that of outgoing waves.}
\label{fig_scatteringprocess}
\end{figure}
In view of Proposition \ref{S4_prop_stationarystate} and (\ref{S4_eq_generalizedeigenfunction_formula}), for a generalized eigenfunction $\psi$ to the equation $U\psi = e^{-i\lambda} \psi$, the uniqueness of the outgoing data 
\begin{gather*}
\begin{split} 
&\alpha^{\sharp}  (\lambda ) = [\alpha_{1}^{\sharp} (\lambda )  , \ldots , \alpha_{N}^{\sharp} (\lambda ) ]^{\mathsf{T}}  , \quad \lambda \in {\bf T} ,\\
&\alpha_j^{\sharp} (\lambda)=e^{-i\lambda} \psi (a_{j,1}^{\sharp})=U \chi_{\mathsf{int}}^* v_{\infty} (a_{j,1}^{\sharp} )+   U \varphi_0 (a_{j,1}^{\sharp} ), \quad j=1,\ldots,N,
\end{split}
\end{gather*} 
associated with the incoming data 
$$
\alpha ^{\flat} = [ \alpha^{\flat}_1 , \ldots , \alpha^{\flat}_N ]^{\mathsf{T}} = [\psi (a_{1,0}^{\flat}), \ldots, \psi (a_{N,0}^{\flat}) ]^{\mathsf{T} } ,
$$
 is as follows.

\begin{prop}
Suppose that an incoming data $\alpha ^{\flat} \in {\bf C}^N $ and $\lambda \in {\bf T}$ are fixed.
Then the outgoing data $\alpha^{\sharp} (\lambda ) $ is uniquely determined by $\alpha^{\flat} $.
The mapping $\alpha^{\flat} \mapsto  \alpha^{\sharp} (\lambda)$ is a linear operator on ${\bf C}^N$ and is unitary.
\label{S4_prop_uniqueness_data}
\end{prop}

Proof.
We consider a solution $\psi \in \ell^{\infty} (A)$ to the equation $U \psi = e^{-i\lambda} \psi $ with the incoming data $\alpha^{\flat} $ as in Proposition \ref{S4_prop_stationarystate} with (\ref{S4_eq_generalizedeigenfunction_formula}).
See also Figure \ref{fig_scatteringprocess}.
Letting $v=\chi_{\mathsf{int}} \psi \in {\bf C}^{A_{\mathsf{int}}} $, we have $(E  -e^{-i\lambda} )v=-f_0$ where $f_0 =\chi_{\mathsf{int}} U \varphi_0$ in the similar way of the proof of Proposition \ref{S4_prop_stationarystate}.
Then $v$ is given by $v=-(E-e^{-i\lambda})^{-1} f_0 $ if $e^{-i\lambda} \in S^1 \setminus \sigma_p (E) $.
The spectral representation (\ref{S4_eq_spectralrepresentation00}), $P_{E,\mu}E=EP_{E,\mu}$, $P_{E,\mu} D_{E,\mu} = D_{E,\mu} P_{E,\mu} = D_{E,\mu} $, and $D_{E,\mu}^{m(\mu)}=0$ imply the representation of the resolvent matrix as 
\begin{gather*}
\begin{split}
(E -e^{-iz})^{-1} & =-\sum_{\mu\in\sigma_p (E )} \sum_{s=0}^{m(\mu)-1} (e^{-iz} -\mu )^{-s-1} (E - \mu )^s P_{E,\mu} \\
& = -\sum_{\mu\in\sigma_p (E )} \sum_{s=0}^{m(\mu)-1} (e^{-iz} -\mu )^{-s-1} P_{E,\mu} (E - \mu )^s P_{E,\mu} ,
\end{split}
\end{gather*}
for $e^{-iz} \not\in \sigma_p (E )$.
Recalling $P_{E,\mu} f_0 =0$ if $\mu \in \sigma_p (E)\cap S^1 $ (see Lemma \ref{S4_lem_lemmaseigenfunction}), we obtain 
$$
v =\sum_{\mu\in\sigma_p (E )\setminus S^1} \sum_{s=0}^{m(\mu)-1} (e^{-i\lambda} -\mu )^{-s-1} P_{E,\mu} (E - \mu )^s P_{E,\mu} f _0 , \quad e^{-i\lambda} \in S^1 \setminus \sigma_p (E).
$$
Even if $e^{-i\lambda} \in \sigma_p (E)\cap S^1 $, we can see that the vector 
\begin{equation}
v = \sum_{\mu\in\sigma_p (E )\setminus S^1} \sum_{s=0}^{m(\mu)-1} (e^{-i\lambda} -\mu )^{-s-1} P_{E,\mu} (E - \mu )^s P_{E,\mu} f_0 ,
\label{S4_eq_resolventsubspace}
\end{equation}
is well-defined and satisfies
\begin{gather*}
\begin{split}
(E-e^{-i\lambda})v &= \sum_{\mu\in\sigma_p (E)\setminus S^1} \sum_{s=0}^{m(\mu)-1} \left( \frac{P_{E,\mu} (E-\mu )^{s+1} P_{E,\mu} f_0 }{(e^{-i\lambda} -\mu )^{s+1}}-\frac{P_{E,\mu} (E-\mu )^s P_{E,\mu}f_0}{(e^{-i\lambda} -\mu )^{s}} \right) \\ 
&=-\sum_{\mu\in\sigma_p (E)\setminus S^1} P_{E,\mu} f_0 = -f_0 ,
\end{split}
\end{gather*}
due to Lemma \ref{S4_lem_lemmaseigenfunction}.
Here we have used the equality $(E-\mu)^{m(\mu)} P_{E,\mu} = D_{E,\mu}^{m(\mu)} =0$.
Then the vector $v$ is uniquely determined by $f_0$ even if $e^{-i\lambda} \in \sigma_p (E)\cap S^1 $ in the orthogonal complement of the eigenspace associated with $e^{-i\lambda}$.
Given an incoming data $\alpha^{\flat}$, the vector $f_0 \in \mathcal{H}_s$ is uniquely determined.
In view of the formula (\ref{S4_eq_resolventsubspace}), $v\in  \mathcal{H}_c^{\perp}$ is uniquely determined by $f_0 \in \mathcal{H}_s$.
By Lemma \ref{S2_lem_eigenvaluesU}, the outgoing data $\alpha^{\sharp} (\lambda )$ is independent of eigenfunctions of $U$.
Thus, for every $j=1,\ldots,N$, the outgoing data 
\begin{equation} 
\alpha_{j}^{\sharp} (\lambda) = U \chi_{\mathsf{int}} ^* v (a_{j,1}^{\sharp} ) +  U \varphi_0 (a_{j,1}^{\sharp} )
\label{S4_eq_outgoingdata}
\end{equation} 
is uniquely determined by $\alpha^{\flat} $.

For a subset $K \subset A$, we define the indicator function $1_K$ of $K$ by $1_{K} \psi (a)=1$ if $a\in K$, and $1_{K} \psi (a)=0 $ if $a\in A\setminus K$.
Let $\Omega^{\flat} = A_{\mathsf{int}} \sqcup (\cup_{j=1}^N \{ a_{j,0}^{\flat} \} )$ and $\Omega^{\sharp} = A_{\mathsf{int}} \sqcup (\cup_{j=1}^N \{ a_{j,1}^{\sharp} \} )$.
Then we have 
$$
U 1 _{\Omega^{\flat}} \psi = 1_{\Omega^{\sharp}} U \psi , \quad \psi \in {\bf C}^A .
$$
For the generalized eigenfunction $U \psi = e^{-i\lambda} \psi $, we have  
\begin{gather*}
\begin{split}
\| \alpha^{\flat} \|^2 _{{\bf C}^N} + \| \chi_{\mathsf{int}} \psi \|^2 _{\ell^2 (A_{\mathsf{int}} )} &= \| 1_{\Omega^{\flat}} \psi \|^2 _{\mathcal{H}} = \| U 1_{\Omega^{\flat}} \psi \|^2 _{\mathcal{H}} = \| 1_{\Omega^{\sharp}} U \psi \|^2 _{\mathcal{H}} \\
&= \| e^{-i\lambda} 1_{\Omega^{\sharp}} \psi \|^2_{\mathcal{H}} = \| \alpha^{\sharp} (\lambda) \|^2 _{{\bf C}^N} + \| \chi_{\mathsf{int}} \psi \|^2 _{\ell^2 (A_{\mathsf{int}})} .
\end{split}
\end{gather*} 
Thus we obtain $\| \alpha^{\flat} \|_{{\bf C}^N} = \| \alpha^{\sharp} (\lambda) \| _{{\bf C}^N} $.
\qed

\medskip

Propositions \ref{S4_prop_stationarystate} and \ref{S4_prop_uniqueness_data} allow us to define the \textit{scattering matrix} for the QW $U $ as follows.
\begin{definition}
We define the scattering matrix $\Sigma  (\lambda)$ for $\lambda\in {\bf T}$ which is a unitary operator on ${\bf C}^N$ by $\Sigma (\lambda) \alpha^{\flat} =\alpha^{\sharp} (\lambda)$.

\label{S4_def_scatteringmatrix}
\end{definition}

The resonance expansion formula of the scattering matrix is a direct consequence of the formula (\ref{S4_eq_outgoingdata}).
\begin{cor}
The scattering matrix $\Sigma(\lambda)$ for $\lambda \in {\bf T}$ is represented by the $N\times N$ unitary matrix such that
\begin{gather*}
\begin{split}
(\Sigma(\lambda)\alpha^{\flat})_j &=  \alpha_{j}^{\sharp} (\lambda) \\
&= \left. U \varphi_0 \right|_{a=a_{j,1}^{\sharp}} + \left.  \sum _{\mu\in \sigma_p (E)\setminus S^1} \sum_{s=0}^{m(\mu)-1} U \chi_{\mathsf{int}}^* \left( \frac{  P_{E,\mu} (E -\mu )^s P_{E,\mu}  }{(e^{-i\lambda} -\mu )^{s+1} } \right) \chi_{\mathsf{int}} U \varphi_0 \right| _{a=a_{j,1}^{\sharp}} ,
\end{split}
\end{gather*}
for $j=1,\ldots,N$, where $\varphi_0$ is defined by (\ref{S4_eq_initialincoming}) depending on $\alpha^{\flat}$.
\label{S4_cor_repscatteringmatrix}
\end{cor}

\textit{Remark.}
We have $ \Sigma (\lambda)=I_N$ which is the $N\times N$ identity matrix for $U=U_0$, since there is no interaction between every tail and the internal graph.

\section{Tunable quantum walks} \label{SS_tunableqw0000}
\subsection{Linearity for an auxiliary parameter} \label{SS_perturbationgwalk}
In the following argument, we consider the \textit{tunable quantum walk} $U_{\epsilon}$ for $\epsilon \in [0,1]$ defined by (\ref{S2_eq_grovermatrix000}), (\ref{S2_eq_tunablegrovermatrix00}) and (\ref{S2_eq_tunablegrovermatrix01}).
At the beginning of this section, we introduce a representation of $U_{\epsilon}$ as $\epsilon \downarrow 0 $.
Let
$$
\Theta^{\flat} = \cup_{j=1}^N A_{v_{j,0}}^{\flat} , \quad  \Theta^{\sharp} = \cup_{j=1}^N A_{v_{j,0}}^{\sharp} ,
$$
and $1_{\Theta^{\bullet}}$ be the indicator function of the set $\Theta^{\bullet}$ for $\bullet = \flat , \sharp $.
By definition, we have $(1-1 _{ \Theta^{\sharp}} )U_{\epsilon} = (1-1_{ \Theta^{\sharp} })U_0 $ and $U_{\epsilon} (1-1_{\Theta^{\flat}})= U_0 (1-1_{\Theta^{\flat} })$.
Thus the equality $U_{\epsilon} -U_0 = 1_{\Theta^{\sharp}} (U_{\epsilon} -U_0 )1_{ \Theta^{\flat}} $ implies that the perturbation $U_{\epsilon} -U_0$ is determined by the matrix $c_{\epsilon}^v =g_{\epsilon}^v$ for $v\in \{ v_{1,0},\ldots,v_{N,0} \} $.

 In the following argument, in order to reduce $U_{\epsilon}$ to a linear perturbation of $U_0$, we introduce the auxiliary parameter 
$$
\kappa := 1-e^{i\pi \epsilon} .
$$
Note that $\kappa \to 0$ if and only if $\epsilon \downarrow 0 $.

Let $\partial V_{\mathsf{int}} = \{ v_{1,0} , \ldots , v_{N,0} \} $.
We verify the notations 
\begin{gather}
n_{\mathsf{i}} (v)= \left\{
\begin{split}
\mathrm{deg} (v) &, \quad v\in V \setminus \partial V_{\mathsf{int}}  , \\ 
\mathrm{deg} (v) -N_j &, \quad v\in  \partial V_{\mathsf{int}} ,
\end{split}
\right. \quad n(v)=\mathrm{deg} (v), \quad v\in V ,
\label{S5_eq_n0}
\end{gather}
where $N_j$ is the number of tails connected to $v=v_{j,0}$.
We have for $v=v_{j,0}\in \partial V_{\mathsf{int}}$
$$
g_{\epsilon}^v = \begin{bmatrix} G_{n_{\mathsf{i}} (v),\epsilon} & 0 \\ 0 & I_{N_j} \end{bmatrix} \left( \frac{1-e^{i\pi \epsilon}}{n(v)}  \begin{bmatrix} J_{n_{\mathsf{i}} (v)} & K_j \\ K_j^{\mathsf{T}} & J_{N_j} \end{bmatrix} +e^{i\pi \epsilon} I_{n(v)} \right) ,
$$
where $K_j$ is the $n_{\mathsf{i}} (v)\times N_j$ all-ones matrix.
By using $J_n^2 = nJ_n$, we continue the computation as 
\begin{gather*}
\begin{split}
g_{\epsilon}^v &= \frac{1-e^{i\pi \epsilon}}{n(v)} J_{n(v)} + e^{i\pi \epsilon} \begin{bmatrix} G_{n_{\mathsf{i}} (v),\epsilon} & 0 \\ 0 & I_{N_j} \end{bmatrix}\\
&= \frac{\kappa}{n(v)} J_{n(v)} + e^{i\pi\epsilon} \begin{bmatrix} n_{\mathsf{i}} (v)^{-1} J_{n_{\mathsf{i}} (v)} -e^{-i\pi\epsilon} (I_{n_{\mathsf{i}} (v)} -n_{\mathsf{i}} (v)^{-1} J_{n_{\mathsf{i}} (v)}) & 0 \\ 0 & I_{N_j} \end{bmatrix} \\
&= g_0^v + \frac{\kappa}{n(v)} \left( J_{n(v)}-n(v) \begin{bmatrix} n_{\mathsf{i}} (v)^{-1} J_{n_{\mathsf{i}} (v)} & 0 \\ 0 & I_{N_j} \end{bmatrix} \right) \\
&=g_0^v +\frac{\kappa}{n(v)} \begin{bmatrix} - n_{\mathsf{i}} (v)^{-1} N_j J_{n_{\mathsf{i}} (v)} & K_j \\ K_j^{\mathsf{T}} & J_{N_j} -n(v) I_{N_j} \end{bmatrix}.
\end{split}
\end{gather*}
Letting 
$$
(g_0^v )^{(1)} = \frac{1}{n(v)} \begin{bmatrix} -  n_{\mathsf{i}} (v)^{-1} N_j J_{n_{\mathsf{i}} (v)} & K_j  \\    K_j^{\mathsf{T}} & J_{N_j} -n(v) I_{N_j} \end{bmatrix}  ,\quad v=v_{j,0} \in \partial V_{\mathsf{int}} ,
$$
we obtain the linear perturbation in the parameter $\kappa = 1-e^{i\pi \epsilon}$
$$
g_{\epsilon}^v = g_0^v+\kappa (g_0^v)^{(1)} .
$$ 
Note that $(g_0^v)^{(1)}$ is a symmetric matrix.

Now we have 
\begin{equation}
U_{\epsilon} = U_0 + \kappa U_0^{(1)} , \quad E_{\epsilon} = E_0 +\kappa E_0^{(1)} , \quad \kappa = 1-e^{i\pi \epsilon} ,
\label{S5_eq_perturbedqws}
\end{equation} 
where $U_0^{(1)} =SC_0^{(1)}$ and $E_0^{(1)} = \chi_{\mathsf{int}} SC^{(1)}_0 \chi_{\mathsf{int}}^* $ in which $C_0^{(1)}$ is the operator of multiplication by the matrix
\begin{gather*}
(c_0^v)^{(1)} = \left\{
\begin{split}
(g_0^v )^{(1)} &, \quad v\in \partial V_{\mathsf{int}} , \\ 
0 &, \quad v\in V\setminus \partial V_{\mathsf{int}} .
\end{split}
\right.
\end{gather*}

\subsection{Spectral mapping theorem} \label{SS_smt}
For the study of eigenvalues of $E_{{\color{red} \epsilon}}$, the spectral mapping theorem (\cite{SeSu}, \cite{MOS}) is one of powerful tools.
The spectral mapping theorem derives an equivalence between the eigenvalues for quantum walks and that of a kind of discrete Laplacian on a finite graph.
Let us summarize the argument of the spectral mapping theorem.

Let ${\bf C}^{V_{\mathsf{int}}} = \{ f:V_{\mathsf{int}} \to {\bf C} \}$.
We introduce the weighted Hilbert space $\ell^2 _{n_{\mathsf{i}}} (V_{\mathsf{int}})$ equipped with the inner product
$$
\langle f,g \rangle_{\ell^2 _{n_{\mathsf{i}}} (V_{\mathsf{int}})} = \sum _{v\in V_{\mathsf{int}}} n_{\mathsf{i}} (v) f(v)\overline{g(v)}, \quad f,g\in {\bf C}^{V_{\mathsf{int}}} ,
$$
recalling the definition of $n_{\mathsf{i}} (v)$ by (\ref{S5_eq_n0}).
We define the operators $d:{\bf C}^{A_{\mathsf{int}}} \to {\bf C}^{V_{\mathsf{int}}}$ and $d^* :{\bf C}^{V_{\mathsf{int}}} \to {\bf C}^{A_{\mathsf{int}}} $ by 
\begin{gather*}
d\psi (v)= \frac{1}{n_{\mathsf{i}} (v)} \sum_{a\in A_v^{\flat}   \cap A_{\mathsf{int}} } \psi (a), \quad v\in V_{\mathsf{int}} , \quad \psi \in {\bf C}^{A_{\mathsf{int}}} , \\
d^* u(a)= u(t(a)), \quad a\in A_{\mathsf{int}} , \quad u\in {\bf C}^{V_{\mathsf{int}}} .
\end{gather*}
It is easy to see that 
\begin{equation}
dd^* =I _{V_{\mathsf{int}}}. \label{S5_eq_ddast}
\end{equation}

The following lemma is a direct consequence of definitions of $E_0$, $E_0^{(1)}$, $d$, $d^*$ and $S$.
Let $B:{\bf C}^{V_{\mathsf{int}}} \to {\bf C}^{V_{\mathsf{int}}}$ be the operator defined by 
\begin{gather}
Bu(v)=\left\{
\begin{split}
n(v_{j,0})^{-1} N_j u(v_{j,0}) &, \quad v=v_{j,0} \in \partial V_{\mathsf{int}} , \\ 
0 &, \quad v\in V_{\mathsf{int}} \setminus \partial V_{\mathsf{int}} ,
\end{split}
\right. \quad u\in {\bf C}^{V_{\mathsf{int}}}.
\label{S5_eq_D}
\end{gather}
\begin{lemma}

We have $E_{\epsilon} =E_0 +\kappa E_0^{(1)}$ where $E_0$ and $E_0^{(1)} $ satisfy
\begin{gather}
E_0 = S(2d^* d-I_{A_{\mathsf{int}}}), \label{S5_eq_E0SMT00} \\
E_0^{(1)} = -Sd^* Bd. \label{S5_eq_E0SMT01}
\end{gather} 
\label{S5_lem_representationEkappa}
\end{lemma} 

The eigenvalues of the QW $E_0$ on $A_{\mathsf{int}}$ are well-studied in view of the spectral mapping theorem.
Actually, it can be reduced to the spectral theory for the discrete Laplacian $T=dSd^*$ given by
\begin{equation}
Tu(v)= \frac{1}{n_{\mathsf{i}} (v)} \sum_{a\in A_v^{\flat} \cap A_{\mathsf{int}}} u(o(a)), \quad v\in V_{\mathsf{int}} , \quad u\in {\bf C}^{V_{\mathsf{int}}} .
\label{S5_eq_discreteLaplacian}
\end{equation}
The main result and its corollary in \cite{MOS} give the complete characterization of eigenvalues and eigenspaces for $E_0 $ as follows.
See also \cite{HKSS} and \cite{SeSu}.
In the following results, let 
\begin{equation}
Y :{\bf C}^{V_{\mathsf{int}}} \times {\bf C}^{V_{\mathsf{int}}} \to {\bf C}^{A_{\mathsf{int}}}, \quad  Y(f,g)=d^* f+Sd^* g, \quad f,g\in {\bf C}^{V_{\mathsf{int}}}.
\label{S5_eq_L}
\end{equation}
We put $\mathcal{Y} =\mathsf{Ran} (Y) \subset {\bf C}^{A_{\mathsf{int}}}$.
The operators $\partial_{\lambda} : {\bf C}^{A_{\mathsf{int}}} \to {\bf C}^{V_{\mathsf{int}}}$ and $\partial_{\lambda}^* : {\bf C}^{V_{\mathsf{int}}} \to {\bf C} ^{A_{\mathsf{int}}} $ for $\lambda =e^{i \, \mathrm{arg} \, \lambda } \in S^1$ are defined by 
\begin{gather}
\partial_{\lambda} u = \left\{
\begin{split}
(\sqrt{2} |\sin (\mathrm{arg} \, \lambda )|)^{-1} d(1-\overline{\lambda }S)u &, \quad \lambda \not\in \{ \pm 1 \} , \\ 
du &, \quad \lambda \in \{ \pm 1 \} , 
\end{split}
\right. \quad u\in {\bf C}^{A_{\mathsf{int}}} ,
\label{S5_eq_delllambda00}
\end{gather}
and
\begin{gather}
\partial_{\lambda}^* f = \left\{
\begin{split}
(\sqrt{2} |\sin (\mathrm{arg} \, \lambda )|)^{-1} (1-\lambda S)d^* f &, \quad \lambda \not\in \{ \pm 1 \} , \\ 
d^* f &, \quad \lambda \in \{ \pm 1 \} , 
\end{split}
\right. \quad f\in {\bf C}^{V_{\mathsf{int}}}.
\label{S5_eq_delllambda}
\end{gather}

Due to \cite[Theorem 3.15 and Corollary 3.17]{MOS} and \cite{HKSS}, we can obtain the following spectral mapping theorem under our settings.
Let $\Phi_{QW} (z)$ be the Joukowsky transform defined by $ \Phi_{QW} (z) =(z+z^{-1} )/2$ for $z\in {\bf C} \setminus \{ 0\} $.

\begin{theorem}
\begin{enumerate}
\item
The multiplicity of $1\in \sigma_p (T)$ is one.
The multiplicity of $-1\in \sigma_p (T)$ is $m_{- 1}  =1 $ if $\Gamma _{\mathsf{int}}$ is bipartite, otherwise we have $m_{-1}=0$, i.e., $-1\not\in \sigma_p (T)$.
We have 
\begin{gather*}
\begin{split}
&\sigma (E_0 |_{ \mathcal{Y}})= \Phi^{-1}_{QW}  (\sigma (T)), \\
&  \sigma (E_0 |_{ \mathcal{Y}^{\perp} }) = \left\{
\begin{split} 
\emptyset &, \quad \Gamma_{\mathsf{int}} \ \text{is a tree},\\
\{ 1 \} &, \quad \Gamma_{\mathsf{int}} \  \text{has only one cycle and its length is odd}, \\
\{ \pm 1 \} &, \quad \text{otherwise},
\end{split}
\right.
\end{split}
\end{gather*}
where the multiplicities of $1$ and $-1$ (if $-1\in \sigma_p (E_0 |_{ \mathcal{Y}^{\perp}})$) as eigenvalues of $ E_0 |_{ \mathcal{Y}^{\perp}} $ are $M_{1}=\max \{ 0, (\# A_{\mathsf{int}} /2) -\# V_{\mathsf{int}} +1 \}$ and $M_{-1} = \max \{ 0, (\# A_{\mathsf{int}} /2) - \# V_{\mathsf{int}} + m_{-1} \} $, respectively. 

\item
The eigenfunctions of an eigenvalue $\lambda\in  \Phi_{QW}^{-1}  (\sigma (T))$ generating $\mathcal{Y}$ are given by 
$$
\mathsf{Ker} (E_0 |_{ \mathcal{Y}} -\lambda )= \{ \partial _{\lambda}^* f \ ; \ f\in \mathsf{Ker} ( T- \Phi_{QW}   (\lambda)) \} .
$$

\item
The eigenspace associated with the eigenvalue $1$ with multiplicity $M_{1}$ and $-1$ with multiplicity $M_{-1}$ except those corresponding to $ \Phi_{QW}^{-1} (\pm 1)\in \sigma_p (E_0 |_{ \mathcal{Y}} )$ are described by 
$$
\mathsf{Ker} (E_0 |_{ \mathcal{Y}^{\perp}} \mp 1)= \mathsf{Ker} (d)\cap \mathsf{Ker} (S\pm 1), 
$$
respectively.

\end{enumerate}

\label{S5_thm_SMTMOS}
\end{theorem}

\textit{Remark.}
We call $\mathcal{Y}$ the \textit{inherited subspace} while $\mathcal{Y}^{\perp}$ the \textit{birth subspace}.
Thus $\sigma_p (E_0 |_{\mathcal{Y} })$ and $\sigma_p (E_0 |_{ \mathcal{Y} ^{\perp}})$ are the set of \textit{inherited eigenvalues} and the set of \textit{birth eigenvalues} associated with $T$, respectively.

\subsection{Exposition of our main purpose}
Now we can describe clearly our main subject.
As has been seen in \S \ref{SS_perturbationgwalk} and \S \ref{SS_smt}, the QW $U_{\epsilon}$ is a perturbation of the free QW $U_0$ as well as $E_{\epsilon}$ is a perturbation of $E_0$.
Roughly speaking, we have ``$U_{\epsilon} \to U_0$" and ``$E_{\epsilon} \to E_0$" as $\epsilon \downarrow 0 $ in a suitable topology.

Take $\lambda \in {\bf T}$.
Let $\Sigma_0 (\lambda)=I_N$ and $\Sigma _{\epsilon} (\lambda)$ be the scattering matrices for $U_0$ and $U_{\epsilon}$ defined as Definition \ref{S4_def_scatteringmatrix}, respectively.
According to classical mechanics intuition, the limit 
$$
\lim_{\epsilon \downarrow 0} \Sigma_{\epsilon} (\lambda) = \Sigma_0 (\lambda)
$$ 
can be expected.
In the following argument, we are going to show that this intuition is correct for almost all $\lambda \in {\bf T}$ as well as there are some exceptional $\lambda \in {\bf T}$.
For these exceptional $\lambda$, i.e., \textit{resonant energies}, we can see that there exists $\lambda _{\epsilon}\in {\bf T}$ such that $\lim_{\epsilon\downarrow 0} \lambda _{\epsilon} = \lambda$ and an anomalous asymptotic behavior like 
$$
\lim_{\epsilon\downarrow 0} \Sigma_{\epsilon} (\lambda _{\epsilon}) = \Sigma_0 (\lambda)+\Sigma_0^{(1)} (\lambda)\not= I_N ,
$$ 
holds true.
There are some resonances near such kind of exceptional $\lambda$.
The set of the exceptional $\lambda$ is a subset of $\sigma_p (U_0)$.

Our main tool is the perturbation theory for matrices \cite{Ka2,Ka}.
Namely, we need to study the following issues.
\begin{enumerate}
\item
Asymptotic expansions of eigenvalues of $E_{\epsilon}$ which give the asymptotic behaviors of resonances near the eigenvalues of $E_0$.

\item
Asymptotic expansions of eigenprojections of $E_{\epsilon}$.

\end{enumerate}

In order to study these issues, we have to apply the \text{reduction process} of generalized eigenspaces which gives a decomposition of a generalized eigenspace into some subspaces.
Particularly, the issue (2) is a consequence of the reduction process.

Applying the results for (1) and (2) to the resonance expansion of $\Sigma _{\epsilon} (\lambda)\alpha^{\flat}$
$$
(\Sigma_{\epsilon} (\lambda)\alpha^{\flat})_j = \left. U_{\epsilon} \varphi_0 \right|_{a=a_{j,1}^{\sharp}} + \left.  \sum _{\mu_{\epsilon} \in \sigma_p (E_{\epsilon}) \setminus S^1} \sum_{s=0}^{m(\mu_{\epsilon})-1} U_{\epsilon} \chi_{\mathsf{int}}^* \left( \frac{  P_{\mu_{\epsilon}} (E_{\epsilon} -\mu_{\epsilon} )^s P_{\mu_{\epsilon}}  }{(e^{-i\lambda} -\mu_{\epsilon} )^{s+1} } \right) \chi_{\mathsf{int}} U_{\epsilon} \varphi_0 \right| _{a=a_{j,1}^{\sharp}} ,
$$
for $P_{\mu_{\epsilon}} = P_{E_{\epsilon},\mu _{\epsilon}}$ and $j=1,\ldots,N$, we are going to show the desired result.

Note that we apply the perturbation theory to $E_0 + \kappa E_0^{(1)}$ (see (\ref{S5_eq_perturbedqws})) which is a linear perturbation of $E_0$ in the auxiliary parameter $\kappa\in {\bf C}$ with $|\kappa| \ll 1$.
After that, for the proof of main theorems, we are going to go back to the original parameter $\epsilon>0$ by using the relation $\kappa =1-e^{i\pi \epsilon}$. 
The details of the perturbation theory is reviewed in the next section.


\section{Perturbation theory for matrices} \label{section_tkato}
\subsection{Linear perturbations for eigenvalues}

Here we review some results of perturbation theory for the matrix
$$
E(\kappa):= E_0 +\kappa E_0^{(1)}, \quad \kappa \in {\bf C} , \quad |\kappa| \ll 1 ,
$$
following the arguments in \cite{Ka2,Ka} for the sake of readability of this paper.
We omit details of proofs for some propositions and lemmas.

The eigenprojection is defined by (\ref{S4_eq_defeigenprojection}) for the matrix $E$ and $\mu\in \sigma_p (E)$.
In the following argument, we often use the notation $\mathcal{C} (\mu)$ as a counterclockwise loop such that neither self-intersections nor other eigenvalues inside this loop exist. 
Now we recall the definitions of the operators $P_{\mu} $ and $S_{\mu}$ for $\mu\in \sigma_p (E_0)$ as 
\begin{gather}
\begin{split}
&P_{\mu} = -\frac{1}{2\pi i} \oint _{\mathcal{C} (\mu)} (E_0 -z)^{-1} dz,  \\
&S_{\mu} =\frac{1}{2\pi i} \oint _{\mathcal{C} (\mu)} (z-\mu )^{-1} (E_0 -z)^{-1} dz.
\end{split} \label{Ap_eq_defprojection00}
\end{gather}
In the similar way, we define $P_{\mu (\kappa)} $ and $S_{\mu(\kappa)}$ for $\mu (\kappa)\in \sigma _p (E(\kappa))$ by 
\begin{gather*}
\begin{split}
&P_{\mu (\kappa)} = -\frac{1}{2\pi i} \oint _{\mathcal{C} (\mu (\kappa))} (E(\kappa) -z)^{-1} dz, \\
&S_{\mu (\kappa)} = \frac{1}{2\pi i} \oint _{\mathcal{C} (\mu (\kappa))} (z-\mu (\kappa))^{-1}  (E(\kappa) -z)^{-1} dz.
\end{split}
\end{gather*}

For the eigennilpotent matrix $D_{\mu(\kappa)}=(E(\kappa)-\mu (\kappa))P_{\mu(\kappa)} $ and the operator $S_{\mu(\kappa)} $, we can see that the Laurent expansion
\begin{equation}
(E(\kappa)-z)^{-1} = -(z-\mu(\kappa))^{-1} P_{\mu(\kappa)} -\sum_{n=1}^{m(\mu(\kappa))-1} (z-\mu(\kappa))^{-n-1} D_{\mu(\kappa)}^n + \sum _{n=0}^{\infty} (z-\mu(\kappa))^n S_{\mu(\kappa)} ^{n+1}
\label{Ap_eq_resolventformulaproj00}
\end{equation}
holds true in a small neighborhood of $\mu(\kappa)$ and another representation
$$
(E(\kappa)-z)^{-1} =-\sum_{\mu(\kappa) \in \sigma_p (E(\kappa))} \sum_{n=0}^{m(\mu(\kappa))-1} (z-\mu(\kappa) )^{-n-1} (E(\kappa )-\mu(\kappa) )^n P_{\mu(\kappa)} 
$$
also holds true.
Then the regular part of $(E(\kappa)-z)^{-1}$ 
$$
S_{\mu(\kappa)} (z) = \sum _{n=0}^{\infty} (z-\mu (\kappa) )^n S_{\mu(\kappa)}^{n+1}
$$
in a small neighborhood of $z=\mu(\kappa)$ satisfies 
\begin{equation}
S_{\mu (\kappa)} (z)= -\sum _{\zeta(\kappa) \in \sigma_p (E(\kappa)) \setminus \{ \mu(\kappa) \}} \sum_{n=0}^{m(\zeta(\kappa))-1} (z-\zeta(\kappa) )^{-n-1} (E(\kappa)-\zeta(\kappa) )^n P_{\zeta(\kappa)} ,
\label{Ap_eq_reducedresolvent}
\end{equation}
and
\begin{equation}
S_{\mu (\kappa)} (\mu (\kappa))= S_{\mu (\kappa)}. 
\label{Ap_eq_reducedresolvent2}
\end{equation}

Let us list some formulas.
These formulas are consequences of the Laurent expansion of $(E(\kappa)-z)^{-1}$ in a small neighborhood of each eigenvalue.
We omit the details of the proofs.
\begin{prop}
Let $\mu(\kappa),\zeta (\kappa)\in \sigma_p (E(\kappa))$.
We have 
\begin{gather*}
\begin{split}
&P_{\mu(\kappa)} P_{\zeta(\kappa)} = P_{\zeta(\kappa)}P_{\mu(\kappa)}=\delta _{\mu(\kappa),\zeta(\kappa)} P _{\mu(\kappa)}, \\ 
&P_{\mu(\kappa)} S_{\mu(\kappa)} = S_{\mu(\kappa)} P_{\mu(\kappa)} =0,\\
&\sum _{\mu(\kappa)\in \sigma_p (E(\kappa))} P_{\mu(\kappa)} =I_{A_{\mathsf{int}}} , \\ 
&P_{\mu(\kappa)} D_{\zeta (\kappa)} = D_{\zeta (\kappa)} P_{\mu(\kappa)} = \delta_{\mu(\kappa),\zeta(\kappa)} D_{\zeta (\kappa)}, \\ 
&D_{\mu(\kappa)} D_{\zeta(\kappa)}=D_{\zeta(\kappa)}D_{\mu(\kappa)} =0 \quad \text{if} \quad \mu(\kappa) \not= \zeta (\kappa),\\
&(E(\kappa)-z)S_{\mu(\kappa)} (z)=S_{\mu(\kappa)} (z) (E(\kappa)-z)=I_{A_{\mathsf{int}}}-P_{\mu(\kappa)}, \quad z\in {\bf C},
\end{split}
\end{gather*}
where $I_{A_{\mathsf{int}}}$ is the identity on ${\bf C}^{A_{\mathsf{int}}} $.

\label{Ap_prop_formulaseigennilpotent}
\end{prop}

The continuity of eigenvalues at $\kappa =0$ can be characterized by the following fact.

\begin{prop}
For sufficiently small $\delta >0$, the matrix $(E(\kappa)-z)^{-1}$ is analytic in $(z,\kappa)\in ({\bf C} \setminus \cup_{|\kappa|\leq \delta} \sigma _p (E(\kappa) ))\times \{ \kappa \in {\bf C} \ ; \ |\kappa |<\delta \}$ and $ (E(0)-z)^{-1} = (E_0 -z)^{-1}$.
\label{Ap_prop_continuityresolventop}
\end{prop}

Proof.
This fact follows from the equality
$$
(E(\kappa)-z)^{-1} =(E_0  -z)^{-1} \left( 1-((z-z_0 ) I_{A_{\mathsf{int}}} -\kappa E_0^{(1)} )(E_0 -z_0 )^{-1}   \right) ^{-1} , \quad |z-z_0 | \ll 1,
$$
for $z_0 \in {\bf C} \setminus \cup_{|\kappa|\leq \delta} \sigma _p (E(\kappa) )$.
\qed

\medskip

\textit{Remark.}
Proposition \ref{Ap_prop_continuityresolventop} holds true also for analytic perturbations $A (\kappa)$ for a square matrix $A_0 =A(0)$ even if $A_0$ is not unitary.

\medskip

Proposition \ref{Ap_prop_continuityresolventop} shows that every eigenvalue of $E(\kappa)$ is continuous in $\kappa$, and every eigenvalue converges to one of eigenvalues of $E_0$ as $\kappa \to 0$.
We call the subset of eigenvalues 
$$
\sigma _p (E(\kappa) ,\mu ) = \left\{ \mu (\kappa) \in \sigma_p (E(\kappa)) \ ; \    \lim_{\kappa \to 0} \mu (\kappa) = \mu  \right\} , 
$$ 
the $\mu$-\textit{group} for an eigenvalue $\mu\in\sigma_p (E_0)$. 
We see that
$$
\sigma_p (E(\kappa) )=\bigsqcup _{\mu\in\sigma_p (E_0 )} \sigma_p (E(\kappa) ,\mu ). 
$$

In view of Proposition \ref{Ap_prop_continuityresolventop}, we can define the \textit{total projection} associated with $\mu\in \sigma_p (E_0)$ by
$$
P _{\mu}^{\mathsf{total} }(\kappa) =-\frac{1}{2\pi i} \oint _{\mathcal{C} (\mu)} (E(\kappa)-z)^{-1} dz , \quad |\kappa| \ll 1.
$$
Namely, $P_{\mu}^{\mathsf{total}} (\kappa)$ is the sum of eigenprojections 
$$
P_{\mu}^{\mathsf{total}} (\kappa) = \sum_{\mu (\kappa) \in \sigma _p (E(\kappa),\mu)} P_{\mu (\kappa)} .
$$
Summarizing these facts, we have the following proposition.

\begin{prop}
The total projection $P_{\mu}^{\mathsf{total}} (\kappa)$ for every $\mu \in \sigma _p (E_0)$ is analytic in $\kappa$ for sufficiently small $|\kappa| \ll 1 $.
The sum of the algebraic multiplicities of eigenvalues in the $\mu$-group coincides with that of $\mu$.
\label{Ap_prop_continuitytotalprojection}
\end{prop}

Proof.
This is a direct consequence of Proposition \ref{Ap_prop_continuityresolventop}.
\qed

\medskip

\textit{Remark.}
Even though Proposition \ref{Ap_prop_continuitytotalprojection} holds true, every eigenprojection $P_{\mu(\kappa)} $ for $\mu(\kappa) \in \sigma_p (E(\kappa) )$ possibly has a pole at $\kappa =0 $.
This pole is at most of finite order.
On the other hand, each eigenvalue $\mu(\kappa) \in \sigma_p (E(\kappa) )$ is analytic in $\kappa$ with $|\kappa| \ll 1$.
Therefore, we have to consider the reduction process in order to show an asymptotic behavior of $P_{\mu(\kappa)}$ in a subspace of the generalized eigenspace.

\subsection{Reduction processes}
Let us turn to the asymptotic behavior of eigenvalues $\mu(\kappa) \in \sigma _p (E(\kappa))$ as $\kappa \to 0 $ and the associated decomposition of eigenspaces.
Since $E_0$ is unitary, eigenvalues of $E_0$ are \textit{semi-simple}, i.e., we have $D_{\mu} =(E_0 - \mu )P_{\mu} =0 $ for all $\mu\in \sigma_p (E_0 )$.
Precisely, in this case, we see that $E_0$ is diagonalizable.

At the beginning of this subsection, we show an asymptotic expansion of the total projection.
We put
\begin{gather*}
\begin{split}
&P_{\mu}^{(1)} = -P_{\mu} E_0^{(1)} S_{\mu} - S_{\mu} E_0^{(1)} P_{\mu} , \\
&P_{\mu}^{(2)} = P_{\mu} E_0^{(1)} S_{\mu} E_0^{(1)} S_{\mu} +S_{\mu} E_0^{(1)} P_{\mu} E_0^{(1)} S_{\mu} +S_{\mu} E_0^{(1)} S_{\mu} E_0^{(1)} P_{\mu} , \\
&\begin{split}
P_{\mu}^{(3)} =& \, -P_{\mu} E_0^{(1)} S_{\mu} E_0^{(1)} S_{\mu} E_0^{(1)} S_{\mu} -S_{\mu} E_0^{(1)} P_{\mu} E_0^{(1)} S_{\mu} E_0^{(1)} S_{\mu} -S_{\mu} E_0^{(1)} S_{\mu} E_0^{(1)} P_{\mu} E_0^{(1)} S_{\mu}\\
&\, -S_{\mu} E_0^{(1)} S_{\mu} E_0^{(1)} S_{\mu} E_0^{(1)} P_{\mu} .
\end{split}
\end{split}
\end{gather*}
\begin{prop}
For $\mu\in \sigma_p (E_0)$, we have 
$$
P_{\mu}^{\mathsf{total}} (\kappa) = P_{\mu} +\kappa P_{\mu}^{(1)} +\kappa^2 P_{\mu}^{(2)} +\kappa^3 P_{\mu}^{(3)} +o(\kappa^3 ), \quad \kappa \to 0 .
$$

\label{Ap_prop_totalproj_asymptotics}
\end{prop}

Proof.
This formula is a consequence of the resolvent equations
\begin{gather*}
\begin{split}
&(E(\kappa)-z)^{-1} \\  
&= (E_0 -z)^{-1} - \kappa (E_0 -z)^{-1} E_0^{(1)} (E(\kappa)-z)^{-1} \\
&= (E_0 -z)^{-1} -\kappa (E_0 -z)^{-1} E_0^{(1)} (E_0 -z)^{-1}+\kappa^2 (E_0 -z)^{-1} E_0^{(1)} (E_0 -z)^{-1} E_0^{(1)} (E_0 -z)^{-1} \\ 
& \quad -\kappa^3 (E_0 -z)^{-1} E_0^{(1)} (E_0 -z)^{-1} E_0^{(1)} (E_0 -z)^{-1} E_0^{(1)} (E_0 -z)^{-1}+o(\kappa^3 ),
\end{split}
\end{gather*}
for suitable $z$, and the equality (\ref{Ap_eq_resolventformulaproj00}).
\qed

\medskip

By using the above formula, we can obtain the asymptotic expansion of eigenvalues of $E(\kappa)$ with its error of $o(\kappa)$.
We put 
\begin{gather*}
\begin{split}
&\widetilde{E}^{(1)} _{\mu} = P_{\mu} E_0^{(1)} P_{\mu} , \\
&\widetilde{E}_{\mu}^{(1,1)} = -P_{\mu} E_0^{(1)} \left( P_{\mu} E_0^{(1)} S_{\mu} + S_{\mu} E_0^{(1)} P_{\mu} \right) ,\\
&\widetilde{E}_{\mu}^{(1,2)} =P_{\mu} E_0^{(1)} \left( P_{\mu} E_0^{(1)} S_{\mu} E_0^{(1)} S_{\mu} +   S_{\mu} E_0^{(1)} P_{\mu} E_0^{(1)} S_{\mu} + S_{\mu} E_0^{(1)} S_{\mu} E_0^{(1)} P_{\mu} \right) ,
\end{split}
\end{gather*}
for $\mu \in \sigma_p (E_0 )$.

\begin{lemma}
Take $\mu (\kappa) \in \sigma_p (E(\kappa),\mu)$ for $\mu\in \sigma_p (E_0)$.
We have 
\begin{equation}
\mu (\kappa) = \mu + \kappa \mu^{(1)}  +o(\kappa), \quad \kappa \to 0,
\label{Ap_eq_differentiableevs00}
\end{equation} 
where $\mu^{(1)}$ is one of eigenvalues of $\widetilde{E} _{\mu}^{(1)} | _{\mathsf{R}(\mu)}$ with $ \mathsf{R}(\mu)=\mathsf{Ran} (P_{\mu}) $.

\label{Ap_lem_differentiableevs}
\end{lemma}

Proof.
We consider the matrix $\widetilde{E}^{(1)} _{\mu} (\kappa) := \kappa^{-1} (E(\kappa) -\mu ) P_{\mu}^{\mathsf{total}} (\kappa) $.
Then $\mu^{(1)} (\kappa) \in \sigma_p (\widetilde{E}_{\mu}^{(1)} (\kappa) ) $ if and only if $\mu +\kappa \mu^{(1)} (\kappa) \in \sigma_p (E(\kappa)|_{\mathsf{Ran} (P_{\mu}^{\mathsf{total}}(\kappa))} )$.
Moreover, we have 
\begin{equation}
\mathsf{Ker} ( \widetilde{E}^{(1)} _{\mu} (\kappa)  -z ) = \mathsf{Ker} (P_{\mu}^{\mathsf{total}} (\kappa) (E(\kappa) -(\mu+\kappa z)) P_{\mu}^{\mathsf{total}} (\kappa)), \quad z\in {\bf C} .
\label{Ap_eq_eigenspacereduced01}
\end{equation}
Then $\mu (\kappa) \in \sigma_p (E(\kappa),\mu )$ is of the form $\mu (\kappa) = \mu +\kappa \mu^{(1)} (\kappa)$ for $\mu ^{(1)} (\kappa) \in \sigma_p (\widetilde{E}_{\mu}^{(1)}(\kappa) )$.
In view of Propositions \ref{Ap_prop_formulaseigennilpotent} and \ref{Ap_prop_totalproj_asymptotics}, we see that 
\begin{equation}
\widetilde{E}_{\mu}^{(1)} (\kappa) = \widetilde{E}^{(1)}_{\mu} +\kappa \widetilde{E}_{\mu}^{(1,1)}+\kappa^2 \widetilde{E}_{\mu}^{(1,2)} +o(\kappa^2), 
\label{Ap_eq_Tkappatildeasymptotic}
\end{equation}
as $\kappa \to 0 $. 
Here we have used the equation $(E_0 - \mu )P_{\mu} =0$ and $(E_0 - \mu )S_{\mu} =I_{A_{\mathsf{int}}} -P_{\mu}$.
Applying Proposition \ref{Ap_prop_continuityresolventop} with its remark to $(\widetilde{E}_{\mu}^{(1)} (\kappa)-z)^{-1}$ and $(\widetilde{E}^{(1)} _{\mu} -z)^{-1}$, we obtain the continuity of the eigenvalue $\mu^{(1)} (\kappa)$ at $\kappa =0$.
Letting $\mu^{(1)} =\lim_{\kappa\to 0} \mu^{(1)} (\kappa) $, the asymptotic behavior (\ref{Ap_eq_differentiableevs00}) follows.
\qed

\medskip

\begin{figure}[b] 
\centering
\includegraphics[width=8cm, bb= 0 0 380 127]{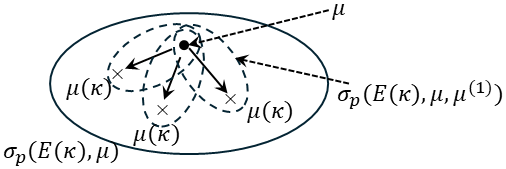}
\caption{ Let $\mu \in \sigma_p (E_0)$.
We have the disjoint union $\sigma_p (E(\kappa) , \mu )=\sqcup_{\mu^{(1)}\in \sigma_p (\widetilde{E} _{\mu}^{(1)} |_{\mathsf{R}(\mu)})} \sigma_p (E(\kappa),\mu,\mu^{(1)})$.}
\label{fig_evgroup}
\end{figure} 
\begin{figure}[t] 
\centering
\includegraphics[width=15cm, bb=0 0 747 269]{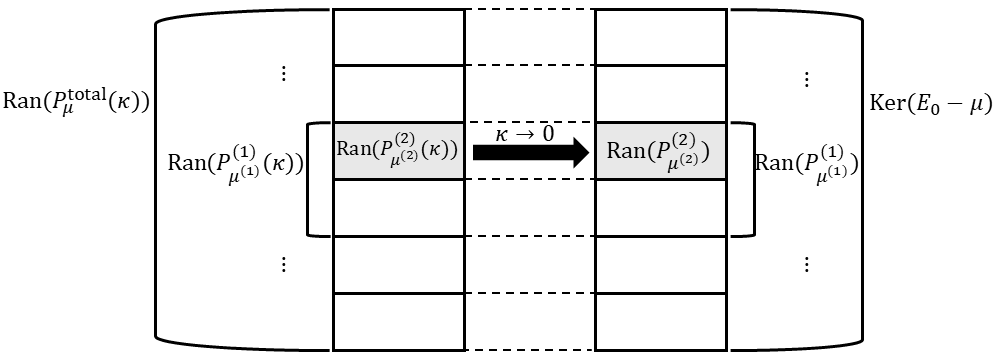}
\caption{ The second order reduction process gives a decomposition of the subspace $\mathsf{Ran} (P_{\mu}^{\mathsf{total}} (\kappa))$ of $\sigma_p (E(\kappa),\mu)$ into some subspaces $\mathsf{Ran} \big( P_{\mu^{(2)}}^{(2)} (\kappa) \big)$. 
Note that $P _{\mu^{(j)}}^{(j)} (\kappa) $ for $j=1,2$ converges to $P_{\mu^{(j)}}^{(j)}$ for $j=1,2$, respectively, under a suitable assumption. }
\label{fig_reductions}
\end{figure}

Due to Lemma \ref{Ap_lem_differentiableevs}, the $\mu$-group $ \sigma_p (E(\kappa) , \mu )$ can be split into $\mu+\kappa \mu^{(1)}$-groups $\sigma _p (E(\kappa),\mu,\mu^{(1)} )$ for every eigenvalue $\mu^{(1)}\in \sigma_p ( \widetilde{E} _{\mu}^{(1)} | _{\mathsf{R} (\mu )} )$:
\begin{gather*}
\begin{split}
&\sigma_p (E(\kappa),\mu,\mu^{(1)} )= \left\{ \mu(\kappa) \in \sigma_p (E(\kappa)) \ ; \ \mu(\kappa) \ \text{satisfies} \ \mu(\kappa) = \mu + \kappa \mu^{(1)} +o(\kappa), \ \kappa \to 0 \right\} , \\
&\sigma_p (E(\kappa),\mu)= \bigsqcup_{\mu^{(1)}\in \sigma_p ( \widetilde{E} _{\mu}^{(1)} | _{\mathsf{R} (\mu )} )} \sigma_p (E(\kappa),\mu,\mu^{(1)} ), 
\end{split}
\end{gather*}
See Figures \ref{fig_evgroup} and \ref{fig_reductions}.
Letting  
\begin{gather}
P_{\mu^{(1)}}^{(1)} (\kappa) = -\frac{1}{2\pi i} \oint _{\mathcal{C} (\mu^{(1)})} (\widetilde{E}^{(1)}_{\mu} (\kappa) -z)^{-1} dz \quad \text{on} \quad \mathsf{Ran} ( P_{\mu}^{\mathsf{total}} (\kappa) ), \label{Ap_eq_higherprojection00} \\
P^{(1)}_{\mu^{(1)}} =-\frac{1}{2\pi i} \oint _{\mathcal{C} (\mu^{(1)})} (\widetilde{E}_{\mu}^{(1)}-z)^{-1} dz\quad \text{on} \quad \mathsf{Ran} ( P_{\mu} ),
\label{Ap_eq_higherprojection}
\end{gather}
$P_{\mu^{(1)}} ^{(1)} (\kappa)$ is the total projection associated with $\mu^{(1)}$-group eigenvalues $\sigma_p (\widetilde{E}_{\mu}^{(1)} (\kappa) , \mu^{(1)} )$ on the subspace $\mathsf{Ran} ( P_{\mu}^{\mathsf{total}} (\kappa) )$.
Due to the equivalence of equations 
$$
\widetilde{E}_{\mu}^{(1)} (\kappa) u=\mu^{(1)} (\kappa) u \quad \text{and} \quad E(\kappa)P_{\mu}^{\mathsf{total}} u=(\mu+\kappa \mu^{(1)} (\kappa) )P_{\mu}^{\mathsf{total}}u, 
$$
the operator $P_{\mu^{(1)}}^{(1)} (\kappa)$ is also the total projection associated with $\sigma_p (E(\kappa),\mu,\mu^{(1)} )$.

\begin{lemma}

We have 
$$
\sum _{\mu^{(1)}\in \sigma_p (\widetilde{E}_{\mu}^{(1)}|_{\mathsf{R} (\mu)})} P _{\mu^{(1)}} ^{(1)} (\kappa) = P_{\mu}^{\mathsf{total}} (\kappa) , \quad \sum _{\mu^{(1)}\in \sigma_p (\widetilde{E}_{\mu}^{(1)}|_{\mathsf{R} (\mu)})} P _{\mu^{(1)}} ^{(1)} = P_{\mu},
$$
for $\mu\in \sigma_p (E_0 )$.
The total projection $P_{\mu^{(1)}}^{(1)} (\kappa)$ associated with $\sigma_p (E(\kappa),\mu,\mu^{(1)} )$ is continuous at $\kappa =0$, i.e., $P_{\mu^{(1)}}^{(1)} (\kappa) \to P^{(1)}_{\mu^{(1)}}$ as $\kappa \to 0 $.

\label{Ap_lem_continuitytotalproj_1}
\end{lemma}

Proof.
The decompositions of the projections $P_{\mu}^{\mathsf{total}} (\kappa)$ and $P_{\mu}$ follow from the definitions of the projections $P_{\mu^{(1)}}^{(1)} (\kappa)$ and $P_{\mu^{(1)}}^{(1)}$.
We can obtain the continuity of $P_{\mu^{(1)}}^{(1)} (\kappa)$ by applying Proposition \ref{Ap_prop_continuitytotalprojection} to $(\widetilde{E}_{\mu}^{(1)} (\kappa) -z)^{-1}$ and $(\widetilde{E}_{\mu}^{(1)} -z)^{-1 } $.
\qed

\medskip

In order to consider the second order reduction process, we tentatively assume that $\widetilde{E}_{\mu}^{(1)} =P_{\mu} E_0^{(1)} P_{\mu} $ is also semi-simple.
Actually, we are going to prove that $\widetilde{E}_{\mu}^{(1)}$ is semi-simple in Lemma \ref{S5_lem_semisimpleE0(1)}.
Let
$$
\widetilde{E}^{(2)} _{\mu^{(1)}}=P_{\mu^{(1)}}^{(1)} \widetilde{E}^{(1,1)} _{\mu} P_{\mu^{(1)}}^{(1)} , \quad \mathsf{R}(\mu^{(1)})=\mathsf{Ran} \big( P _{\mu^{(1)}}^{(1)} \big) .
$$
Then we can repeat the reduction process (the same procedure for the proof of Lemma \ref{Ap_lem_differentiableevs}) for $\mu^{(1)} (\kappa) \in \sigma_p (\widetilde{E}_{\mu}^{(1)} (\kappa))$ in order to see that $\sigma_p (E(\kappa) , \mu , \mu^{(1)} )$ is split into some subsets as
$$
\sigma_p (E(\kappa) , \mu , \mu^{(1)} )= \bigsqcup _{\mu^{(2)} \in \sigma_p \big( \widetilde{E}^{(2)} _{\mu^{(1)}} |_{\mathsf{R}(\mu^{(1)})} \big)} \sigma_p ( E(\kappa) , \mu , \mu^{(1)} , \mu^{(2)} ), 
$$ 
where
\begin{gather}
\sigma_p ( E(\kappa) , \mu , \mu^{(1)} , \mu^{(2)} ) = \{ \mu (\kappa) \in \sigma_p (E(\kappa)) \ ; \ \mu (\kappa) \ \text{satisfies (\ref{Ap_eq_asymptotic2ndorder})} \}, \label{Ap_eq_2ndreduction_ev} \\
\mu (\kappa) = \mu +\kappa \mu^{(1)} +\kappa^2 \mu^{(2)} +o(\kappa^2), \quad \kappa \to 0 .
\label{Ap_eq_asymptotic2ndorder}
\end{gather}
We obtain the following result.
Let 
\begin{gather}
\widetilde{E}_{\mu^{(1)}}^{(2)} (\kappa) = \kappa^{-1} \big( \widetilde{E}_{\mu}^{(1)} (\kappa) -\mu^{(1)} \big) P_{\mu^{(1)}}^{(1)} (\kappa) , \label{Ap_eq_higherprojection001}  \\
P_{\mu^{(2)}}^{(2)} (\kappa) = -\frac{1}{2\pi i} \oint _{\mathcal{C} (\mu^{(2)})} \big( \widetilde{E}^{(2)}_{\mu^{(1)}} (\kappa) -z \big) ^{-1} dz \quad \text{on} \quad \mathsf{Ran} \big( P _{\mu^{(1)}}^{(1)} (\kappa) \big) , \label{Ap_eq_higherprojection01} 
\end{gather}
\begin{gather}
P^{(2)}_{\mu^{(2)}} =-\frac{1}{2\pi i} \oint _{\mathcal{C} (\mu^{(2)})} \big( \widetilde{E}_{\mu^{(1)}}^{(2)}-z \big) ^{-1} dz \quad \text{on} \quad \mathsf{Ran} \big( P _{\mu^{(1)}}^{(1)} \big) , \label{Ap_eq_higherprojection002} \\ 
S_{\mu^{(2)}}^{(2)}=\frac{1}{2\pi i} \oint _{\mathcal{C} (\mu^{(2)})} (z-\mu^{(2)} )^{-1}   \big( \widetilde{E}_{\mu^{(1)}}^{(2)}-z \big)^{-1} dz \quad  \text{on} \quad \mathsf{Ran} \big( P _{\mu^{(1)}}^{(1)} \big) ,
\label{Ap_eq_higherprojection02}
\end{gather}
and
\begin{gather*}
\begin{split}
&P_{\mu^{(2)}}^{(2,1)} = -P^{(2)}_{\mu^{(2)}} \widetilde{E}^{(2,1)} _{\mu^{(1)}} S^{(2)}_{\mu^{(2)}} - S_{\mu^{(2)}}^{(2)} \widetilde{E}^{(2,1)}_{\mu^{(1)}} P^{(2)}_{\mu^{(2)}}  , \\
&P_{\mu^{(2)}}^{(2,2)} = P^{(2)}_{\mu^{(2)}} \widetilde{E}^{(2,1)} _{\mu^{(1)}} S^{(2)}_{\mu^{(2)}} \widetilde{E}^{(2,1)} _{\mu^{(1)}} S^{(2)}_{\mu^{(2)}} +S^{(2)}_{\mu^{(2)}} \widetilde{E}^{(2,1)} _{\mu^{(1)}} P^{(2)}_{\mu^{(2)}} \widetilde{E}^{(2,1)} _{\mu^{(1)}} S^{(2)}_{\mu^{(2)}} +S^{(2)}_{\mu^{(2)}} \widetilde{E}^{(2,1)} _{\mu^{(1)}} S^{(2)}_{\mu^{(2)}} \widetilde{E}^{(2,1)} _{\mu^{(1)}} P^{(2)}_{\mu^{(2)}} ,
\end{split}
\end{gather*}
with 
\begin{gather*}
\begin{split}
&\widetilde{E}_{\mu^{(1)}}^{(2,1)} =-P _{\mu^{(1)}}^{(1)} \widetilde{E}_{\mu}^{(1,1)} \left( P_{\mu^{(1)}}^{(1)} \widetilde{E}_{\mu}^{(1,1)} S_{\mu^{(1)}}^{(1)} + S_{\mu^{(1)}}^{(1)} \widetilde{E}_{\mu}^{(1,1)} P_{\mu^{(1)}}^{(1)}  \right) ,\\
&S_{\mu^{(1)}}^{(1)} = \frac{1}{2\pi i} \oint _{\mathcal{C} (\mu^{(1)})} (z-\mu^{(1)})^{-1} (\widetilde{E}^{(1)} _{\mu} -z)^{-1} dz.
\end{split}
\end{gather*}

\begin{prop}
Suppose that $\widetilde{E}_{\mu}^{(1)}$ is semi-simple.
Every eigenvalue $\mu(\kappa)\in \sigma_p (E(\kappa),\mu,\mu^{(1)})$ has an asymptotic expansion (\ref{Ap_eq_asymptotic2ndorder}) for $\mu^{(2)} \in \sigma_p (\widetilde{E}^{(2)} _{\mu^{(1)}}|_{\mathsf{R}(\mu^{(1)})})$.
We have 
\begin{gather*}
\sum _{\mu^{(2)}\in \sigma_p \big( \widetilde{E}_{\mu}^{(2)} |_{\mathsf{R} (\mu^{(1)})} \big)} P^{(2)}_{\mu^{(2)}} (\kappa) = P_{\mu^{(1)}}^{(1)} (\kappa) , \quad \sum _{\mu^{(2)}\in \sigma_p \big( \widetilde{E}_{\mu^{(1)}}^{(2)} |_{\mathsf{R} (\mu^{(1)})} \big)} P^{(2)}_{\mu^{(2)}} = P_{\mu^{(1)}}^{(1)},  \\
P_{\mu^{(2)}}^{(2)} (\kappa) = P _{\mu^{(2)} }^{(2)} +\kappa P_{\mu^{(2)}}^{(2,1)} +\kappa^2 P_{\mu^{(2)}}^{(2,2)}+ o(\kappa^2 ), \quad \kappa \to 0 .
\end{gather*}

\label{Ap_prop_asymptotic2ndorder}
\end{prop}

\textit{Remark.}
It is generally unknown how many times the reduction process can be repeated.
However, since the eigenvalues of $E(\kappa)$ and $E_0$ have finite multiplicities, an eigenvalue of $E_0$ can split into at most finite number of eigenvalues of $E(\kappa)$.
Then the complete decomposition of $\mathsf{Ran} (P_{\mu}^{\mathsf{total}} (\kappa))$ can be derived by at most a finite number of reduction processes.

\section{Resonant scattering} \label{section_maintheorem}
\subsection{Perturbation of eigenvalues} \label{subsection_perturbationeigenvalue}

By (\ref{Ap_eq_asymptotic2ndorder}), $\mu(\kappa)\in \sigma_p (E(\kappa))$ splitting away from $\mu\in\sigma_p (E_0)$ has an asymptotic expansion as $\kappa\to 0$.
From now on, we characterize the coefficients $\mu^{(1)}$ and $\mu^{(2)}$ by using the spectrum of $T=dSd^*$ and some geometric information of the underlying graph $\Gamma_{\mathsf{int}}$.

Let us consider an eigenvalue $\mu$ of $E(\kappa)$ such that $\mu$ is invariant with respect to $\kappa$ with $|\kappa| \ll 1 $.
Thus $\mu$ must satisfy $ |\mu| =1$ since we have $\mu \in \sigma_p (E_0) \subset S^1 $.
Theorem \ref{S5_thm_SMTMOS} implies that the associated eigenspace $\mathsf{Ker} ( E(\kappa) -\mu )$ is independent of $\kappa$ if the eigenvalue $\mu$ is invariant with respect to $\kappa$ as follows. 
Now let us recall the subspace $\mathcal{Y}=\mathsf{Ran } (Y)$ defined for the operator $Y$ as in (\ref{S5_eq_L}).

\begin{lemma}
Let $\delta >0$ be sufficiently small.
If $\mu \in \sigma_p (E(\kappa))$ for all $\kappa$ with $0\leq |\kappa| \leq \delta$, the subspace $\mathsf{Ker} (E(\kappa)-\mu )$ is given by 
$$
\mathsf{Ker} (E(\kappa)-\mu ) = \{ u\in \mathcal{Y} \ ; \ u= \partial_{\mu}^* f, \ f\in \mathsf{Ker} (T-\Phi_{QW} (\mu)), \ f|_{\partial V_{\mathsf{int}}}=0 \} ,
$$
for $\mu \in S^1 \setminus \{ \pm 1 \}$ and
$$
\mathsf{Ker} (E(\kappa)-\mu ) =  \mathsf{Ker} (E_0|_{\mathcal{Y}^{\perp}} -\mu ) =\mathsf{Ker} (d)\cap \mathsf{Ker} (S+\mu),
$$
for $\mu\in \{ \pm 1 \}$.
Especially, $\mathsf{Ker} (E(\kappa)-\mu)$ is invariant with respect to $\kappa$.
\label{S7_lem_kerindependencekappa}
\end{lemma}

Proof.
Recall the definition of $U(\kappa)$ and $U_0$ as in (\ref{S5_eq_perturbedqws}).
Note that $\mu\in \cap_{0\leq |\kappa| \leq \delta} \sigma_p (U(\kappa))$ due to Lemma \ref{S2_lem_eigenvaluesU}.
Then the corresponding eigenfunctions vainish in $A\setminus A_{\mathsf{int}} $ and $u \in \mathsf{Ker} (E(\kappa)-\mu)$ if and only if $u=\chi_{\mathsf{int}} \psi$ for $\psi \in \mathsf{Ker} (U(\kappa)-\mu )$.
Take an eigenfunction $\psi \in \mathcal{H}$.
Since we have $U(\kappa)\psi (a_{j,1}^{\sharp})=0$ and $\psi |_{A^{\flat}_{v_{j,0}} \setminus A_{\mathsf{int}}} =0$, we can see $\sum_{b\in A^{\flat}_{v_{j,0}} \cap A_{\mathsf{int}}} \psi (b)=0$, i.e., 
\begin{equation}
d\psi (v_{j,0 } )=0, \quad j=1,\ldots,N.
\label{S7_eq_kerindependence001}
\end{equation}
We also see
\begin{equation}
(S+\mu )\psi (a)=0, \quad a\in A^{\sharp}_{v_{j,0}},
\label{S7_eq_kerindependence002}
\end{equation}
by using the equation $U(\kappa)\psi = \mu\psi$ at $a$ and the equality (\ref{S7_eq_kerindependence001}).
Recalling $u=\chi_{\mathsf{int}} \psi$ and (\ref{S7_eq_kerindependence001}), we have
\begin{gather*}
\begin{split}
E(\kappa) u &= E_0  u +\kappa E_0^{(1)} u \\
&=E_0 u -\kappa Sd^* Bdu \\
&=E_0 u.
\end{split}
\end{gather*}
Then $u=\chi_{\mathsf{int}}\psi$ satisfies $E_0 u= \mu u$, (\ref{S7_eq_kerindependence001}) and (\ref{S7_eq_kerindependence002}) which are independent of $\kappa$.

Let us consider the case $\mu \in S^1 \setminus \{ \pm 1 \}$.
Theorem \ref{S5_thm_SMTMOS} implies that $u = \partial_{\mu}^* f$ for $f\in \mathsf{Ker } (T-\Phi_{QW} (\mu))$.
It follows that $f(v_{j,0})=0 $ for $j=1,\ldots,N$ from (\ref{S7_eq_kerindependence002}) as
\begin{gather*}
\begin{split}
0&=(S+\mu )\partial _{\mu}^* f(a)\\
&=\frac{1}{\sqrt{2} |\sin (\mathrm{arg} \, \mu )|} (1-\mu ^2 )d^* f(\overline{a})\\
&=\frac{1}{\sqrt{2} |\sin (\mathrm{arg} \, \mu )|} (1-\mu ^2 ) f(v_{j,0}),
\end{split}
\end{gather*}
for $a\in A^{\sharp}_{v_{j,0}} $.

Let us turn to the case $\mu\in \{ \pm 1 \}$.
Note that the multiplicity of $1\in \sigma_p (T)$ is one and that of $-1\in \sigma_p (T)$ is also one (if $\Gamma_{\mathsf{int}}$ is bipartite) due to Theorem \ref{S5_thm_SMTMOS}.
Then we can split $V_{\mathsf{int}}$ into two subsets: $V_{\mathsf{int}} = V_1 \sqcup V_2$ and $o(a)\in V_1$ for an edge $a\in A_{\mathsf{int}}$ if and only if $t(a)\in V_2$.
The subspace $ \mathsf{Ker} (T-\mu)$ is spanned by the function
\begin{gather*}
f(v)= \left\{
\begin{split}
&1 , \quad v\in V_{\mathsf{int}}, \quad \mu=1, \\
&\left\{
\begin{split}
1 &, \quad v\in V_1,\\
-1 &, \quad v\in V_2,\\
\end{split}
\right. \quad \mu=-1.
\end{split}
\right.
\end{gather*}
In these cases, $u= \partial_{\mu}^* f= d^* f$ does not satisfy the condition (\ref{S7_eq_kerindependence001}) since we see
$$
\left| \sum_{b\in A^{\flat}_{v_{j,0}} \cap A_{\mathsf{int}}} u (b) \right| = \left| \sum_{b\in A^{\flat}_{v_{j,0}} \cap A_{\mathsf{int}}}  f(v_{j,0}) \right| = n_{\mathsf{i}} (v_{j,0})> 0.
$$
On the other hand, $u\in \mathsf{Ker} (d)\cap \mathsf{Ker} (S+\mu)$ trivially satisfies (\ref{S7_eq_kerindependence001}) and (\ref{S7_eq_kerindependence002}).

Now we have proved 
\begin{gather*}
\begin{split}
&\mathsf{Ker} (E(\kappa)-\mu) \\
&\subset \left\{
\begin{split}
\{ u\in \mathcal{Y} \ ; \ u= \partial_{\mu}^* f, \ f\in \mathsf{Ker} (T-\Phi_{QW} (\mu)), \ f|_{\partial V_{\mathsf{int}}}=0 \} &, \quad \mu\in S^1 \setminus \{ \pm 1\} ,\\
\mathsf{Ker} (d)\cap \mathsf{Ker} (S+\mu) &, \quad \mu \in \{ \pm 1 \} .
\end{split}
\right.
\end{split}
\end{gather*}
The converse is clearly valid.
Then we obtain the lemma.
\qed

\medskip

In view of Lemma \ref{S7_lem_kerindependencekappa}, we can define persistent eigenvalues as follows.

\begin{definition}
Suppose that $E(\kappa)$ has an eigenvalue $\mu\in \sigma_p (E(\kappa))$ which is invariant with respect to $\kappa$ for sufficiently small $|\kappa|$.
We call such an eigenvalue a \textit{persistent eigenvalue}.
We define the \textit{persistent eigenspace} $\mathcal{E}_{per}  (\mu) \subset \ell^2 (A_{\mathsf{int}})$ as the subspace spanned by the eigenvectors of a persistent eigenvalue $\mu$ of $E(\kappa)$.
We call every vector in $\mathcal{E}_{per} (\mu)$ a \textit{persistent state} associated with a persistent eigenvalue $\mu$.
\label{S5_def_darkstate}
\end{definition}

\medskip

Lemma \ref{S7_lem_kerindependencekappa} allows us to write $\mathcal{E}_{per} (\mu)$ for a persistent eigenvalue $\mu$ as 
\begin{gather}
\mathcal{E}_{per} (\mu) = \left\{
\begin{split}
\mathsf{Ran} (\partial_{\mu}^* | _{\mathcal{V}_{per} (\mu)})&, \quad \mu\in S^1 \setminus \{ \pm 1\},\\
\mathsf{Ker} (d ) \cap \mathsf{Ker} (S+\mu) &, \quad \mu \in \{ \pm 1 \} ,
\end{split}
\right.
\label{S5_eq_resultpersistentes}
\end{gather}
where 
\begin{equation}
\mathcal{V}_{per} (\mu) =  \left\{ f\in \mathsf{Ker} (T-  \Phi_{QW}  (\mu)) \ ;  \ f|_{\partial V_{\mathsf{int}}} =0 \right\} .
\label{S7_eq_defvper}
\end{equation}
If $g_1,g_2 \in \mathsf{Ker} (T-\Phi_{QW} (\mu) )$ satisfy $\langle g_1 , g_2 \rangle _{\ell^2_{n_{\mathsf{i}}} (V_{\mathsf{int}})} =0$, we can see $\langle \partial _{\mu}^* g_1 , \partial_{\mu}^* g_2 \rangle _{\ell^2 (A_{\mathsf{int}})} =0$ by
\begin{gather}
\begin{split}
\langle \partial_{\mu}^* g_1 , \partial_{\mu}^* g_2 \rangle_{\ell^2 (A_{\mathsf{int}})} 
&=\frac{1}{2\sin^2 (\mathrm{arg} \, \mu )} \langle d(1-\mu S -\overline{\mu} S +1)d^* g_1 ,g_2 \rangle _{\ell^2 _{n_{\mathsf{i}}} (V_{\mathsf{int}})} \\
&=\frac{1}{2\sin^2 (\mathrm{arg} \, \mu )} \langle (2-2\cos^2 (\mathrm{arg} \, \mu )) g_1 ,g_2 \rangle _{\ell^2_{n_{\mathsf{i}}} (V_{\mathsf{int}})} \\
&= \langle  g_1 ,g_2 \rangle _{\ell^2_{n_{\mathsf{i}}} (V_{\mathsf{int}})} = 0 ,
\end{split}
\label{S7_eq_computationparmuast}
\end{gather}
for $\mu \not= \pm 1$ and $\langle \partial_{\mu}^* g_1 , \partial_{\mu}^* g_2 \rangle_{\ell^2 (A_{\mathsf{int}})} = \langle d^* g_1 , d^* g_2 \rangle_{\ell^2 (A_{\mathsf{int}})} =\langle  g_1 ,g_2 \rangle _{\ell^2_{n_{\mathsf{i}}} (V_{\mathsf{int}})} = 0 $ for $\mu = \pm 1$.
Here we have used the equality $(\mu+\overline{\mu} )Tg_j = 2\cos^2 (\mathrm{arg} \, \mu) g_j$.
Then, Theorem \ref{S5_thm_SMTMOS} and Lemma \ref{S7_lem_kerindependencekappa} imply the decomposition
\begin{equation}
\mathsf{Ran} (P_{\mu})= \mathcal{H}_1 (\mu) \oplus \mathcal{E}_{per} (\mu)  ,
\label{S7_eq_RanPmudecomposition}
\end{equation}
with 
\begin{equation}
\mathcal{H} _1 (\mu):= \mathsf{Ker} (E_0|_{\mathcal{Y}} -\mu ) \cap \mathcal{E}_{per} (\mu)^{\perp} = \mathsf{Ran} \big( \partial_{\mu}^* | _{\mathsf{Ker} (T-\Phi_{QW} (\mu)) \cap \mathcal{V}_{per} (\mu)^{\perp}} \big) .
\label{S7_eq_kerE0mudecomposition00} 
\end{equation}
In our main theorem, we are interested in the subspace $\mathcal{H}_1 (\mu)$.

\begin{lemma}
For any $\mu\in \sigma_p (E_0)$, the subspaces $\mathcal{H}_1 (\mu)$ and $\mathcal{E}_{per} (\mu) $ are invariant subspaces of $\widetilde{E}_{\mu}^{(1)}=P_{\mu} E_0^{(1)} P_{\mu} $.

\label{S7_lem_invariantsubspaces}
\end{lemma}

Proof.
By the definition of $\widetilde{E}_{\mu}^{(1)} $, $\mathsf{Ran} (P_{\mu})$ is invariant for $ \widetilde{E}_{\mu}^{(1)}$.
Let us show $\mathsf{Ran} (\widetilde{E}_{\mu}^{(1)}|_{\mathcal{H}_1 (\mu)})\subset \mathcal{H}_1 (\mu) $.
Let $u\in \mathcal{H}_1 (\mu)$ and $v\in \mathcal{E}_{per} (\mu) $.
There extsts $f\in \mathsf{Ker} (T-\Phi_{QW} (\mu))\cap \mathcal{V}_{per} (\mu)^{\perp}$ such that $u=\partial_{\mu}^* f$. 
For the case $\mu \not\in \{ \pm 1 \}$, there exists $g\in \mathcal{V}_{per} (\mu)$ such that $v=\partial_{\mu}^* g$.
It follows that $\langle E_0^{(1)}u,v\rangle_{\ell^2 (A_{\mathsf{int}})} =0$ from
\begin{gather*}
\begin{split}
\langle E_0^{(1)} u,v \rangle _{\ell^2 (A_{\mathsf{int}} )} &= \langle E_0^{(1)} \partial_{\mu}^* f,\partial_{\mu}^* g \rangle _{\ell^2 (A_{\mathsf{int}} )} \\
&=\frac{\mu}{2\sin^2 (\mathrm{arg} \, \mu )} \langle B(T-\overline{\mu})f,(T-\mu)g \rangle _{\ell^2 _{n_{\mathsf{i}} }(V_{\mathsf{int}})} \\
&= -\frac{\mu}{2} \langle Bf,g\rangle  _{\ell^2 _{n_{\mathsf{i}} }(V_{\mathsf{int}})} =0,
\end{split}
\end{gather*}
where we have used the equality $(T-\overline{\mu} ) f = i \sin (\mathrm{arg} \, \mu) f$ and $(T-\mu ) g = -i \sin (\mathrm{arg} \, \mu ) g$.
For the case $\mu \in \{ \pm 1\}$, we have for $v\in \mathsf{Ker} (d)\cap \mathsf{Ker} (S+\mu)$
$$
\langle E_0^{(1)}u,v\rangle _{\ell^2 (A_{\mathsf{int}})} = - \langle Bdu,dSv \rangle _{\ell^2 _{n_{\mathsf{i}} }(V_{\mathsf{int}})} = \mu \langle Bdu,dv \rangle _{\ell^2 _{n_{\mathsf{i}} }(V_{\mathsf{int}})} =0.
$$
Then we have $\widetilde{E}_{\mu}^{(1)} u \in \mathcal{H}_1 (\mu)$.
$\mathsf{Ran} (\widetilde{E}_{\mu}^{(1)}|_{\mathcal{E}_{per} (\mu)} ) \subset \mathcal{E}_{per} (\mu)$ is proved in the similar way.
\qed

\medskip

Let $\{ f_1 , \ldots , f_p \} $ and $\{ g_1 , \ldots , g_q \}$ be a pair of orthonormal bases of $\mathsf{Ker} (T-  \Phi_{QW}  (\mu))\cap \mathcal{V}_{per} (\mu)^{\perp}$ and $\mathcal{V}_{per} (\mu)$, respectively.

\begin{lemma}
Let $\mu\in \sigma_p (E_0 )$.
The following facts hold true.
\begin{enumerate}
\item
 Letting $u_j = \partial_{\mu}^* f_j$ for $j=1,\ldots,p$, it follows that $\{ u_1 , \ldots , u_p \}$ is an orthonormal basis of $\mathcal{H}_1 (\mu) $.

\item
 Letting $v_k= \partial_{\mu}^* g_k$ for $k=1,\ldots,q$, it follows that $\{ v_1 , \ldots,v_q \}$ is an orthonormal basis of $\mathcal{E}_{per} (\mu)$ for $\mu \in S^1 \setminus \{ \pm 1\}$.

\item
 $ f_1 |_{\partial V_{\mathsf{int}}} ,\ldots, f_p |_{\partial V_{\mathsf{int}}} $ are linearly independent.
\end{enumerate}
\label{S5_lem_orthonormalbasis}
\end{lemma}

Proof.
We obtain $\langle u_j,u_k \rangle _{\ell^2 (A_{\mathsf{int}})} = \langle f_j,f_k \rangle _{\ell^2_{n_{\mathsf{i}}} (V_{\mathsf{int}})} =\delta_{j,k}$ and $\langle v_j,v_k \rangle _{\ell^2 (A_{\mathsf{int}})} = \langle g_j,g_k \rangle _{\ell^2_{n_{\mathsf{i}}} (V_{\mathsf{int}})} =\delta_{j,k}$ in the same way as (\ref{S7_eq_computationparmuast}).
This equality, Theorem \ref{S5_thm_SMTMOS}, Lemma \ref{S7_lem_kerindependencekappa} and (\ref{S7_eq_RanPmudecomposition}) imply the assertions (1) and (2).

Suppose that $c_1 f_1 |_{\partial V_{\mathsf{int}}} +\cdots +c_p f_p |_{\partial V_{\mathsf{int}}} =0$ for some constants $c_1 , \ldots ,  c_p$.
Letting $f=c_1 f_1  +\cdots +  c_p f_p $, we have $f \in \mathsf{Ker} (T-  \Phi_{QW}  (\mu)) \cap \mathcal{V}_{per}  (\mu)$.
Since $\{ f_1 , \ldots , f_p \} $ is an orthonormal basis of $\mathsf{Ker} (T- \Phi_{QW}  (\mu))\cap \mathcal{V}_{per}  (\mu)^{\perp}$, we obtain $f \in \mathcal{V}_{per} (\mu) \cap \mathcal{V}_{per}  (\mu)^{\perp} $.
Then $f =0$ on $V_{\mathsf{int}}$, so that $c_1 =\cdots =  c_p =0 $.
The assertion (3) has been proved.
\qed

\medskip

Let $ P_{\mu,1} $ and $P_{\mu,2}$ be the orthogonal projections onto $\mathcal{H}_1 (\mu)$ and $\mathcal{E}_{per} (\mu)$, respectively. 
Note that $P_{\mu} =  P_{\mu,1}+P_{\mu,2}$ and $P_{\mu} P_{\mu,j}  = P_{\mu,j}  P_{\mu} = P_{\mu,j}  $ hold true.
The semi-simplicity of eigenvalues of $\widetilde{E}_{\mu}^{(1)}$ are guaranteed in the next lemma.
Here we define the matrix $M^{(1)} _{\mu} $ by
$$
M_{\mu}^{(1)}= \left[ m_{\mu,j,k}^{(1)} \right]_{j,k=1}^p , \quad
m_{\mu,j,k}^{(1)} = -\langle B f_j,f_k \rangle_{\ell^2_{n_{\mathsf{i}}} (V_{\mathsf{int}})} ,
$$
 where $\{ f_1,\ldots,f_p \}$ is given as above.
Note that $M_{\mu}^{(1)}$ is Hermitian.

\begin{lemma}
Let $\mu \in\sigma_p (E_0 ) $.
For $\widetilde{E}_{\mu}^{(1)}=P_{\mu} E_0^{(1)} P_{\mu}$, the following assertions hold true.
\begin{enumerate}
\item
As in Lemma \ref{S5_lem_orthonormalbasis}, we choose an orthonormal basis $\{ u_1 , \ldots ,u_p \}$ of $\mathcal{H}_1 (\mu)$ with $u_j = \partial_{ \mu }^* f_j$ where $\{ f_1,\ldots,f_p \}$ is an orthonormal basis of $\mathsf{Ker} ( T- \Phi_{QW}  (\mu)) \cap \mathcal{V}_{per}^{\perp}$.
Then the matrix of $\widetilde{E}_{\mu}^{(1)} $ on $\mathcal{H}_1 (\mu)$ is given by $\mu \gamma M^{(1)}_{\mu}$ where $\gamma =1 /2$ for $\mu \in \sigma_p (E_0 |_{ \mathcal{Y}} )\setminus \{ \pm 1 \}$ and $\gamma  =1$ for $\mu \in  \{ \pm 1 \}$.
In particular, we have 
$$
\sigma_p (\widetilde{E}_{\mu}^{(1)}  |_{\mathsf{Ran} ( P_{\mu,1})}) = \sigma_p ( \mu \gamma  M_{\mu}^{(1)}).
$$
The eigenvalues of $M_{\mu}^{(1)} $ are semi-simple and negative. 
There exists a constant $c \geq 1$ which depends only on $\Gamma_{\mathsf{int}}$ and $ \partial V_{\mathsf{int}}$ such that 
\begin{equation}
- \left(  \max_{v_{j,0} \in \partial V_{\mathsf{int}}} \frac{ N_j }{n(v_{j,0})} \right)  \leq M_{\mu}^{(1)} <-c ^{-1}  \left( \min_{v_{j,0} \in \partial V_{\mathsf{int}}} \frac{N_j}{n(v_{j,0})} \right)  .
\label{S5_eq_normMmu1}
\end{equation}

\item
The matrix of $\widetilde{E}_{\mu}^{(1)} $ on $\mathcal{E}_{per} (\mu)$ is $0$.

\item
Especially, $\widetilde{E}_{\mu}^{(1)}$ is semi-simple.

\end{enumerate}

\label{S5_lem_semisimpleE0(1)}
\end{lemma}

Proof. 
Let us compute $\langle E_0^{(1)} u_j , u_k \rangle_{\ell^2 (A_{\mathsf{int}})} $ for $\mu \in \sigma_p (E_0 | _{ \mathcal{Y}} ) \setminus \{ \pm 1 \} $ as
\begin{gather}
\begin{split}
\langle E_0^{(1)} u_j , u_k \rangle_{\ell^2 (A_{\mathsf{int}})} &=\langle \partial_{\mu} E_0^{(1)} \partial_{\mu}^*  f_j , f_k \rangle_{\ell^2_{n_{\mathsf{i}}} (V_{\mathsf{int}})} \\
&=\frac{\mu}{2\sin^2 (\mathrm{arg} \, \mu )} \langle (T-\overline{\mu} )  B (T-\overline{\mu})  f_j,f_k \rangle _{\ell^2_{n_{\mathsf{i}}} (V_{\mathsf{int}} )} \\
&= \frac{\mu}{2} m_{\mu,j,k}^{(1)},
\end{split}
\label{S7_eq_calcsubspace00}
\end{gather}
where we have used the equality $(T-\mu)  f_k = -  i \sin (\mathrm{arg} \, \mu)  f_k $ and $(T-\overline{\mu})  f_j  = i \sin (\mathrm{arg} \, \mu ) f_j $ for $\mu \in \sigma_p (E_0 | _{ \mathcal{Y}} ) \setminus \{ \pm 1 \} $, and
$$
\langle E_0^{(1)} u_j , u_k \rangle_{\ell^2 (A_{\mathsf{int}})} = \langle \partial_{\mu} E_0^{(1)} \partial_{\mu}^*  f_j , f_k \rangle_{\ell^2_{n_{\mathsf{i}}} (V_{\mathsf{int}})} 
= -\langle  B  f_j, Tf_k \rangle _{\ell^2_{n_{\mathsf{i}}} (V_{\mathsf{int}} )} = \mu m_{\mu,j,k}^{(1)},
$$
for $\mu\in \{ \pm 1 \} $.
Then the matrix of $\widetilde{E}_{\mu}^{(1)}$ on $\mathcal{H}_1 (\mu)$ associated with the orthonormal basis $\{ u_1, \ldots ,u_p \}$ is given by $\mu \gamma M_{\mu} ^{(1)}$.
 Since the matrix $ M_{\mu}^{(1)} $ is Hermitian, eigenvalues of $ M_{\mu}^{(1)} $ are semi-simple.

Let us check that $M_{\mu}^{(1)}$ is negative definite.
Actually, $-M_{\mu}^{(1)}$ is a Gram matrix: letting $F=B^{1/2} [f_1,\ldots,f_p]$, we have $-M_{\mu}^{(1)} =  F^* F \geq 0$.
Suppose that $-M_{\mu}^{(1)} w=0$ for $w=[w_1 ,\ldots , w_p]^{\mathsf{T}} \in {\bf C}^p$.
In view of $0=\langle F^* F w,w\rangle_{{\bf C}^s} = \|  F w\|^2_{\ell^2_{n_{\mathsf{i}}} (V_{\mathsf{int}})} $, we have $ F w=0$.
This implies $ w_1 f_1 | _{\partial V_{\mathsf{int}}} + \cdots +  w_p f_p | _{\partial V_{\mathsf{int}}} =0$.
Then $w=0$ follows from the assertion (3) in Lemma \ref{S5_lem_orthonormalbasis}.
We obtain $ -M_{\mu}^{(1)} >0$.

The estimates (\ref{S5_eq_normMmu1}) are shown as follows.
We have 
$$
\|  Fw\|^2_{\ell^2_{n_{\mathsf{i}}} (V_{\mathsf{int}})} \leq  \left(  \max_{v_{j,0} \in \partial V_{\mathsf{int}}} \frac{ N_j }{n(v_{j,0})} \right)  \| [  f_1,\ldots,f_p]  w\|^2_{\ell^2_{n_{\mathsf{i}}} (V_{\mathsf{int}})} = \left(  \max_{v_{j,0} \in \partial V_{\mathsf{int}}} \frac{ N_j }{n(v_{j,0})} \right)  \| w\|^2_{{\bf C}^s} .
$$
Here we have used $ [f_1 , \ldots ,f_p]^* [f_1 , \ldots ,f_p] =[\langle f_j , f_k \rangle _{\ell^2_{n_{\mathsf{i}}} (V_{\mathsf{int}})} ]_{j,k=1}^p  = I_p $.
This estimate and the inequality $\| -M_{\mu}^{(1)} \|_{{\bf B} ({\bf C}^s )} \leq \|  F \|_{{\bf B} ({\bf C}^s ; \ell^2_{n_{\mathsf{i}}} (V_{\mathsf{int}}))}^2 $ imply the first inequality of (\ref{S5_eq_normMmu1}).
For the second inequality of (\ref{S5_eq_normMmu1}), let us show that there exists a constant $c \geq 1$ such that 
$$
\| g\| _{\ell^2_{n_{\mathsf{i}}} (V_{\mathsf{int}})} \leq c\| 1_{\partial V_{\mathsf{int}}} g \|_{\ell^2_{n_{\mathsf{i}}} (V_{\mathsf{int}})}, \quad g\in \mathsf{Ker} (T-  \Phi_{QW}  (\mu))\cap \mathcal{V}_{per}^{\perp}.
$$
Here $1_{\partial V_{\mathsf{int}}}$ is the characteristic function of $\partial V_{\mathsf{int}}$.
If there does not exist such a constant, we can take a sequence $\{ g^{(n)} \} _{n=1}^{\infty} \subset \mathsf{Ker} (T-  \Phi_{QW}  (\mu))\cap \mathcal{V}_{per}^{\perp}$ such that $\| g^{(n)} \| _{\ell^2_{n_{\mathsf{i}}} (V_{\mathsf{int}})} =1$ and $\| 1_{\partial V_{\mathsf{int}}} g^{(n)} \|_{\ell^2_{n_{\mathsf{i}}} (V_{\mathsf{int}})} \to 0$ as $n\to \infty $.
In view of Bolzano-Weierstrass theorem, we can take a subsequence $\{ g^{(n_k)} \} _{k=1}^{\infty}$ such that $g^{(n_k)}$ converges to a function $g\in \mathsf{Ker} (T-  \Phi_{QW} (\mu))\cap \mathcal{V}_{per}^{\perp}$.
However, we also have $1_{\partial V_{\mathsf{int}}} g=0$ so that $g\in \mathcal{V}_{per}$.
Then we see $g=0$ which contradicts $\| g\| _{\ell^2_{n_{\mathsf{i}}} (V_{\mathsf{int}})} = \lim_{k\to \infty} \| g^{(n_k)} \| _{\ell^2_{n_{\mathsf{i}}} (V_{\mathsf{int}})} =1$.
Now we compute
\begin{gather*}
\begin{split}
\langle -M_{\mu}^{(1)} w,w \rangle _{{\bf C}^s} &= \|  B^{1/2} [ f_1,\ldots,f_p ]w \|^2 _{\ell^2_{n_{\mathsf{i}}} (V_{\mathsf{int}})} \\
&\geq  \left( \min_{v_{j,0} \in \partial V_{\mathsf{int}}} \frac{N_j}{n(v_{j,0})} \right)  \| 1_{\partial V_{\mathsf{int}}} [ f_1,\ldots, f_p  ]w\|^2_{\ell^2_{n_{\mathsf{i}}} (V_{\mathsf{int}})} \\
&\geq c^{-1}  \left( \min_{v_{j,0} \in \partial V_{\mathsf{int}}} \frac{N_j}{n(v_{j,0})} \right)  \|  [ f_1,\ldots,f_p ]w\|^2_{\ell^2 _{n_{\mathsf{i}}} (V_{\mathsf{int}})} \\
&= c^{-1}  \left( \min_{v_{j,0} \in \partial V_{\mathsf{int}}} \frac{N_j}{n(v_{j,0})} \right)  \|  w\|^2_{{\bf C}^s}
\end{split}
\end{gather*}
which shows the desired estimate.
The assertion (1) has been proved.

Let us turn to the assertion (2).
For an orthonormal basis $\{ v_1 , \ldots , v_q  \}$ of $\mathcal{E}_{per} (\mu)$, we can see $\langle E_0^{(1)} v_j ,v_k \rangle _{\ell^2 (A_{\mathsf{int}} )}=0$ by some calculations similar to (\ref{S7_eq_calcsubspace00}). 
Here we have used Lemma \ref{S5_lem_orthonormalbasis} in order to see that there exists an orthonormal basis $\{ g_1 ,\ldots,g_q\}$ of $\mathcal{V}_{per} (\mu)$ such that $v_j = \partial_{\mu}^* g_j$ for $j=1,\ldots,q$.
Then we obtain (2).

Due to Lemma \ref{S7_lem_invariantsubspaces} and the assertions (1)-(2), the matrix of $\widetilde{E}_{\mu}^{(1)}$ on $\mathsf{Ran} (P_{\mu})$ is given by 
$$
\mathrm{diag} \left[ \mu \gamma  M_{\mu}^{(1)} , 0 \right] \quad \text{on} \quad \mathcal{H}_1 (\mu) \oplus \mathcal{E} _{per} (\mu) ,
$$
as the block matrix representation.
Note that $\mu\gamma M_{\mu}^{(1)}$ is a normal matrix.
Then $\widetilde{E}_{\mu}^{(1)}$ is semi-simple.
\qed

\medskip

Due to Lemma \ref{S5_lem_semisimpleE0(1)}, the semi-simplicity of $\widetilde{E}_{\mu}^{(1)}$ has been confirmed and we can apply Proposition \ref{Ap_prop_asymptotic2ndorder} to $E(\kappa)$.
However, in the remaining part, it is difficult to study the resonances in detail without any restriction on $P_{\mu(\kappa)}$ for $\mu (\kappa) \in \sigma_p (E(\kappa))$.
Then we introduce the third assumption as follows.
In the remaining part, we always assume that Assumptions 1-3 hold true.

\medskip

{\bf Assumption 3.}
Let $\mu (\kappa) \in\sigma_p ( E(\kappa) ,\mu ) $ for $\mu\in \sigma_p (E_0 |_{ \mathcal{Y}})$.
Recall $\mathsf{R}(\mu)=\mathsf{Ran} (P_{\mu})$ and $\mathsf{R}(\mu^{(1)})=\mathsf{Ran} \big( P _{\mu^{(1)}}^{(1)} \big)$.
\begin{enumerate}
\item
The generalized eigenspace $\mathsf{Ran} (P_{\mu(\kappa) })$ satisfies 
$$
\mathsf{Ran} (P_{\mu(\kappa) }) =  \mathsf{Ran} \big( P_{\mu^{(2)}}^{(2)} ( \kappa(\epsilon) ) \big) , 
$$
for one of pairs of eigenvalues $\mu^{(1)} \in \sigma_p (\widetilde{E} ^{(1)} _{\mu} | _{\mathsf{R}(\mu)})$ and $\mu^{(2)} \in \sigma_p \big( \widetilde{E} ^{(2)} _{\mu^{(1)}} | _{\mathsf{R}(\mu^{(1)})} \big) $.

\item
The eigenspace $\mathsf{Ran} (P_{\mu})=\mathsf{Ker} (E_0 -\mu)$ satisfies 
$$
\mathsf{Ker} (E_0 -\mu) = \bigoplus _{\mu^{(1)}\in \sigma_p (\widetilde{E} ^{(1)} _{\mu} | _{\mathsf{R}(\mu)})} \left( \bigoplus  _{\mu^{(2)} \in \sigma_p \big( \widetilde{E} ^{(2)} _{\mu^{(1)}} | _{\mathsf{R}(\mu^{(1)})} \big) } \mathsf{Ran} \big( P_{\mu^{(2)}}^{(2)} \big) \right) .
$$ 

\end{enumerate}

\medskip

By the second order reduction process of perturbed eigenvalues (Proposition \ref{Ap_prop_asymptotic2ndorder}), we can see the asymptotic expansion of second order for $\mu(\kappa) \in \sigma_p (E(\kappa),\mu,\mu^{(1)},\mu^{(2)})$ with sufficiently small $|\kappa| \ll 1$ as 
\begin{equation}
 \mu(\kappa ) = \mu+ \kappa \mu^{(1)} +\kappa^2 \mu^{(2)} +o(\kappa^2 ), \quad \kappa \to 0 ,
\label{S5_eq_asymptotic_evmukappa001}
\end{equation}
where
$$
\mu^{(1)}\in \sigma_p (\widetilde{E}_{\mu}^{(1)} |_{\mathsf{R} (\mu)}),  \quad \mu^{(2)}\in \sigma_p \big( \widetilde{E}_{\mu^{(1)}}^{(2)} |_{\mathsf{R} (\mu^{(1)})} \big) .  
$$

\begin{lemma} 
Suppose that  $\mu (\kappa) \in \sigma_p (E(\kappa),\mu,\mu^{(1)} ,\mu^{(2)})$ for $\mu\in \sigma_p (E_0)$ satisfies $| \mu (\kappa) |<1$.
Then we have $\mu^{(1)}\in \sigma_p \big( P_{\mu^{(2)}}^{(2)} \widetilde{E}_{\mu}^{(1)} P_{\mu^{(2)}}^{(2)} |_{\mathsf{Ran} (P_{\mu,1})} \big) $, $\mu^{(2)}\in \sigma_p \big( P_{\mu^{(2)}}^{(2)} \widetilde{E}_{\mu^{(1)}}^{(2)} P_{\mu^{(2)}}^{(2)} | _{\mathsf{Ran} (P_{\mu,1})} \big) $  and $\mu^{(1)}$ is of the form $\mu \gamma  \eta^{(1)} $ for a negative eigenvalue $\eta^{(1)}$ of $M_{\mu}^{(1)} $ and the constant $\gamma$ taken as in Lemma \ref{S5_lem_semisimpleE0(1)}.

\label{S5_lem_mu1covergence}
\end{lemma}

Proof. 
Under the assumption $|\mu(\kappa)|<1$, $\mu(\kappa)$ is never persistent eigenvalue.
Then we see $ \mathsf{Ran} \big( P_{\mu^{(2)}}^{(2)} \big)  \subset \mathcal{H}_1 (\mu)$ by Assumption 3 and the asymptotic behaviors $P _{\mu(\kappa)} = P^{(2)} _{\mu^{(2)}} +\kappa P _{\mu^{(2)}}^{(2,1)} +O(\kappa^2)$ and $\mu(\kappa)=\mu+\kappa \mu^{(1)} +O(\kappa^2)$ as $\kappa \to 0$ (see Proposition \ref{Ap_prop_asymptotic2ndorder}).
We show 
\begin{equation}
D_{\mu(\kappa)} = P _{\mu(\kappa)} (E(\kappa)-\mu(\kappa))  P _{\mu(\kappa)} =-\kappa^2 \mu^{(2)} P _{\mu^{(2)}}^{(2)} +O(\kappa^3).
\label{S7_eq_Dconverge00}
\end{equation}
Actually, we have 
\begin{gather}
\begin{split}
&P _{\mu(\kappa)} (E(\kappa)-\mu(\kappa))  P _{\mu(\kappa)} \\
&= \kappa P _{\mu^{(2)}}^{(2)} (E_0^{(1)} -\mu^{(1)})  P _{\mu^{(2)}}^{(2)} +\kappa^2 \left( - \mu^{(2)} P_{\mu^{(2)}}^{(2)} +P_{\mu^{(2)}}^{(2,1)} (E_0 -\mu ) P_{\mu^{(2)}}^{(2,1)}  \right)  +O(\kappa^3 ),
\end{split}
\label{S7_eq_Dconverge01}
\end{gather}
where we have used the equalities $(E_0 -\mu ) P_{\mu^{(2)}}^{(2)} = P_{\mu^{(2)}}^{(2)} ( E_0 -\mu )=0 $.
Recall $ \mathsf{Ran} \big(  P_{\mu^{(2)}}^{(2)} \big) \subset \mathsf{Ran} \big(  P_{\mu^{(1)}}^{(1)} \big) $. 
In view of Lemma \ref{S5_lem_semisimpleE0(1)}, $\widetilde{E} _{\mu}^{(1)}$ is semi-simple. 
Then we have $P_{\mu^{(2)}}^{(2)} (E_0^{(1)} - \mu^{(1)} )  P_{\mu^{(2)}}^{(2)} =0$.
By definition, it follows $S_{\mu^{(2)}}^{(2)} = P _{\mu^{(1)}}^{(1)} S_{\mu^{(2)}}^{(2)} = S_{\mu^{(2)}}^{(2)}  P _{\mu^{(1)}}^{(1)} $.
Thus we can see 
$$
P_{\mu^{(2)}}^{(2,1)} (E_0 -\mu ) P_{\mu^{(2)}}^{(2,1)} = P _{\mu^{(2)}}^{(2)} \widetilde{E} _{\mu^{(1)}}^{(2,1)} S_{\mu^{(2)}}^{(2)} (E_0 -\mu ) S_{\mu^{(2)}}^{(2)} \widetilde{E} _{\mu^{(1)}}^{(2,1)} P _{\mu^{(2)}}^{(2)} =0.
$$
Therefore, we obtain (\ref{S7_eq_Dconverge00}) from (\ref{S7_eq_Dconverge01}).

Now we consider the equality 
$$
P_{\mu(\kappa)} \kappa^{-1} ( E(\kappa)- E_0) P_{\mu(\kappa)} = P _{\mu(\kappa)} \kappa^{-1} (\mu(\kappa) - E_0 )P _{\mu(\kappa)} + \kappa^{-1} D_{\mu (\kappa)} .
$$
As $\kappa \to 0$, the left-hand side converges to $P_{\mu^{(2)}} ^{(2)} E_0^{(1)} P_{\mu^{(2)}} ^{(2)} $.
In view of (\ref{S7_eq_Dconverge00}), the right-hand side is
\begin{gather*}
\begin{split}
&P _{\mu(\kappa)} \kappa^{-1} (\mu(\kappa) - E_0 )P _{\mu(\kappa)} + \kappa^{-1} D_{\mu (\kappa)}\\
&=- \kappa^{-1} P_{\mu^{(2)}}^{(2)} (E_0-\mu )P_{\mu^{(2)}}^{(2)} + \mu^{(1)} P _{\mu^{(2)}}^{(2)} - P _{\mu^{(2)}}^{(2)} (E_0 -\mu ) P _{\mu^{(2)}}^{(2,1)} -    P _{\mu^{(2)}}^{(2,1)} (E_0 -\mu )P _{\mu^{(2)}}^{(2)} +O(\kappa) \\
&=\mu^{(1)} P _{\mu^{(2)}}^{(2)} +O(\kappa),
\end{split}
\end{gather*}
as $\kappa \to 0$.
Then we obtain $P _{\mu^{(2)}}^{(2)} E_0^{(1)} P _{\mu^{(2)}}^{(2)} = \mu^{(1)} P _{\mu^{(2)}}^{(2)}$.
Since we have $ \mathsf{Ran} \big( P_{\mu^{(2)}}^{(2)} \big) \subset \mathcal{H}_1 (\mu)$, Lemma \ref{S5_lem_semisimpleE0(1)} implies $\mu^{(1)}=\mu \gamma \eta^{(1)}$ for a negative eigenvalue $\eta^{(1)} \in \sigma_p ( M_{\mu}^{(1)})$.
Especially, $\mu^{(1)} $ does not vanish. 

The generalized eigenspace of $\mu^{(2)}$ is coincides with $\mathsf{Ran} \big( P_{\mu^{(2)}}^{(2)} \big) \subset \mathcal{H}_1 (\mu)$.
Then we have $ \mu^{(2)}\in \sigma_p \big( P_{\mu^{(2)}}^{(2)} \widetilde{E}_{\mu^{(1)}}^{(2)} P_{\mu^{(2)}}^{(2)} | _{\mathsf{Ran} (P_{\mu,1})} \big) $.
\qed

\subsection{Resonant scattering}
Now we are ready to discuss the resonant scattering.
Here we go back to the parameter $\epsilon$ by using the relation $\kappa=\kappa (\epsilon)=1-e^{i\pi \epsilon}$.
Due to the notations $U(\kappa)=U_0+\kappa U_0^{(1)}$, $E(\kappa)=E_0 +\kappa E_0^{(1)}$, $\mu (\kappa)\in \sigma_p (E(\kappa))$, et al., as in \S \ref{SS_tunableqw0000}-\S \ref{subsection_perturbationeigenvalue}, we note that
$$
U_{\epsilon} = U(\kappa (\epsilon)), \quad E_{\epsilon}=E(\kappa (\epsilon)), \quad \mu _{\epsilon} = \mu (\kappa (\epsilon) ), 
$$
and so on.
In view of Corollary \ref{S4_cor_repscatteringmatrix}, we see that the scattering matrix $\Sigma_{\epsilon} (\lambda)$ is given by $\Sigma_{\epsilon} (\lambda)\alpha^{\flat} = \alpha _{\epsilon} ^{\sharp} (\lambda)=[\alpha_{\epsilon,1} ^{\sharp} (\lambda), \ldots, \alpha_{\epsilon,N}^{\sharp} (\lambda)]^{\mathsf{T}}$ with
\begin{gather}
\begin{split}
&\alpha_{\epsilon,j}^{\sharp} (\lambda)-U_{\epsilon} \varphi_0 |_{a=a_{j,1}^{\sharp}} \\
&= \left. \sum_{\mu_{\epsilon}\in \sigma_p (E_{\epsilon})\setminus S^1} \sum _{s=0}^{m(\mu_{\epsilon})-1} U_{\epsilon} \chi_{\mathsf{int}}^* \left( \frac{P_{\mu_{\epsilon}} (E_{\epsilon} -\mu_{\epsilon})^s P_{\mu_{\epsilon}} }{(e^{-i\lambda} -\mu_{\epsilon})^{s+1} } \right) \chi_{\mathsf{int}} U_{\epsilon} \varphi_0 \right| _{a=a_{j,1}^{\sharp}}  ,
\end{split}
\label{S5_eq_perturbedscatteringmatrix}
\end{gather}
for an incoming data $\alpha^{\flat}$ and the associated outgoing data $\alpha_{\epsilon}^{\sharp} (\lambda)$.

 Let us derive the asymptotic behavior of the denominator on the right-hand side of (\ref{S5_eq_perturbedscatteringmatrix}).
There exists a negative eigenvalue $\eta^{(1)} \in \sigma_p (M_{\mu}^{(1)})$ such that $\mu^{(1)} =\mu \gamma  \eta^{(1)}$ in view of Lemmas \ref{S5_lem_semisimpleE0(1)} and \ref{S5_lem_mu1covergence}.
The asymptotic behavior (\ref{S5_eq_asymptotic_evmukappa001}) is rewritten as 
\begin{gather}
\begin{split}
\mu_{\epsilon} &= \mu + \left( -i\pi \epsilon + \frac{\pi^2 \epsilon^2}{2}\right) \gamma \mu \eta^{(1)} + \left( -i\pi \epsilon + \frac{\pi^2 \epsilon^2}{2}\right)^2  \mu^{(2)} +o(\epsilon^2) \\ 
&= \mu \left( 1-i\pi  \gamma \eta^{(1)} \epsilon \right) + \epsilon^2 \pi^2 \left( \frac{ \gamma \mu \eta^{(1)}}{2} - \mu^{(2)} \right) +o(\epsilon^2 ) \\ 
&= \mu e^{-i\pi \gamma \eta^{(1)} \epsilon } + \frac{\epsilon^2 \pi^2}{2} \left(  \mu \gamma \eta^{(1)} (\gamma \eta^{(1)}+1) -2 \mu^{(2)} \right) +o(\epsilon^2 ) .
\end{split}
\label{S5_eq_puiseux02}
\end{gather}
Similarly, we also see that 
\begin{equation}
U_{\epsilon} = U_0 -i\pi \epsilon U_0^{(1)}+O(\epsilon^2), \quad E_{\epsilon} = E_0 -i\pi \epsilon E_0^{(1)}+O(\epsilon^2).
\label{S7_eq_puiseux003} 
\end{equation}
We derive the asymptotic behavior of $ P_{\mu_{\epsilon}} (E_{\epsilon}-\mu_{\epsilon})^s P_{\mu_{\epsilon}}$ for $0\leq s\leq m(\mu_{\epsilon})-1$ as follows.
\begin{lemma}
Let $0\leq s\leq m(\mu_{\epsilon})-1$.
For $\mu_{\epsilon}\in \sigma_p (E(\kappa(\epsilon)),\mu,\mu^{(1)},\mu^{(2)})\setminus S^1$, we have
\begin{equation}
P_{\mu_{\epsilon}}(E_{\epsilon}-\mu_{\epsilon})^s P _{\mu_{\epsilon}}=(\pi^2 \mu^{(2)}\epsilon^2 )^s P _{\mu^{(2)}}^{(2)} +O(\epsilon^{2s+1}),
\label{S7_eq_PEPasymptotic}
\end{equation}
as $\epsilon\downarrow 0$.
\label{S7_lem_asymptoticPEPs}
\end{lemma}

Proof.
Recall $\kappa=\kappa (\epsilon)=1-e^{i\pi \epsilon}$.
By using (\ref{S7_eq_Dconverge00}), we obtain $ P_{\mu_{\epsilon}}(E_{\epsilon}-\mu_{\epsilon}) P _{\mu_{\epsilon}} = -\kappa^2 \mu^{(2)} P_{\mu^{(2)}}^{(2)} +O(\kappa^3)$. 
Finally, applying $\kappa = 1-e^{i\pi \epsilon} =-i\pi \epsilon + (\pi^2 \epsilon^2 )/2 +O(\epsilon^3)$, we have proved the lemma.
\qed

\medskip

Let us note a technical comment for $\mu^{(1)}$ and $\mu^{(2)} $.
The constants $\mu \gamma \eta^{(1)} (\gamma \eta^{(1)}+1)$ and $\mu \gamma \eta^{(1)} (\gamma \eta^{(1)}+1)-2\mu^{(2)}$ which have crucial roles appear in the main result (Theorem \ref{S5_thm_resonantscattering}). 
Especially, $\mu \gamma \eta^{(1)} (\gamma \eta^{(1)}+1)$ and $\mu \gamma \eta^{(1)} (\gamma \eta^{(1)}+1)-2\mu^{(2)}$ appear in the denominators of the principal terms of the resonance expansion of the scattering matrix $\Sigma _{\epsilon} (\lambda)$.
We can see $\mu\gamma\eta^{(1)} (\gamma\eta^{(1)}+1)\not= 0$ by (\ref{S5_eq_normMmu1}).
On the other hand, Assumption 3 guarantees $\mu \gamma \eta^{(1)} (\gamma \eta^{(1)}+1)-2\mu^{(2)} \not= 0$ as follows.

\begin{prop}
Let $\mu_{\epsilon} \in \sigma_p (E_{\epsilon}) \setminus S^1$.
We have $\mu \gamma \eta^{(1)} (\gamma \eta^{(1)}+1)-2\mu^{(2)} \not= 0$.

\label{S7_prop_reductionassumption}
\end{prop}

Proof.
Recall
\begin{gather*}
\begin{split}
&\alpha_{\epsilon,j}^{\sharp} (\lambda)-U_0 \varphi_0 | _{a=a_{j,1}^{\sharp}} \\
&= \sum_{\mu_{\epsilon}\in \sigma_p (E_{\epsilon})\setminus S^1} \sum_{s=0}^{m(\mu_{\epsilon})-1}  \left.   U_{\epsilon} \chi _{\mathsf{int}} \left( \frac{ P_{\mu_{\epsilon}}(E_{\epsilon}-\mu_{\epsilon})^s P _{\mu_{\epsilon}} }{(e^{-i\lambda}-\mu_{\epsilon})^{s+1}} \right) \chi_{\mathsf{int}} U_{\epsilon} \varphi_0 \right| _{a=a_{j,1}^{\sharp}}+O(\epsilon) ,
\end{split}
\end{gather*}
which follows from (\ref{S5_eq_perturbedscatteringmatrix}).

Suppose that $\mu \gamma \eta^{(1)} (\gamma \eta^{(1)}+1)-2\mu^{(2)} = 0$ for some eigenvalues $\mu^{(1)}$ and $\mu^{(2)} $.
We put
\begin{gather*}
\begin{split}
\widetilde{\sigma}_p^{(0)} (E(\kappa(\epsilon)),\mu,\mu^{(1)}) = \left\{ \mu_{\epsilon} \in \sigma_p (E(\kappa(\epsilon)),\mu,\mu^{(1)}) \ ; \ \begin{split} &\mu_{\epsilon} \ \text{satisfies} \\ & \mu \gamma \eta^{(1)} (\gamma \eta^{(1)}+1)-2\mu^{(2)} = 0 \\ &\text{in its asymptotic expansion.} \end{split} \right\} .
\end{split}
\end{gather*}
Take $\lambda_{\epsilon}\in {\bf T}$ such that $e^{-i\lambda_{\epsilon}}=\mu e^{-i\pi \gamma \eta^{(1)} \epsilon}\in S^1$ for one of $\mu_{\epsilon} \in \widetilde{\sigma}_p^{(0)} (E(\kappa(\epsilon)),\mu,\mu^{(1)})$.
Note that $e^{-i\lambda_{\epsilon}} \to \mu$ as $\epsilon \downarrow 0$.
By the assumption and the asymptotic behavior (\ref{S5_eq_puiseux02}), we have 
\begin{gather}
e^{-i\lambda_{\epsilon}} - \mu' _{\epsilon}  = \left\{
\begin{split}
o(\epsilon^2) & , \quad \mu'_{\epsilon} = \mu _{\epsilon}\in \widetilde{\sigma}_p^{(0)} (E(\kappa(\epsilon)),\mu,\mu^{(1)}) ,\\
c_2 \epsilon^2 +o(\epsilon^2) &, \quad \mu'_{\epsilon} \in \sigma_p (E(\kappa(\epsilon)),\mu,\mu^{(1)}) \setminus  \widetilde{\sigma}_p^{(0)} (E(\kappa(\epsilon)),\mu,\mu^{(1)})  , \\
c_1 \epsilon +O(\epsilon^2) &, \quad  \mu'_{\epsilon} \in \sigma_p (E(\kappa(\epsilon)),\mu) \setminus  \sigma_p (E(\kappa(\epsilon)),\mu,\mu^{(1)}),  
\end{split}
\right.
\label{S7_eq_diverge00}
\end{gather} 
for some nonzero constants $c_1,c_2$. 
The asymptotic behavior (\ref{S7_eq_puiseux003}) and Lemma \ref{S7_lem_asymptoticPEPs} imply that every term of $ \alpha_{\epsilon,j}^{\sharp} (\lambda)-U_0 \varphi_0 | _{a=a_{j,1}^{\sharp}}$ satisfies
$$
 -\pi^2 \epsilon^2 \sum_{s=0}^{m(\mu'_{\epsilon})-1} \left. (e^{-i\lambda}-\mu'_{\epsilon})^{-s-1} U_0^{(1)} \chi_{\mathsf{int}} \left( (\pi^2 \mu^{(2)}\epsilon^2 )^s P _{\mu^{(2)}}^{(2)} +O(\epsilon^{2s+1}) \right) \chi_{\mathsf{int}}^* U_0^{(1)} \varphi_0 \right| _{a=a_{j,1}^{\sharp}} +O(\epsilon).
$$
The coefficient of every term on the right-hand side is of the form
$$
c_3 (e^{-i\lambda}-\mu'_{\epsilon})^{-s-1} \epsilon^{2s+2}, \quad c_3 \not = 0.
$$
The asymptotic behavior (\ref{S7_eq_diverge00}) implies that $|e^{-i\lambda}-\mu'_{\epsilon}|^{-s-1} \epsilon^{2s+2} $ for $ \mu' _{\epsilon}\in \widetilde{\sigma}_p^{(0)} (E(\kappa(\epsilon)),\mu,\mu^{(1)}) $ diverges to $\infty$ as $\epsilon \downarrow 0$ as well as other terms are bounded or converge to zero.
However, the scattering matrix $\Sigma_{\epsilon}(\lambda)$ is unitary for all $\lambda \in {\bf T}$ (see also Proposition \ref{S4_prop_uniqueness_data}).
Then the above argument is contradictory.
We obtain $\mu \gamma \eta^{(1)} (\gamma \eta^{(1)}+1)-2\mu^{(2)} \not= 0$
\qed

\medskip

Now let us show the main results under Assumptions 1-3.
The non-resonant scattering is as follows.
\begin{theorem}

For $e^{-i\lambda} \in S^1 \setminus \sigma_p (E_0)$, we have $\Sigma_{\epsilon} (\lambda)=I_N +O(\epsilon)$ as $\epsilon\downarrow 0$.

\label{S5_thm_nonresonantscattering}
\end{theorem}

Proof.
Due to (\ref{S5_eq_asymptotic_evmukappa001}), we have 
\begin{gather*}
\begin{split}
& \alpha_{\epsilon,j}^{\sharp} (\lambda)-U_0 \varphi_0 |_{a=a_{j,1}^{\sharp}} \\
 & =\left. \sum_{\mu_{\epsilon} \in \sigma_p (E_{\epsilon}) \setminus S^1 }  \sum_{s=0}^{m(\mu_{\epsilon})-1} U_{\epsilon} \chi_{\mathsf{int}}^* \left( \frac{  P_{\mu _{\epsilon}}  (E_{\epsilon} -\mu_{\epsilon} )^s  P_{\mu _{\epsilon}}  }{(e^{-i\lambda} -\mu+i\pi \epsilon \mu^{(1)}+O(\epsilon^2 ) )^{s+1}} \right) \chi_{\mathsf{int}} U_{\epsilon} \varphi_0    \right| _{a=a_{j,1}^{\sharp}} +O(\epsilon) .
\end{split}
\end{gather*}
We have $ \lim_{\epsilon \downarrow 0 } P_{_{\epsilon}} = P_{\mu^{(2)}}^{(2)} $ in view of Proposition \ref{Ap_prop_asymptotic2ndorder} and Assumption 3.
Due to the assumption $e^{-i\lambda} \in S^1 \setminus \sigma_p (E_0)$, it follows that $(e^{-i\lambda} -\mu+i\pi \epsilon \mu^{(1)}+O(\epsilon^2 ))^{-1}$ is bounded for $\epsilon >0$.
We also have $U_{\epsilon} \chi_{\mathsf{int}}^* P_{_{\epsilon}} (E_{\epsilon} -\mu_{\epsilon} )^s P_{_{\epsilon}} \chi_{\mathsf{int}} U_{\epsilon} \varphi_0 | _{a=a_{j,1}^{\sharp}} =O(\epsilon^2 )$ from (\ref{S7_eq_puiseux003}) and $P_{\mu^{(2)}}^{(2)} (E_0 - \mu )=(E_0 -\mu)P_{\mu^{(2)}}^{(2)}=0$.
Then we have proved $\alpha_{\epsilon,j}^{\sharp} (\lambda) -U_0 \varphi_0 |_{a=a_{j,1}^{\sharp}} =O(\epsilon)$.
\qed

\medskip

On the other hand, the resonant scattering for $U_{\epsilon}$ is as follows.
\begin{theorem}

Take a resonance $\mu_{\epsilon} \in \sigma_p (  E(\kappa(\epsilon)) , \mu , \mu ^{(1)}  )$ with $\mu^{(1)} = \mu \gamma  \eta ^{(1)} $.
Let $e^{-i\lambda_{\epsilon}} = \mu e^{-i\pi \gamma \eta^{(1)} \epsilon  } \in S^1 $.
We have $ \Sigma_{\epsilon} (\lambda_{\epsilon})=I_N + \Sigma _0^{(1)} (\mu )+  O(\epsilon)$ as $\epsilon \downarrow 0$ where 
\begin{gather*}
\begin{split}
&\Sigma_0^{(1)} (\mu) \alpha ^{\flat} | _{a=a_{j,1}^{\sharp}}  = \sum_{\mu^{(1)} \in \sigma_p (\widetilde{E}_{\mu}^{(1)}|_{\mathsf{R} (\mu)})} \left( \sum_{\mu^{(2)} \in \sigma_p \big( \widetilde{E}_{\mu^{(1)}}^{(2)}|_{\mathsf{R} (\mu^{(1)})} \big)}  \widetilde{\Sigma}_0^{(1)} (\mu,\mu^{(1)},\mu^{(2)})   \right) ,\\
& \widetilde{\Sigma}_0^{(1)} (\mu,\mu^{(1)},\mu^{(2)})= \frac{2(1-\rho (\mu^{(1)},\mu^{(2)})^{m(\mu^{(2)})} )}{\mu \gamma \eta^{(1)} ( \gamma \eta^{(1)} + 1)} \left.  U_{0}^{(1)} \chi_{\mathsf{int}}^* P_{\mu^{(2)}}^{(2)} \chi_{\mathsf{int}} U_{0}^{(1)} \varphi_0  \right| _{a=a_{j,1}^{\sharp}} , 
\end{split}
\end{gather*}
where the sum on the right-hand side is taken over $\mu^{(2)}$ such that $P_{\mu^{(2)}}^{(2)}$ satisfies $\mathsf{Ran} \big( P_{\mu^{(2)}}^{(2)} \big) \subset  \mathcal{H}_1 (\mu)$, and we put
$$
 \rho (\mu^{(1)},\mu^{(2)}) = -\frac{2\mu^{(2)}}{\mu \gamma \eta^{(1)} ( \gamma \eta^{(1)} + 1)-2\mu^{(2)}} .
$$

\label{S5_thm_resonantscattering}
\end{theorem}

Proof.
Recalling Assumption 3, we often use $\mathsf{Ran} (P_{\mu_{\epsilon} }) =  \mathsf{Ran} \big( P_{\mu^{(2)}}^{(2)} ( \kappa(\epsilon) ) \big) $.
 As has been seen in the proof of Theorem \ref{S5_thm_nonresonantscattering}, every term of $\alpha_{\epsilon,j}^{\sharp} (\lambda_{\epsilon} ) -U_0 \varphi_0 |_{a=a_{j,1}^{\sharp}}  $ corresponding to $\mu'_{\epsilon} \in \sigma_p (E_{\epsilon}) \setminus \sigma_p (E(\kappa(\epsilon),\mu)$ is negligible as $\epsilon \downarrow 0 $.
Then we have 
\begin{gather*}
\begin{split}
& \alpha_{\epsilon,j}^{\sharp} (\lambda_{\epsilon} ) -U_0 \varphi_0 |_{a=a_{j,1}^{\sharp}} \\
&=   \sum _{\mu'_{\epsilon} \in \sigma_p ( E(\kappa(\epsilon)) , \mu ) } \sum _{s=0}^{m(\mu'_{\epsilon})-1} \left. U_{\epsilon} \chi_{\mathsf{int}}^* \left( \frac{  P_{\mu'_{\epsilon}} (E_{\epsilon} -\mu'_{\epsilon})^s P_{\mu'_{\epsilon}}   }{(e^{-i\lambda_{\epsilon}} -\mu'_{\epsilon})^{s+1} } \right) \chi_{\mathsf{int}} U_{\epsilon} \varphi_0 \right| _{a=a_{j,1}^{\sharp}} +O(\epsilon ) ,
\end{split}
\end{gather*} 
as $\epsilon \downarrow 0 $.
For $\mu'_{\epsilon} \in \sigma_p (E(\kappa(\epsilon)),\mu ) \setminus \sigma_p ( E(\kappa(\epsilon)),\mu , \mu^{(1)})$, we can see that 
\begin{gather*}  
\begin{split}
\left. U_{\epsilon} \chi_{\mathsf{int}}^* \left( \frac{  P_{\mu'_{\epsilon}} (E_{\epsilon} -\mu'_{\epsilon})^s P_{\mu'_{\epsilon}}   }{(e^{-i\lambda_{\epsilon}} -\mu'_{\epsilon})^{s+1} } \right) \chi_{\mathsf{int}} U_{\epsilon} \varphi_0 \right| _{a=a_{j,1}^{\sharp}}  &= \frac{O(\epsilon^2 )}{i\pi \mu \epsilon (\gamma' \zeta^{(1)} - \gamma \eta^{(1)})/2 +O(\epsilon^2 )} \\
&= O(\epsilon ),
\end{split}
\end{gather*}
where $\mu'_{\epsilon} = \mu + \kappa \gamma' \mu \zeta^{(1)} +O(\kappa^2)$, $\kappa = 1-e^{i\pi\epsilon}$, with $\gamma' \zeta^{(1)} \not= \gamma \eta^{(1)}$.
Then we have 
\begin{gather}
\begin{split}
&\alpha_{\epsilon,j}^{\sharp} (\lambda_{\epsilon} ) -U_0 \varphi_0 |_{a=a_{j,1}^{\sharp}} \\
&=   \sum _{\mu'_{\epsilon} \in \sigma_p ( E_{\epsilon}, \mu , \mu^{(1)} ) } \sum _{s=0}^{m(\mu'_{\epsilon})-1} \left. U_{\epsilon} \chi_{\mathsf{int}}^* \left( \frac{  P_{\mu'_{\epsilon}} (E_{\epsilon} -\mu'_{\epsilon})^s P_{\mu'_{\epsilon}}   }{(e^{-i\lambda_{\epsilon}} -\mu'_{\epsilon})^{s+1} } \right) \chi_{\mathsf{int}} U_{\epsilon} \varphi_0 \right| _{a=a_{j,1}^{\sharp}} +O(\epsilon ) ,
\end{split}
\label{S5_eq_asymptoticproof000}
\end{gather} 

Now we note that
$$
\sigma_p ( E(\kappa(\epsilon)),\mu,\mu^{(1)})= \bigsqcup _{\mu^{(2)}\in \sigma_p \big( \widetilde{E}_{\mu^{(1)}} ^{(2)} | _{\mathsf{R} (\mu^{(1)}) } \big) } \sigma_p ( E(\kappa(\epsilon)),\mu,\mu^{(1)},\mu^{(2)}).
$$
In view of (\ref{S5_eq_puiseux02}), we have for $\mu'_{\epsilon}\in \sigma_p ( E(\kappa(\epsilon)),\mu,\mu^{(1)},\mu^{(2)})$
\begin{gather}
\begin{split}
( e^{-i\lambda_{\epsilon}} - \mu'_{\epsilon} )^{s+1} =&\, \left( -\frac{ \pi^2 \epsilon^2}{2} \left(   \mu \gamma \eta^{(1)} ( \gamma \eta^{(1)} +1) -2 \mu^{(2)} \right) +o(\epsilon^2 ) \right)^{s+1} \\
=&\, \left( -\frac{\pi^2 \epsilon^2}{2} \left(  \mu \gamma \eta^{(1)} ( \gamma \eta^{(1)} +1) -2 \mu^{(2)} \right)  \right)^{s+1} +o(\epsilon^{2s+2} ),
\end{split}
\label{S7_eq_asymptotic00main}
\end{gather} 
and $\mu \gamma \eta^{(1)} (\gamma \eta^{(1)} +1) -2 \mu^{(2)}$ does not vanish due to Proposition \ref{S7_prop_reductionassumption}.
The asymptotic behaviors (\ref{S7_eq_puiseux003}), (\ref{S7_eq_asymptotic00main}) and Lemma \ref{S7_lem_asymptoticPEPs} imply the expansion formula
\begin{gather}
\begin{split}
&\left. U_{\epsilon} \chi_{\mathsf{int}}^* \left( \frac{  P_{\mu'_{\epsilon}} (E_{\epsilon} -\mu'_{\epsilon})^s P_{\mu'_{\epsilon}}   }{(e^{-i\lambda_{\epsilon}} -\mu'_{\epsilon})^{s+1} } \right) \chi_{\mathsf{int}} U_{\epsilon} \varphi_0 \right| _{a=a_{j,1}^{\sharp}} \\
&= \left. - U_0^{(1)} \chi_{\mathsf{int}} \left(  \frac{(-2)^{s+1} (\mu^{(2)} )^s P _{\mu^{(2)}}^{(2)} +O(\epsilon)}{( \mu \gamma \eta^{(1)} (\gamma \eta^{(1)} +1) -2 \mu^{(2)}  )^{s+1} +o(1 )}  \right)  \chi_{\mathsf{int}}^* U_0^{(1)} \varphi_0 \right| _{a=a_{j,1}^{\sharp}} +O(\epsilon)\\
&=\left. - U_0^{(1)} \chi_{\mathsf{int}} \left(  \frac{(-2)^{s+1} (\mu^{(2)} )^s P _{\mu^{(2)}}^{(2)} }{( \mu \gamma \eta^{(1)} (\gamma \eta^{(1)} +1) -2 \mu^{(2)}  )^{s+1} }  \right)  \chi_{\mathsf{int}}^* U_0^{(1)} \varphi_0 \right| _{a=a_{j,1}^{\sharp}} +O(\epsilon),
\end{split}
\label{S7_eq_computemain0001}
\end{gather}
for $0\leq s\leq m(\mu'_{\epsilon})-1=m(\mu^{(2)})-1$ with $\mu'_{\epsilon}\in \sigma_p ( E(\kappa(\epsilon)),\mu,\mu^{(1)},\mu^{(2)})$.
 We also have
$$
\sum_{s=0}^{m(\mu^{(2)})-1} \frac{(-2)^{s+1} (\mu^{(2)} )^s}{( \mu \gamma \eta^{(1)} (\gamma \eta^{(1)} +1) -2 \mu^{(2)}  )^{s+1}}= \frac{-2( 1-\rho (\mu^{(1)},\mu^{(2)} )^{m(\mu^{(2)})})}{\mu \gamma \eta^{(1)} (\gamma \eta^{(1)}+1)} .
$$
Plugging this equality into (\ref{S7_eq_computemain0001}), we obtain
\begin{gather}
\begin{split}
&\left. U_{\epsilon} \chi_{\mathsf{int}}^* \left( \frac{  P_{\mu'_{\epsilon}} (E_{\epsilon} -\mu'_{\epsilon})^s P_{\mu'_{\epsilon}}   }{(e^{-i\lambda_{\epsilon}} -\mu'_{\epsilon})^{s+1} } \right) \chi_{\mathsf{int}} U_{\epsilon} \varphi_0 \right| _{a=a_{j,1}^{\sharp}} \\
&=\frac{2( 1-\rho (\mu^{(1)},\mu^{(2)} )^{m(\mu^{(2)})})}{\mu \gamma \eta^{(1)} (\gamma \eta^{(1)}+1)} \left. U_0^{(1)} \chi_{\mathsf{int}} P_{\mu^{(2)}}^{(2)} \chi_{\mathsf{int}}^* U_0^{(1)} \varphi_0 \right| _{a=a_{j,1}^{\sharp}}+O(\epsilon).
\end{split}
\label{S7_eq_computemain0002}
\end{gather}
The equalities (\ref{S5_eq_asymptoticproof000}) and (\ref{S7_eq_computemain0002}) imply the theorem.
\qed

\section{Examples} \label{section_numerical}
\subsection{Cycles and complete graphs}
\begin{figure}[b]
\centering
\includegraphics[width=4cm, bb=0 0 222 121]{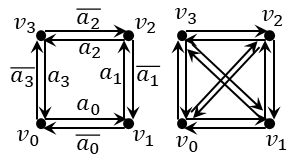}
\caption{Left: $C_4$, the cycle with four vertices. / Right: $K_4$, the complete graph with four vertices.}
\label{fig_cycle4}
\end{figure}
 Theorem \ref{S5_thm_resonantscattering} implies that the resonances of $U_{\epsilon}$ near $\sigma_{ess} (U_{\epsilon})$ contributes to an anomalous scattering at resonant energies. 
Finally, let us show some numerical visualizations of the resonant scattering in typical examples $\Gamma_{\mathsf{int}} = C_4 $ which is the cycle with its length $4$ or $\Gamma_{\mathsf{int}} =K_4$ which is the complete graph with four vertices (see Figure \ref{fig_cycle4}).

For $C_4$, let us list the spectral properties of $E_0$ as follows.
Namely, we put for $C_4 = (V_{\mathsf{int}} ,A_{\mathsf{int}} )$ with
\begin{gather*}
\begin{split}
&V_{\mathsf{int}} = \{ v_0 , v_1 , v_2 , v_3 \} , \\
&A_{\mathsf{int}} = \{ a_0 = (v_0 ,v_1 ), \, a_1 = (v_1 ,v_2 ) , \, a_2 = (v_2 ,v_3 ) , \, a_3 = (v_3 ,v_0), \overline{a}_0 ,  \overline{a}_1 ,  \overline{a}_2 , \overline{a}_3  \} .
\end{split}
\end{gather*}
In this case, letting $u=[u(a_0),u(a_1),u(a_2),u(a_3), u(\overline{a}_0),u(\overline{a}_1),u(\overline{a}_2),u(\overline{a}_3)]^{\mathsf{T}} \in {\bf C}^{A_{\mathsf{int}}}$, we have
$$
E_0 u= \begin{bmatrix} Q_4 & 0 \\ 0 & Q_4^* \end{bmatrix} u ,  \quad
Q_4 = \begin{bmatrix} 0 & 0 & 0 & 1 \\ 1 & 0 & 0 & 0 \\ 0 & 1 & 0 & 0 \\ 0 & 0 & 1 & 0 \end{bmatrix} .
$$
Therefore, we can see $\sigma_p (E_0)= \{ \pm 1 , \pm i \} $ and every eigenvalue has the geometric multiplicity $2$.
The eigenspaces are given as follows:
\begin{gather}
\begin{split}
&\mathsf{Ker} (E_0 \mp 1)= \mathsf{Span} \left( \frac{1}{2} [1,\pm 1,1,\pm 1,0,0,0,0] ^{\mathsf{T}} , \frac{1}{2} [0,0,0,0,1,\pm 1,1,\pm 1] ^{\mathsf{T}} \right) , \\
&\mathsf{Ker} (E_0 \mp i) = \mathsf{Span} \left( \frac{1}{2} [1,\mp i,-1,\pm i,0,0,0,0] ^{\mathsf{T}} , \frac{1}{2} [0,0,0,0,1,\pm i,-1,\mp i] ^{\mathsf{T}} \right) .
\end{split}
\label{example_eq_es}
\end{gather}

The corresponding Laplacian $T$ is given by 
$$
Tf(v_j )= \frac{1}{2} \left( f(v_{j+1})+f(v_{j-1})  \right) , \quad f\in {\bf C}^{V_{\mathsf{int}}} ,
$$
under the periodic condition.
We can see that $\sigma_p (T)=\{ 0, \pm 1 \}$ and 
\begin{gather}
\begin{split}
&\mathsf{Ker} (T)= \mathsf{Span} \left( \frac{1}{2} [1,0,-1,0]^{\mathsf{T}}, \frac{1}{2} [0,1,0,-1]^{\mathsf{T}} \right) , \\
&\mathsf{Ker} (T\mp 1)= \mathsf{Span} \left( \frac{1}{2\sqrt{2}} [1,\pm 1,  1, \pm 1]^{\mathsf{T}} \right) .
\end{split}
\label{S8_eq_esT}
\end{gather}
In view of (\ref{example_eq_es}) and (\ref{S8_eq_esT}), we obtain 
\begin{gather}
\mathsf{Ker} (E_0|_{\mathcal{Y}} \mp 1)= \mathsf{Span} \left(  \frac{1}{2\sqrt{2}} [1,\pm 1,  1,\pm 1, \pm 1,1,\pm 1,1]^{\mathsf{T}}  \right) , \label{S8_eq_nonperesE0} \\
\mathsf{Ker} (E_0|_{\mathcal{Y}^{\perp}} \mp 1)= \mathsf{Span} \left(  \frac{1}{2\sqrt{2}} [1,\pm 1,  1,\pm 1, \mp 1,-1,\mp 1,-1]^{\mathsf{T}}  \right) . \label{S8_eq_peresE0}
\end{gather}
Also for the case $\Gamma_{\mathsf{int}} = K_4$, we can see explicit representations of the spectrum and the eigenspaces.
We omit the details.


\subsection{With some tails}
\begin{figure}[b]
\centering
\includegraphics[width=8cm, bb=0 0 426 176]{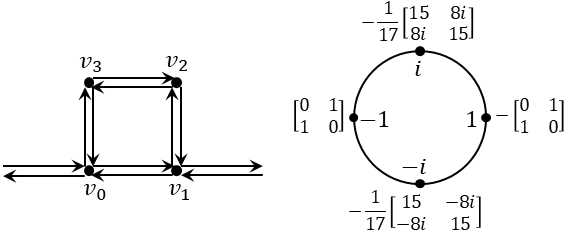}
\caption{The examples of $I_2 + \Sigma_0^{(1)} (\mu)$ as in Theorem \ref{S5_thm_resonantscattering} for the case $C_4$ with two tails.
Note that there are resonant energies $\pm 1$ and $\pm i$.
For $\pm 1$, we can see that the resonant tunneling effect occurs.}
\label{fig_vis_scatteringmatrix_2tails}
\end{figure}
\begin{figure}[t]
\centering
\includegraphics[width=10cm, bb=0 0 530 281]{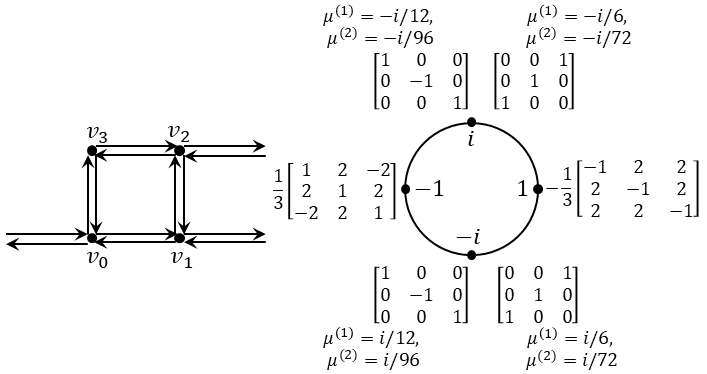}
\caption{The examples of $I_3 + \Sigma_0^{(1)} (\mu)$ as in Theorem \ref{S5_thm_resonantscattering} for the case $C_4$ with three tails.
Note that there are resonant energies $\pm 1$ and $\pm i$.
For $(\mu,\mu^{(1)},\mu^{(2)})=(\pm i , \mp i /6 , \mp i /72)$, we can see that the resonant tunneling effect occurs.}
\label{fig_vis_scatteringmatrix_3tails}
\end{figure}

Before showing numerical results, we derive two concrete examples of the resonant scatterings for $C_4 $ with two tails and three tails.
Note that $\mathcal{V}_{per} (\pm 1)=\mathcal{V}_{per} (\pm i )= \emptyset $ in view of (\ref{S8_eq_esT}).
In order to do this, we need to compute $\mu^{(1)}$ and $\mu^{(2)} $ in the application of Theorem \ref{S5_thm_resonantscattering}.
Actually, in the similar way of Lemma \ref{S5_lem_semisimpleE0(1)}, we have the matrix of the operator
$$
P_{\mu^{(2)}}^{(2)} \widetilde{E}_{\mu^{(1)}}^{(2)} P_{\mu^{(2)}}^{(2)} = \sum_{\zeta\in \sigma_p (E_0)\setminus \{ \mu \}} (\mu - \zeta )^{-1} P_{\mu^{(2)}}^{(2)} E_0^{(1)} P_{\zeta,1} E_0^{(1)} P_{\mu^{(2)}}^{(2)}.
$$
Let $ \{ f_1^{\mu} , \ldots,f_{s(\mu)}^{\mu} \} $ be an orthonormal system of $\mathsf{Ker} (T-\Phi_{QW} (\mu)) $ such that $\{ \partial_{\mu}^* f_1 ^{\mu} , \ldots , \partial_{\mu}^* f_{s(\mu)}^{\mu} \}$ is an orthonormal basis of $\mathsf{Ran} \big( P_{\mu^{(2)}}^{(2)} \big)$.
We take an orthonormal basis $\{ g_1^{\zeta} , \ldots, g_{p(\zeta)}^{\zeta} \} $ of $\mathsf{Ker} (T-\Phi_{QW} (\zeta))$.
The matrix of $P_{\mu^{(2)}}^{(2)} \widetilde{E}_{\mu^{(1)}}^{(2)} P_{\mu^{(2)}}^{(2)} $ on $\mathsf{Ran} \big( P_{\mu^{(2)}}^{(2)} \big) $ is given by 
$$
\sum_{\zeta\in \sigma_p (E_0)\setminus \{ \mu \}} (\mu - \zeta )^{-1} \omega_{\mu} \omega_{\zeta} \omega_{\mu}^{\dag} \omega_{\zeta}^{\dag} \big( M_{\zeta,\mu}^{(2)} \big)^* M_{\zeta,\mu}^{(2)} ,
$$
where 
\begin{gather*}
\omega_{\mu} = \left\{
\begin{split}
\frac{i\mu \sin (\mathrm{arg} \, \mu )}{\sqrt{2} |\sin (\mathrm{arg} \, \mu)|} &, \quad \mu\in S^1 \setminus \{ \pm 1 \} , \\
-1 &, \quad \mu \in \{ \pm 1 \} ,
\end{split}
\right. \quad 
\omega_{\mu} ^{\dag} = \left\{
\begin{split}
\frac{i \sin (\mathrm{arg} \, \mu )}{\sqrt{2} |\sin (\mathrm{arg} \, \mu)|} &, \quad \mu\in S^1 \setminus \{ \pm 1 \} , \\
\mu &, \quad \mu \in \{ \pm 1 \} ,
\end{split}
\right.
\end{gather*}
and
$$
M_{\zeta,\mu}^{(2)} = \left[ \langle Bf_k^{\mu} , g_j^{\zeta} \rangle _{\ell^2 _{n_{\mathsf{i}}}(V_{\mathsf{int}})} \right] _{\substack{1\leq j\leq p(\zeta),\\ 1\leq k\leq s(\mu) }} .
$$
For cases of two tails (see Figure \ref{fig_vis_scatteringmatrix_2tails}) and three tails (see Figure \ref{fig_vis_scatteringmatrix_3tails}), we can get the lists of eigenvalues as in Tables \ref{table_eigenvalues} and \ref{table_eigenvalues2}, respectively.

\begin{table}[t]
\centering
\begin{tabular}{rcrrrr} \hline
   $\mu$ & Multi. (non-per.) & $\mu^{(1)}$ & $\gamma$ & $\eta^{(1)}$ & $\mu^{(2)}$ \\ \hline
   $1$ & $1$ & $-1/6$ & $1$ & $-1/6$ & $-1/72$ \\
   $i$ & $2$ & $-i/12$ & $1/2$ & $-1/6$ & $-(3i \pm 2)/(4\cdot 72) $ \\ 
   $-1$ & $1$  & $1/6$ & $1/2$ & $-1/6$  & $1/72$ \\
   $-i$ & $2$ & $i/12$ & $1$ & $-1/6$ & $(3i \pm 2)/(4\cdot 72)$ \\ \hline
\end{tabular}\\ \vspace*{0.2cm}
\begin{tabular}{rrrr} \hline 
$\mu$ & $\mu^{(1)}$ & $\mu^{(2)}$  & $\mathsf{Ran} \big( P _{\mu^{(2)}}^{(2)} \big)$    \\ \hline 
$\pm 1$ & $\mp 1/6$ & $\mp 1/72$ & $ (2\sqrt{2})^{-1}[1,\pm 1,1, \pm 1]^{\mathsf{T}} $    \\
$\pm i$ & $\mp i/12$ & $\mp (3i+2)/(4\cdot 72)$ & $(2\sqrt{2})^{-1} [1,\mp 1,-1,\pm 1]^{\mathsf{T}}$   \\ 
$\pm i$ & $\mp i/12$ & $\mp (3i-2)/(4\cdot 72)$ & $(2\sqrt{2})^{-1} [1,\pm 1,-1,\mp 1]^{\mathsf{T}}$   \\
 \hline
\end{tabular}
\caption{(Upper table) The list of the triple $(\mu,\mu^{(1)},\mu^{(2)})$ for non-persistent eigenvalues of $C_4$ with two tails (Figure \ref{fig_vis_scatteringmatrix_2tails}) and its multiplicities. 
The eigenvalues $\pm i$ split into two eigenvalues $\mu(\kappa)$ (see $\mu^{(2)}$). / 
(Lower table) The list of eigenvectors in $\mathsf{Ran} (T-\Phi_{QW} (\mu))$ which generate $\mathsf{Ran} \big( P _{\mu^{(2)}}^{(2)} \big)$ passing through $\partial_{\mu}^*$ for $C_4$ with two tails (Figure \ref{fig_vis_scatteringmatrix_2tails}). }
\label{table_eigenvalues}
\end{table}

\begin{table}[t]
\centering
\begin{tabular}{rcrrrr} \hline
   $\mu$ & Multi. (non-per.) & $(\mu^{(1)},\mu^{(2)})$ & $\gamma$ & $\eta^{(1)}$  \\ \hline
   $1$ & $1$ & $(-1/4,-1/96)$ & $1$ & $-1/4$  \\
   $i$ & $2$ & $(-i/6,-i/72)$, $(-i/12,-i/96)$ & $1/2$ & $-1/3$, $-1/6 $ \\ 
   $-1$ & $1$  & $(1/4,1/96)$ & $1/2$ & $-1/4$  &  \\
   $-i$ & $2$ & $(i/6,i/72)$, $(i/12,i/96)$ & $1$ & $-1/3$, $-1/6 $  \\ \hline
\end{tabular} \\ \vspace*{0.2cm}
\begin{tabular}{rrrr} \hline
$\mu$ & $\mu^{(1)}$ & $\mu^{(2)}$  & $\mathsf{Ran} \big( P _{\mu^{(2)}}^{(2)} \big) $   \\ \hline
$\pm 1$ & $\mp 1/4$ & $\mp 1/96$ & $ (2\sqrt{2})^{-1}[1,\pm 1,1, \pm 1]^{\mathsf{T}}$    \\
$\pm i$ & $\mp i/6$ & $ \mp i/72$ & $2^{-1} [1,0,-1,0]^{\mathsf{T}}$   \\ 
$\pm i$ & $\mp i/12$ & $\mp i/96$ & $2^{-1} [0,1,0,- 1]^{\mathsf{T}}$   \\
 \hline
\end{tabular}
\caption{(Upper table) The list of the triple $(\mu,\mu^{(1)},\mu^{(2)})$ for non-persistent eigenvalues of $C_4$ with three tails (Figure \ref{fig_vis_scatteringmatrix_3tails}) and its multiplicities. 
The eigenvalues $\pm i$ split into two eigenvalues $\mu(\kappa)$ (see $\mu^{(1)}$ and $\mu^{(2)}$). /
(Lower table) The list of eigenvectors in $ \mathsf{Ker} ( T-\Phi_{QW} (\mu )) $ which generate $\mathsf{Ran} \big( P _{\mu^{(2)}}^{(2)} \big)$ passing through $\partial_{\mu}^*$ for $C_4$ with three tails (Figure \ref{fig_vis_scatteringmatrix_3tails}). }
\label{table_eigenvalues2}
\end{table}

Let us compute the scattering matrix in the sense of $I_N + \Sigma_0^{(1)} (\mu) $.
We take the standard basis $\{ {\bf e}_1 , \ldots, {\bf e}_N \}$ of ${\bf C}^N$.
In order to compute the $(j,k)$-entry of $\Sigma_0^{(1)} (\mu)$, we consider $\langle \Sigma_0^{(1)} (\mu) {\bf e}_j , {\bf e}_k \rangle_{{\bf C}^N}$.
Now $\iota^{\flat}:{\bf C}^N \to \ell^2 (A)$ and $\iota ^{\sharp} :{\bf C}^N \to \ell^2 (A)$ are defined by 
\begin{gather*}
\iota^{\flat} {\bf u} (a)= \left\{
\begin{split}
u_j &, \quad a=a_{j,0}^{\flat},\\
0&, \quad \text{otherwise},
\end{split}
\right. \quad 
\iota^{\sharp} {\bf u} (a)= \left\{
\begin{split}
u_j &, \quad a=a_{j,1}^{\sharp},\\
0&, \quad \text{otherwise},
\end{split}
\right.
\end{gather*}
for ${\bf u} =[u_1,\ldots,u_N ]^{\mathsf{T}} \in {\bf C}^N $.
Without loss of generality, it is sufficient to consider 
\begin{gather*}
\begin{split}
\left\langle (\iota^{\sharp})^* U_0^{(1)} \chi_{\mathsf{int}}^* P_{\mu^{(2)}}^{(2)} \chi_{\mathsf{int}} U_0^{(1)} \iota^{\flat} {\bf e}_j , {\bf e}_k \right\rangle _{{\bf C}^N} 
&=\left\langle  P_{\mu^{(2)}}^{(2)} SC_0^{(1)}  \iota^{\flat} {\bf e}_j , C_0^{(1)} \iota^{\flat}{\bf e}_k \right\rangle _{\ell^2 (A_{\mathsf{int}})} \\
&= \sum_{l=1}^{s(\mu)} \left\langle SC_0^{(1)} \iota^{\flat} {\bf e}_j , \partial_{\mu}^* f_l^{\mu} \right\rangle _{\ell^2 (A_{\mathsf{int}})} \left\langle \partial _{\mu}^* f_l^{\mu},C_0^{(1)} \iota^{\flat} {\bf e}_k \right\rangle _{\ell^2 (A_{\mathsf{int}})} \\
&= \mu \gamma  \frac{n_{\mathsf{i}} (v_{j,0}) }{n (v_{j,0}) } \frac{n_{\mathsf{i}} (v_{k,0}) }{n (v_{k,0}) } \sum_{l=1}^{s(\mu)} \overline{f_l^{\mu} (v_{j,0})} f_l^{\mu} (v_{k,0}),
\end{split}
\end{gather*}
by using (\ref{S5_eq_perturbedqws}).
For concrete cases, we use an orthonormal system $\{ f_1^{\mu} , \ldots, f_{s(\mu)}^{\mu} \}$ of $\mathsf{Ker} (T-\Phi_{QW} (\mu))$ which generates $\mathsf{Ran} \big( P_{\mu^{(2)}}^{(2)} \big)$ passing through $\partial_{\mu}^*$.
See Tables \ref{table_eigenvalues} and \ref{table_eigenvalues2}.

The explicit results of $I_N + \Sigma_0^{(1)} (\mu)$ for resonant energies $\{ \pm 1 , \pm i \} $ are given in Figures \ref{fig_vis_scatteringmatrix_2tails} and \ref{fig_vis_scatteringmatrix_3tails}.
For the case of two tails, we can see that the resonant tunneling effect occurs at $\mu = \pm 1$.
For the case of three tails, the situation is more complicated than the previous case.
Let $\Gamma _{\mathsf{t},j}$ for $j=1,2,3$ be the tail connected to the vertex $v_{j-1}$.
For example, for $\mu =i$, we take the pattern of the inflow as $\alpha^{\flat} = [1,0,0]^{\mathsf{T}}$, we can observe the resonant tunneling effect. 
On the other hand, for the pattern of inflow $\alpha^{\flat} =[0,1,0]^{\mathsf{T}}$, we see that the complete reflection occurs.

\subsection{Numerical examples}
\begin{figure}[b]
\centering
\includegraphics[width=3.5cm, bb=0 0 202 116]{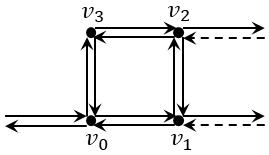} \
\includegraphics[width=3.5cm, bb=0 0 203 114]{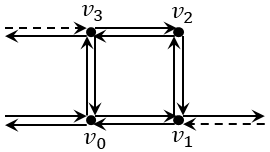} \
\includegraphics[width=3.5cm, bb=0 0 204 116]{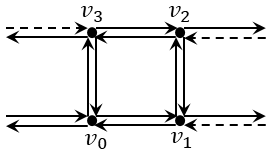}
\caption{$C_4$ with three tails (Left / Middle). 
Suppose that the inflow has its support only on the tail connected to the vertex $v_0$.
There are two types (Left / Middle) in view of symmetry. 
Note that $\sigma _p (E_{\epsilon})$ for every case coincides with each other.
$C_4$ with four tails has a stronger symmetry (Right).
For $K_4$, we also consider the similar situations.}
\label{fig_C4_3tails}
\end{figure}

In the remaining part, we only show numerical examples.
For the visibility of numerical examples, we choose a slightly large parameter $\epsilon =0.25$.
If the inflow vanishes on the incoming tails except for one connected to the vertex $v_0$, there are two types of $\Gamma$ in the sense of symmetry (see Figure \ref{fig_C4_3tails}).
\begin{figure}[b]
\centering
\includegraphics[width=5.5cm, bb=0 0 432 432]{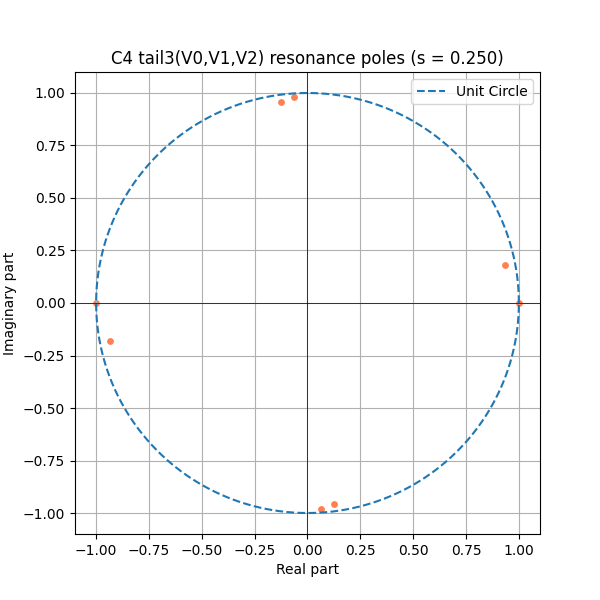
} 
\includegraphics[width=5.5cm, bb=0 0 432 432]{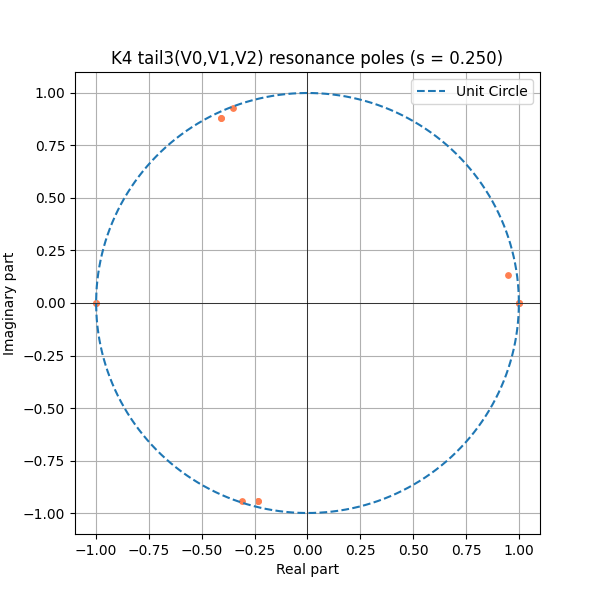} 
\caption{Resonances for $ C_4$ and $K_4 $ with three tails, $\epsilon =0.25$. 
There exist birth eigenvalues $\pm 1$. }
\label{fig_C4_3tails_resonances}
\end{figure}
Figure \ref{fig_C4_3tails_resonances} shows the location of the resonances of $C_4$ and $K_4$ with three tails.
There are birth eigenvalues $\pm 1$ which are persistent with respect to the parameter $\epsilon>0$.
Note that we can see that there are some persistent eigenvalues other than $\pm 1$ for the cases where there is only one tail or there are two tails connected to $v_0$ and $v_2$.

\begin{figure}[t]
\centering
\includegraphics[width=6cm, bb=0 0 420 328]{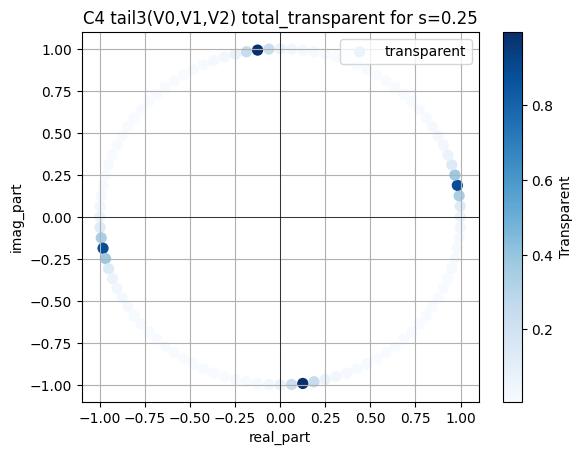
} 
\includegraphics[width=6cm, bb=0 0 420 328]{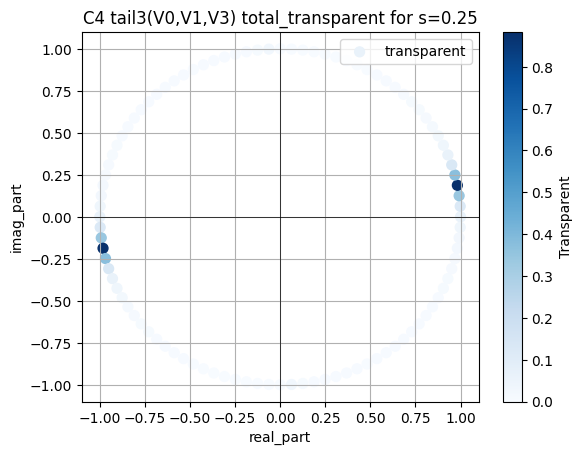
} 
\caption{Left: Total transmission for $C_4$ with three tails, $\epsilon =0.25$ (Left of Figure \ref{fig_C4_3tails}). / Right: Total transmission for $C_4$ with three tails, $\epsilon =0.25$ (Middle of Figure \ref{fig_C4_3tails}). 
Both of these cases, the inflow is taken as $\alpha^{\flat}=[1,0,0]^{\mathsf{T}} $.
}
\label{fig_C4_3tails1_transmission}
\end{figure}
\begin{figure}[t]
\centering
\includegraphics[width=6cm, bb=0 0 420 328]{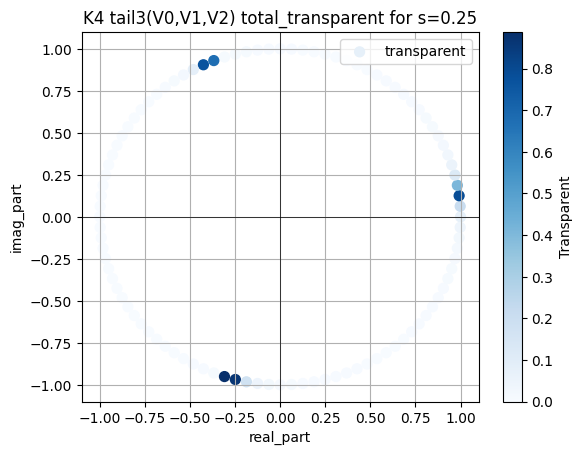
} 
\includegraphics[width=6cm, bb=0 0 420 328]{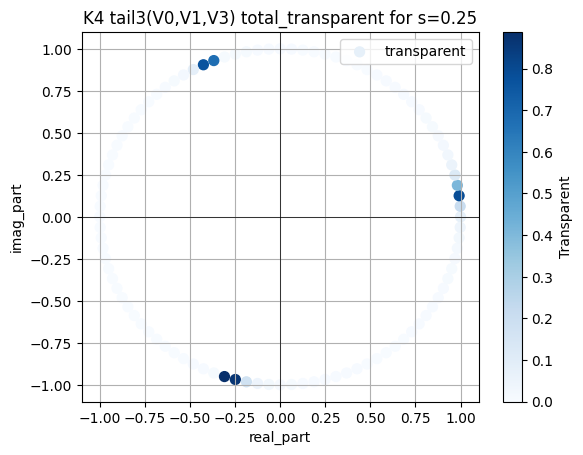
} 
\caption{Left: Total transmission for $K_4$ with three tails, $\epsilon =0.25$ (Left of Figure \ref{fig_C4_3tails}). / Right: Total transmission for $K_4$ with three tails, $\epsilon =0.25$ (Middle of Figure \ref{fig_C4_3tails}). 
Both of these cases, the inflow is taken as $\alpha^{\flat}=[1,0,0]^{\mathsf{T}} $.
}
\label{fig_K4_3tails1_transmission}
\end{figure}
\begin{figure}[t]
\centering
\includegraphics[width=5.5cm, bb=0 0 432 432]{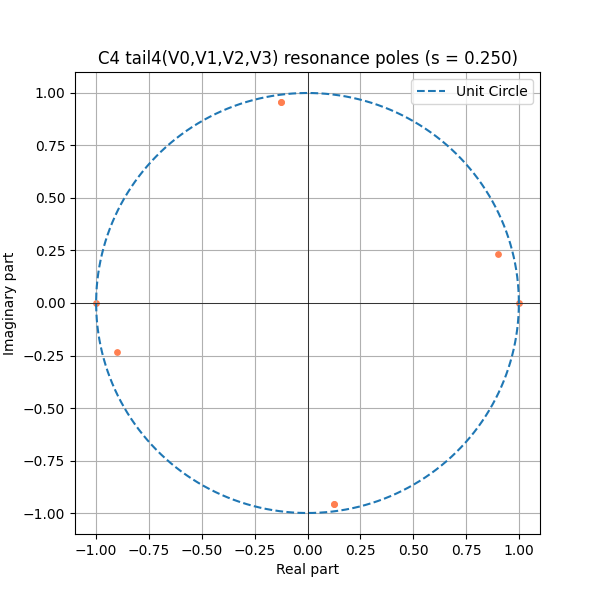
} 
\includegraphics[width=5.5cm, bb=0 0 432 432]{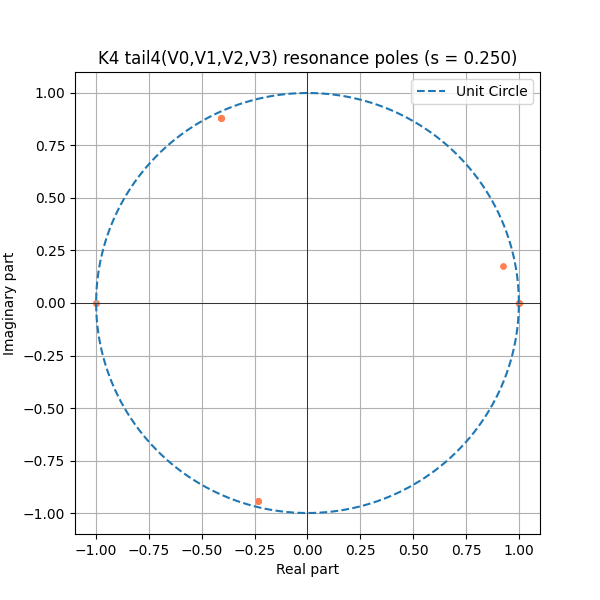} 
\caption{Resonances for $ C_4$ and $K_4 $ with four tails, $\epsilon =0.25$. 
There exist birth eigenvalues $\pm 1$ and degenerate resonances. }
\label{fig_C4_4tails_resonances}
\end{figure}

Figure \ref{fig_C4_3tails1_transmission} shows the ``total transmission" for every $e^{-i\lambda}$.
Here the ``total transmission" is defined by the intensity $|\tau (\lambda)|^2 :=|\alpha_2^{\sharp} (\lambda)|^2 +|\alpha_3^{\sharp} (\lambda)|^2$ if we choose $\alpha_1^{\flat}=1$ and $\alpha_2^{\flat} = \alpha_3^{\flat} =0$ where $\alpha^{\flat}_1 $ is the intensity on $a_{1,0}^{\flat}$ such that $t(a_{1,0}^{\flat})=v_0$.
We can see that $e^{-i\lambda}$ is a resonant energy if $|\tau (\lambda)|\sim 1$.
On the other hand, for non-resonant cases, we see that $|\tau (\lambda)|^2 \sim 0 $.
For the case of Left in Figure \ref{fig_C4_3tails}, we can see the resonant tunneling effect near the resonances of $U_{\epsilon}$.
For the case of Middle in Figure \ref{fig_C4_3tails}, there are resonant tuuneling effects only near $\pm 1$.
Also for $K_4$ with three tails, we can see that our theory is reproduced numerically (Figure \ref{fig_K4_3tails1_transmission}).

The cases of $C_4$ and $K_4$ with four tails have stronger symmetry than the case of three tails.
 As in Figure \ref{fig_C4_4tails_resonances}, we can see some degenerate resonances as well as persistent eigenvalues $\pm 1$ which are birth eigenvalues.

Let us note that, in Figures \ref{fig_C4_3tails1_transmission} and \ref{fig_K4_3tails1_transmission}, the numerical approximations of the scattered wave $\Sigma_{\epsilon} (\lambda) \alpha^{\flat}$ are computed by the time dependent construction (Theorem \ref{S4_prop_stationarystate} or \cite[Theorem 3.1]{HiSe}).

\medskip

{\bf Acknowledgements.}
The authors are supported by the following funds: Grant-in-Aid for JSPS Fellows 22J00430 for KH, JSPS Grant-in-Aid for Scientific Research (C) 24K06761 for HM, and JSPS Grant-in-Aid for Scientific Research (C) 24K06863 for ES.
RI and EY were graduate students of Graduate School of Science and Engineering, Ehime University, and they contributed early studies for this manuscript and visualizations by numerical computations to this paper.
The authors certify that the current affiliations of RI and EY are not involved in this paper.
The authors greatly appreciate the helpful comments by the reviewer for our first manuscript.

\end{document}